\documentclass{article}

\usepackage[verbose=true,letterpaper]{geometry}
\newgeometry{
    textheight=9in,
    textwidth=5.5in,
    top=1in,
    headheight=12pt,
    headsep=25pt,
    footskip=30pt
  }

\usepackage[square,numbers]{natbib}

\usepackage[linesnumbered,ruled,vlined]{algorithm2e}
\SetKwInOut{Input}{Input}
\SetKwInOut{Output}{Output}

\usepackage{graphicx}
\graphicspath{{images/}}
\usepackage{subcaption}
\usepackage[utf8]{inputenc}
\usepackage[T1]{fontenc}
\usepackage{url}
\usepackage{booktabs}
\usepackage{amsfonts}
\usepackage{nicefrac}
\usepackage{microtype}
\usepackage{xcolor}
\usepackage{amsmath}
\usepackage{bm}

\usepackage[hidelinks]{hyperref}
\usepackage{breakurl}

\title{\textbf{Tk-merge: Computationally Efficient Robust Clustering Under General Assumptions}}

\author{
    \vspace*{-1mm}
    \textbf{Luca Insolia} \\
    \vspace*{-1mm}
    \small{Sant’Anna School of Advanced Studies} \\
    \vspace*{-1mm}
    \small{Institute of Economics \& EMbeDS} \\
    \vspace*{-1mm}
    \small{Pisa, 56127, Italy} \\
    \vspace*{-1mm}
    \small{\texttt{Luca.insolia@santannapisa.it}}
    \and
    \vspace*{-1mm}
    \textbf{Domenico Perrotta}\\
    \vspace*{-1mm}
    \small{European Commission}\\
    \vspace*{-1mm}
    \small{Joint Research Centre (JRC)}\\
    \vspace*{-1mm}
    \small{Ispra, 21027, Italy}\\
    \vspace*{-1mm}
    \small{\texttt{domenico.perrotta@ec.europa.eu}}
    }
    
\date{}
    
\begin{document}

\maketitle

\begin{abstract}
    \noindent
    We address general-shaped clustering problems under very weak parametric assumptions with a two-step hybrid robust clustering algorithm based on trimmed k-means and hierarchical agglomeration. 
    The algorithm has low computational complexity and effectively identifies the clusters also in presence of data contamination.
    We also present natural generalizations of the approach as well as an adaptive procedure to estimate the amount of contamination in a data-driven fashion. 
    Our proposal outperforms state-of-the-art robust, model-based methods in our numerical simulations and  real-world applications related to color quantization for image analysis, human mobility patterns based on GPS data, biomedical images of diabetic retinopathy, and functional data across weather stations.
\end{abstract}

\section{Introduction}
\label{sec:intro}
    
    Cluster analysis aims at aggregating ``similar'' objects under same groups according to some similarity measure, and it is widely used as an exploratory tool across different domains. 
    The relevant literature has proposed algorithms characterized by various degrees of sophistication. The simplest ones rely on assumptions, sometimes unexpressed albeit strong, which restrict considerably their applicability to complex data. On the other hand, also the more flexible algorithms are often not practicable in applications, because of their computational and technical complexities. 
    We aim at reaching flexibility without sacrificing algorithmic simplicity, computational efficiency and practicality of use in real case studies.
    
    In the last decades, the growing size and complexity of the data to be processed in any field and the availability of dedicated software and computing power have drastically increased the number of interested cluster analysis users, but have also fragmented its  scientific community that today is mainly polarized around machine learning and statistical modeling. We specify our perspective, motivation and contributions.
 
    \paragraph{Frame of reference}
		As most unsupervised learning problems, in cluster analysis the lack of a response variable that serves as a ground truth to compare prediction results, complicates the matter, posing issues that are often underestimated or simply neglected \citep{hennig2015true,von2005towards}.
		Typically, the clustering partition cannot be an objective one, as it depends on the notion of clusters, the algorithm in use, and the nature of the problem.
    	Moreover, real-world problems are commonly affected by some form of ``data contamination''.
    	Here     	detecting noisy signals and/or outliers is essential, as data contamination might disrupt the performance of classical clustering methods. 
		By performance we mean the effectiveness in detecting the actual classification (when it is known) as well as the level of stability reached by the clustering partition when contaminants are removed.
		Data contamination can be thought of as adversarial (e.g.,~fraudulent transactions, disturbance of a signal transmission)
		or can be modeled as an additional cluster encompassing noise if the contamination structure is known (e.g.,~uniform, point-mass).
		Importantly, outlier detection itself can be a major goal of the analysis as it often provides relevant domain-specific 
		insights
		\cite{cerioli2014robust}.

	\paragraph{Motivation}
        In the context of model-based clustering \citep{bock2002clustering,breiman2001statistical},
		there are two main strategies in dealing with data contamination: mixture models \citep{fraley1998many} and trimming approaches \citep{cuesta1997trimmed}. 
		While the former can be used to tolerate and fit contaminants, the latter aim at their exclusion from the model and represents the focus of the present work.
        In this setting, existing methods are very effective but also computationally intensive and require mild assumptions on the shape of the clusters or their volume
		\citep{garcia2008general}.
		However, trimming methods in the presence of general-shaped components (for example non-elliptical) have not received much attention in the literature, probably because a general and elegant theory providing such flexibility is not easy to derive.

   \paragraph{Contributions}
		We aim to fill this gap by combining and extending existing methodologies in a framework that relies on very weak and general assumptions. 
		Unlike existing robust methods, our proposal identifies general-shaped clusters and tolerates data contamination in a computationally efficient manner.
		It builds upon existing works based on two-step clustering, where a preliminary model-based algorithm is followed by a hierarchical agglomeration phase \citep{melnykov2019clustering,peterson2018merging}.
		It thus inherits their properties but,
		unlike existing hybrid methods that lack robustness (i.e.,~only a pre-processing step is proposed in \citep{peterson2018merging}), it can also detect and discard arbitrary forms of contamination.
		In the first step, we exploit the robustness and computational efficiency of trimmed k-means \citep{cuesta1997trimmed}, which is used to identify an inflated number of clusters and detect outlying units and/or noise.
		The second step performs hierarchical clustering based on the robust ``centroids'' computed in the first step, 
		and we mainly focus on using an easy-to-calculate distance metric and linkage function.
		We also discuss generalizations of this approach to tackle more complex data structures, as well as a ``semi-automated'' procedure to estimate the amount of contamination based on the work of \cite{torti2021semiautomatic}.
		Real-world applications showing the effectiveness and generality of our proposal are presented. These include image, human mobility, biomedical, and weather data.
		
		\paragraph{Structure of the Paper}
        	We detail our proposal in Section~\ref{Sec2: model} after giving the essential background in Section~\ref*{Sec1:related work}, and we compare it with state-of-the-art methods through numerical simulations in Section~\ref{Sec4: experimentation}.
        	Then Section~\ref{Sec5: application} illustrates how it works in the above-mentioned representative real-world applications.
        	Final remarks follow in Section~\ref{Sec6: conclusions}, with additional details in the \hyperref[sec:app]{Appendix}.

    \section{Background and Preliminaries} \label{Sec1:related work}
    
        Consider a dataset $ \bm{X} = (\bm{x}_1, \ldots, \bm{x}_n )^T \in \mathbb{R}^{n \times p} $, where  $n$ is the sample  size and $p$ is the number of features. 
        As customary, we assume that $n$ is reasonably greater than $p$ -- a ratio of $n/p>5$ is a reasonable choice suggested for Gaussian mixtures \citep{rousseeuw1990unmasking} -- 
        and that the $n$ points belong to $K$ clusters.
        We further assume that a proportion of points $m/n$ might not belong to any of the aforementioned clusters and should be detected and remain unclassified. 
        Data contamination encompasses both outliers (i.e.,~~units with atypical values in a few dimensions) and noise (i.e.,~units with atypical values in most dimensions).
        With respect to the assumed clusters' shapes, one can distinguish between spherical, elliptical, and general-shaped components.
    	Moreover, homogeneity (or homoscedasticity) between clusters refers to the fact that they have approximately the same volume and spread of points.
    	On the other hand,
    	heterogeneity (or heteroskedasticity) refers to clusters with unequal dispersion.
        
        In this work we focus on a convenient combination of probabilistic methods -- which rely on parametric models to describe the clusters themselves -- and hierarchical algorithms of wider applicability -- which rely on non-probabilistic reasoning to merge such clusters into meaningful groups.
		
      \paragraph{Probabilistic Methods}
    
		\textit{Model-based clustering} often relies on \textit{finite mixture models} \citep{mclachlan2004finite, melnykov2010finite}.
		Typically their output is not a data partition, but rather a probabilistic statement about the membership of each unit to any cluster. 
		Nevertheless, to keep our setting as simple as possible, we will be focusing on crisp assignments of units to a partition.
		These are powerful and flexible  approaches, capable of modeling highly heterogeneous groups.
		On the other hand, they typically require a high-level knowledge of the problem, to ensure that the functional form of its components is reasonable, and often rely on a computationally intensive expectation-maximization (EM) algorithm \citep{dempster1977maximum}.
 
\begin{figure*}[t!]
            \centering
            \begin{subfigure}[b]{0.24\textwidth}
                \centering
                \includegraphics[width=\textwidth]{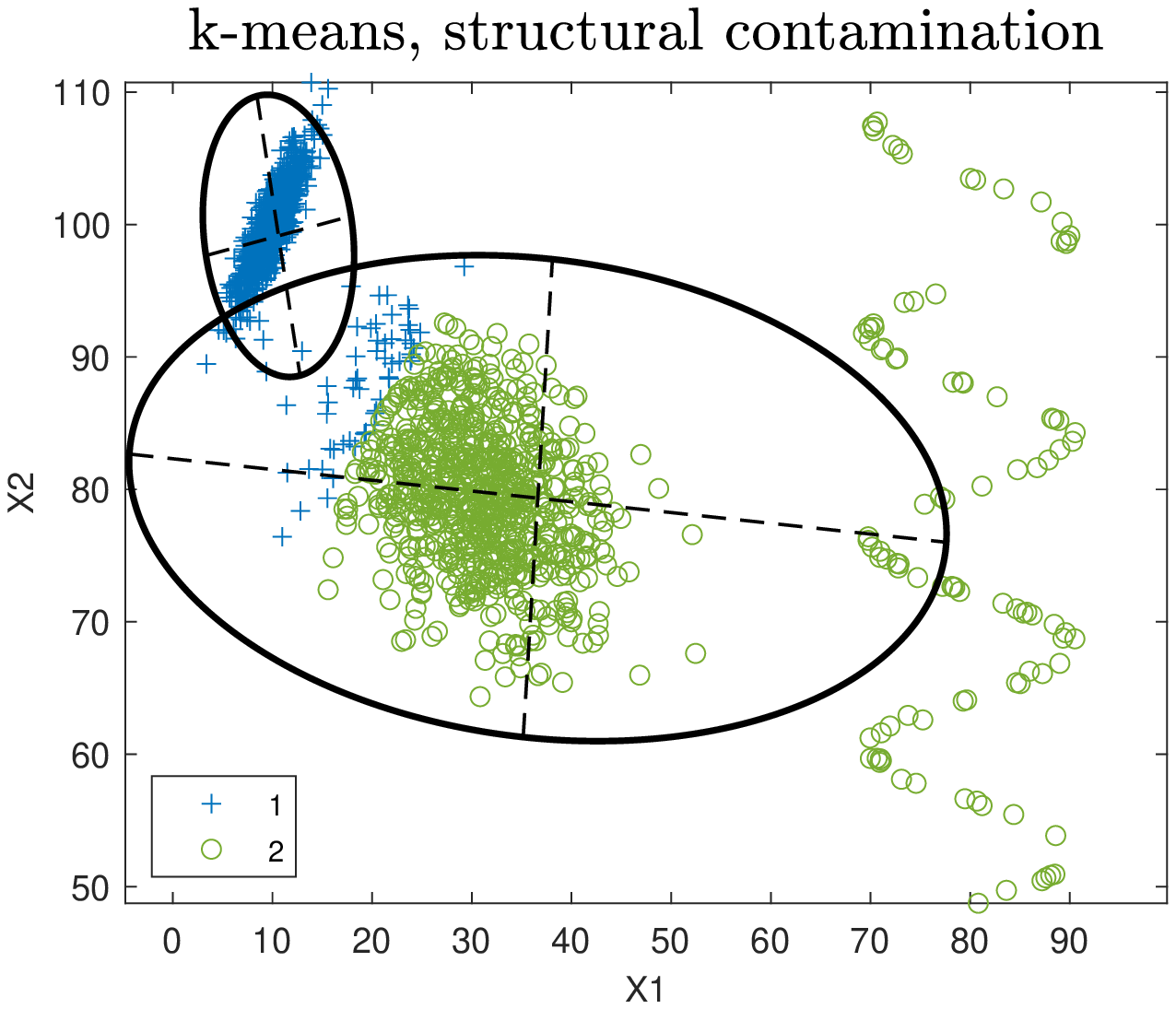}
            \end{subfigure}
            \begin{subfigure}[b]{0.24\textwidth}  
                \centering 
                \includegraphics[width=\textwidth]{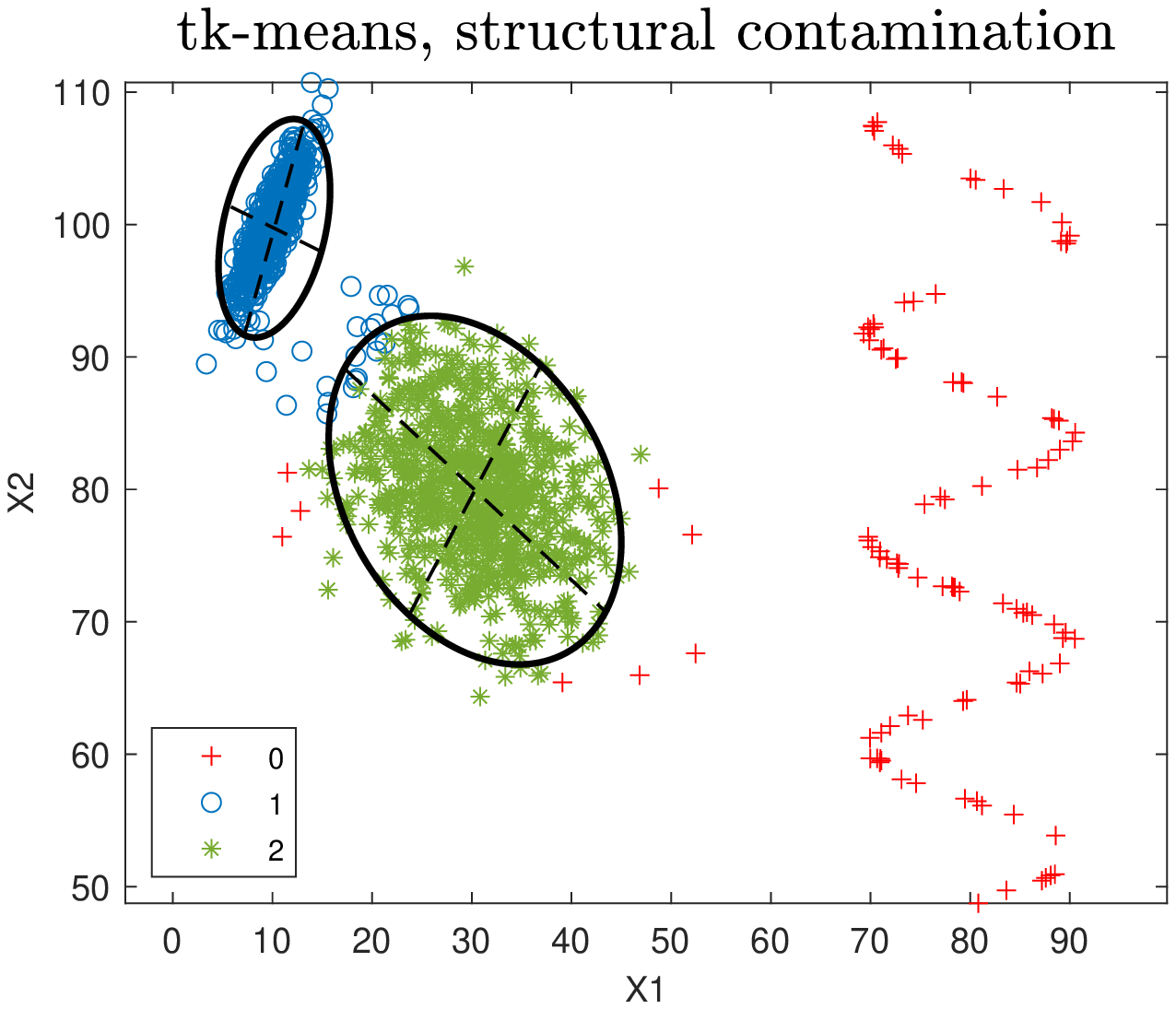}
            \end{subfigure}
            \begin{subfigure}[b]{0.24\textwidth}   
                \centering 
                \includegraphics[width=\textwidth]{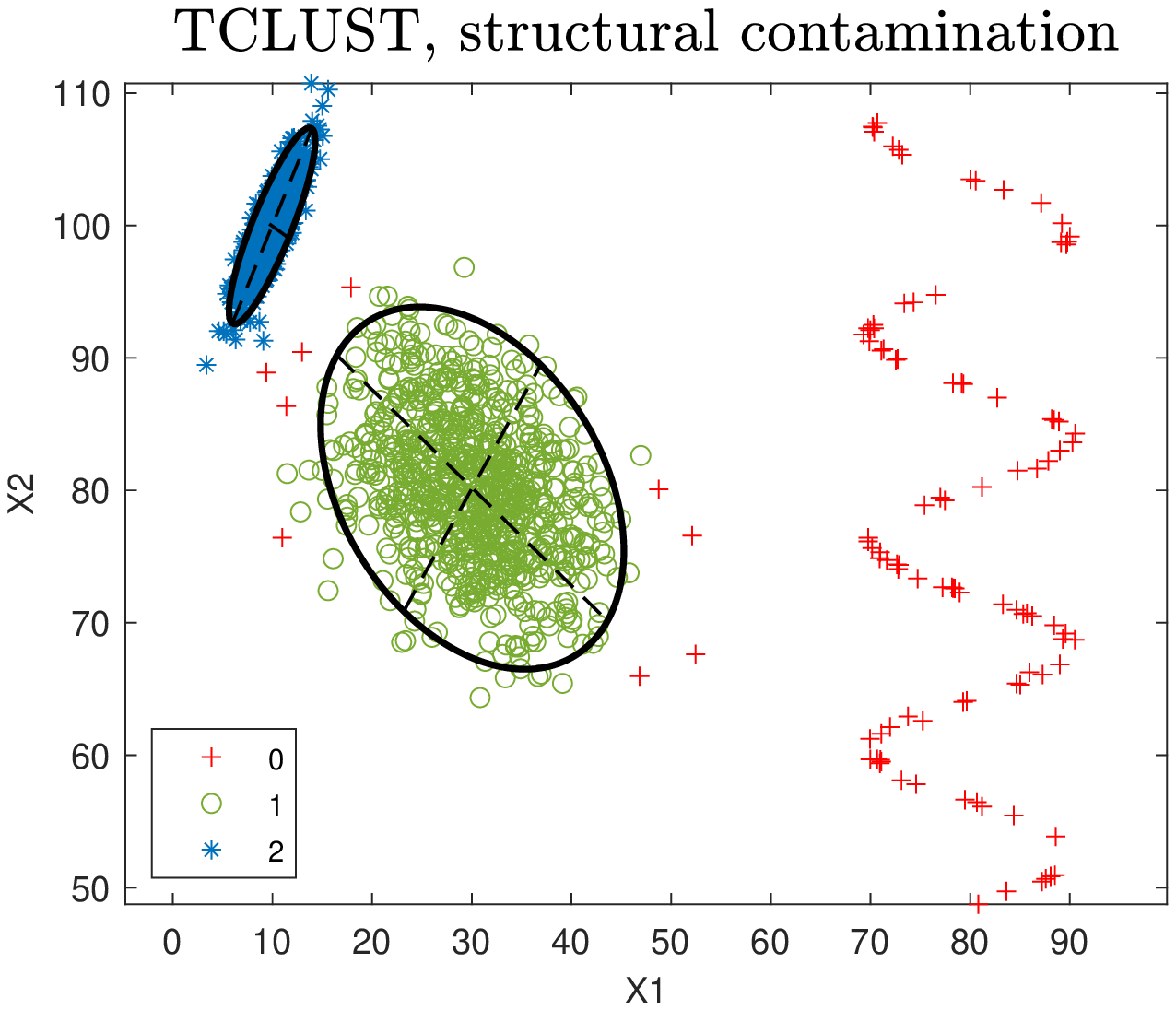}
            \end{subfigure}
            \begin{subfigure}[b]{0.24\textwidth}   
                \centering 
                \includegraphics[width=\textwidth]{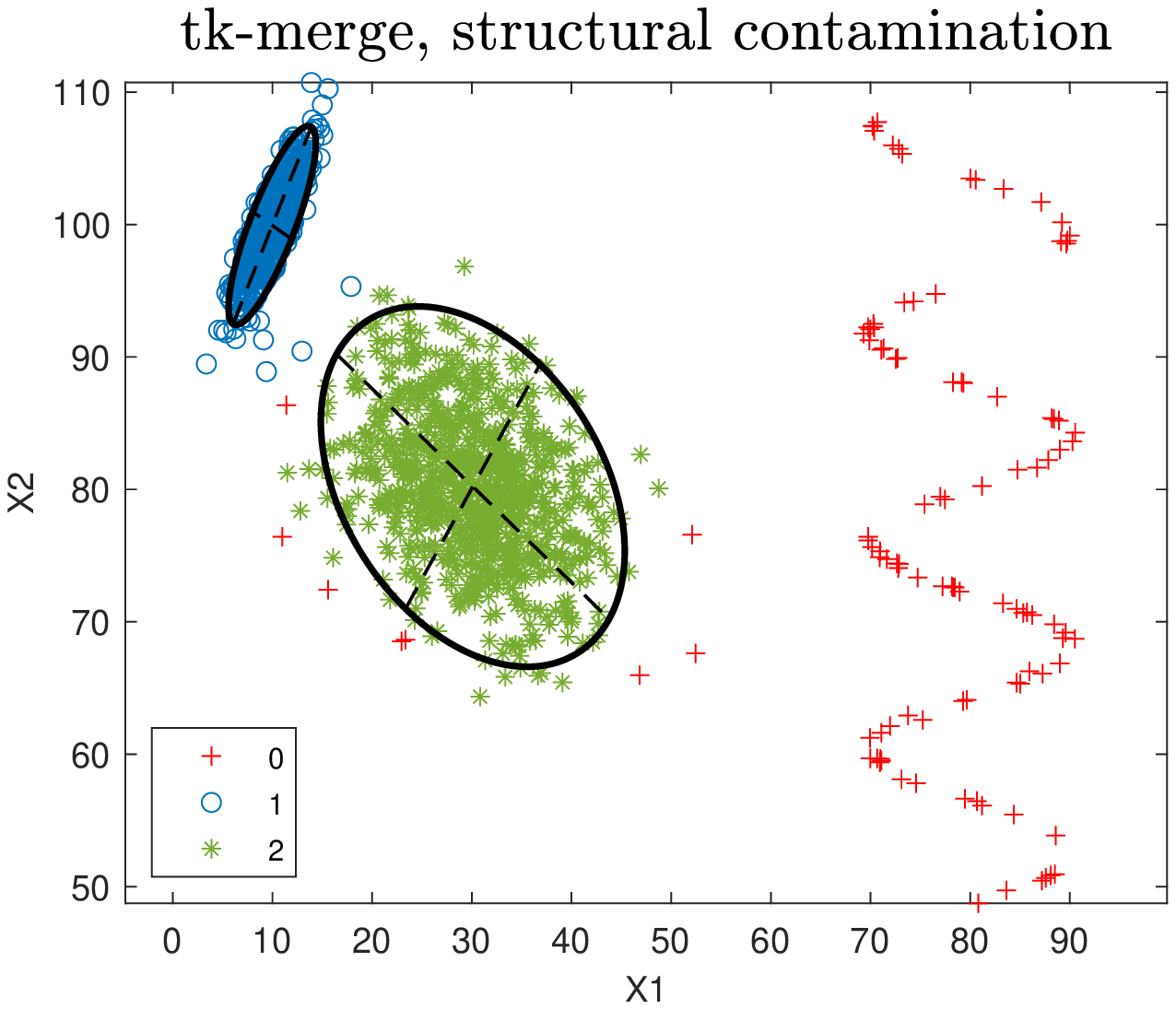}
            \end{subfigure}
            \vskip\baselineskip
            \begin{subfigure}[b]{0.24\textwidth}   
                \centering 
                \includegraphics[width=\textwidth]{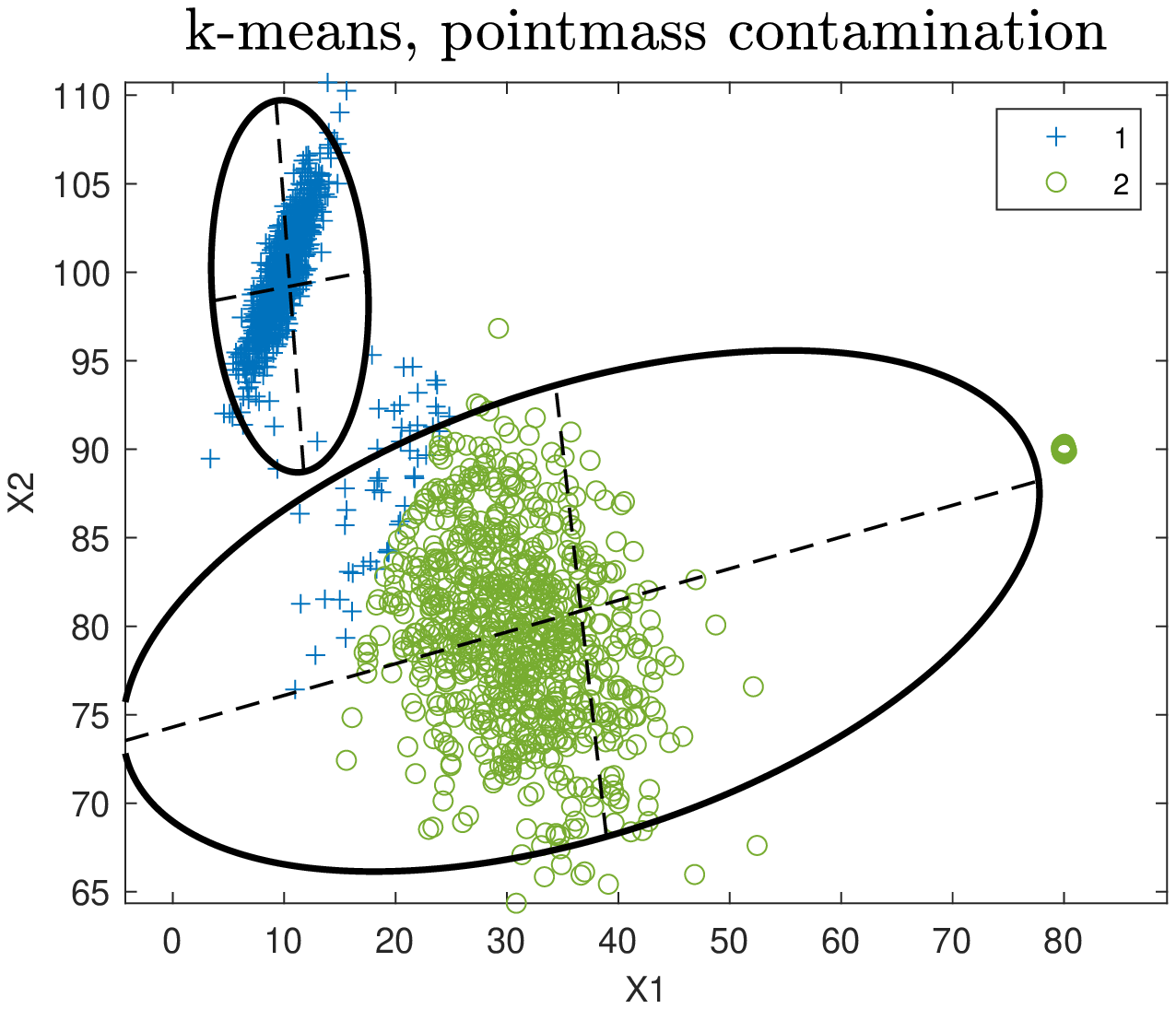}
            \end{subfigure}
            \begin{subfigure}[b]{0.24\textwidth}   
                \centering 
                \includegraphics[width=\textwidth]{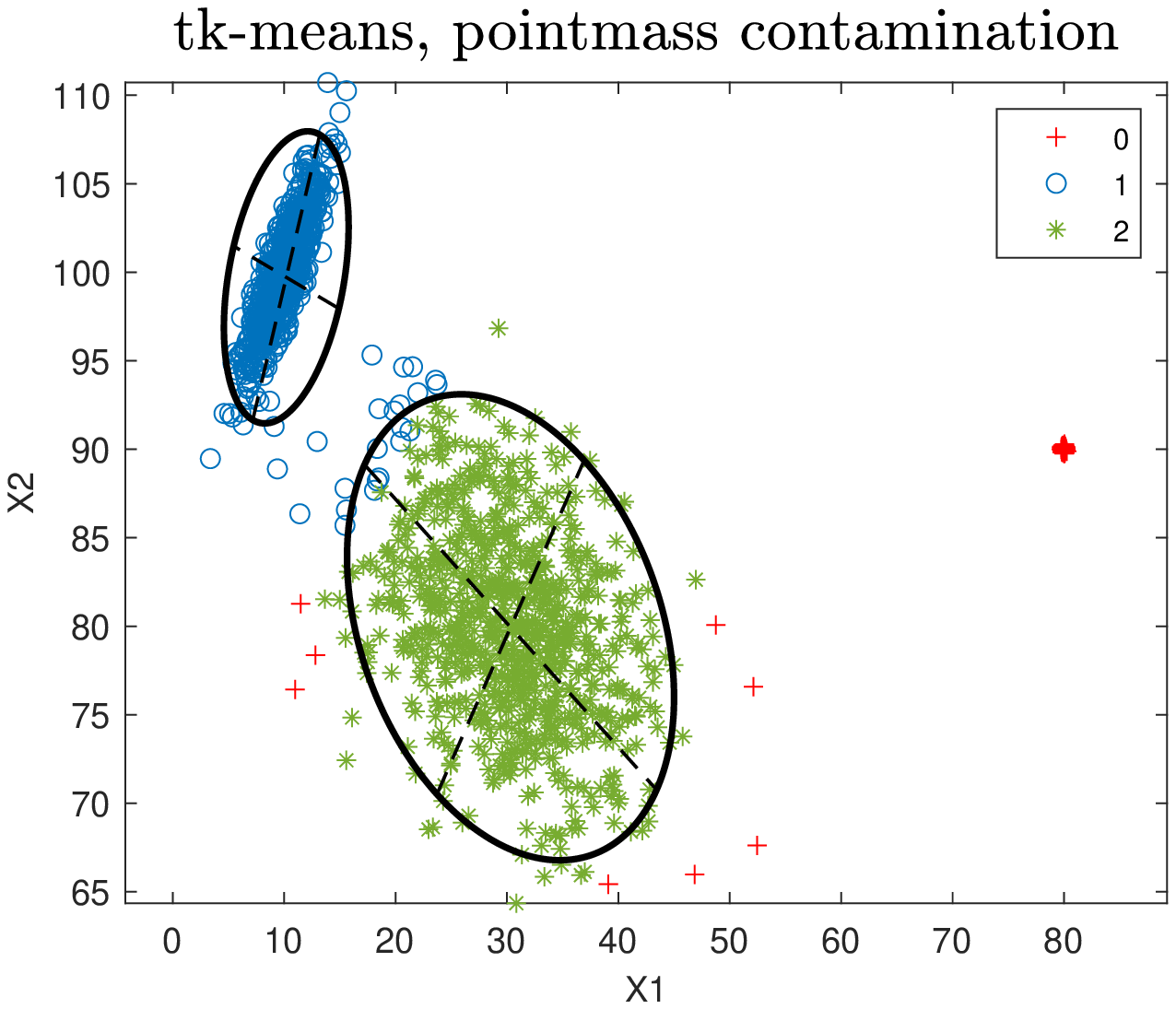}
            \end{subfigure}
            \begin{subfigure}[b]{0.24\textwidth}   
                \centering 
                \includegraphics[width=\textwidth]{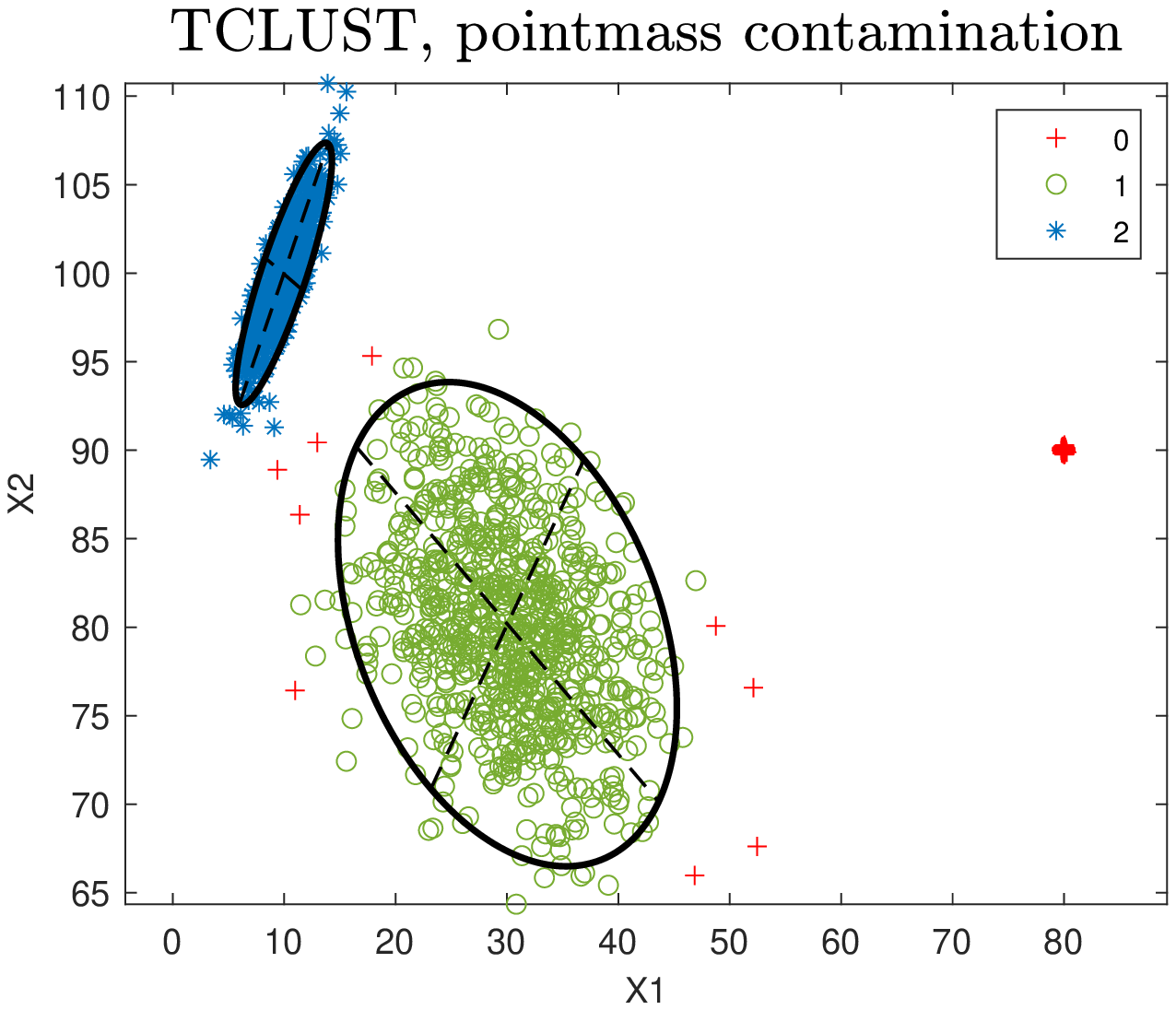}
            \end{subfigure}
            \begin{subfigure}[b]{0.24\textwidth}   
                \centering 
                \includegraphics[width=\textwidth]{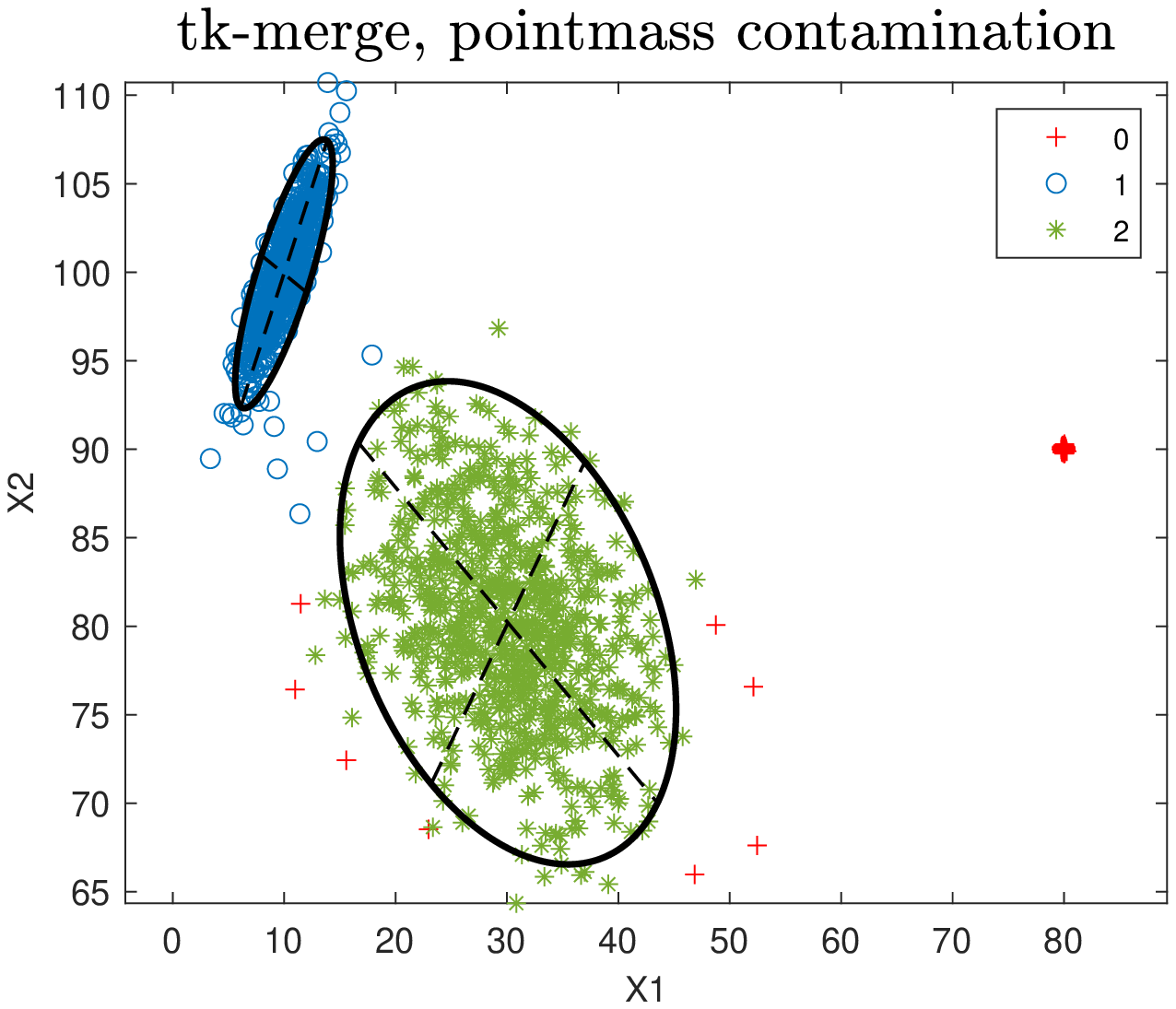}
            \end{subfigure}
            \caption{Clustering partitions for k-means, tk-means, TCLUST, and tk-merge for two different datasets: structural (top panels) and point-mass (bottom panels) contamination.
            The superimposed ellipses are not spherical for tk-means and k-means because their axes are calculated from the empirical mean and covariance of the group units.
            Trimmed observations are labeled as zero and colored in red.}
            \label{fig:cont}
        \end{figure*}

        The well-known \textit{k-means} \citep{celebi2014partitional} is very fast to compute using the Euclidean norm. 
		Its solution minimizes the sum of squared errors within clusters 
		$
            Q_1 = \min_{C_j, \mu_j} 
			\{ \sum_{j=1}^{K} \sum_{i \in C_j} \lVert \bm{x}_i - \bm{\mu}_j \rVert_2 ^2  \} ,
		$
        where $C_j$ indicates the $j$-th cluster, and $ \bm{\mu}_j  \in \mathbb{R}^{p} $ is the corresponding centroid.
        Operationally, obtaining the global solution is an intractable problem, but approximate solutions are computed in a two-step iterative process based on resampling methods.
        Given an initial set of centroids, it first assigns each point to the closest one, and then centroids are updated according to the points belonging to them.
		The resulting partition depends on the initialization itself, but the convergence to a local minimum is guaranteed \cite{vassilvitskii2006k}. 
		Despite its computational efficiency, the Euclidean distance implies homogeneity and sphericity of k-means clusters, which is a quite strong assumption. In addition, the method is very sensitive to data contamination: indeed, even a single point can hinder the whole clustering partition (see Figure~\ref{fig:cont} for an example).
      
        \textit{MCLUST} generalizes k-means assuming Gaussian mixtures with different covariance structures \citep{banfield1993model}.
		This increases model flexibility, since elliptical clusters of different volume, shape and orientation can be identified through different restrictions on their covariance matrices  \citep{celeux1995gaussian}. However, the lack of robustness to unknown and arbitrary forms of data contamination remains.

        \paragraph{Trimming Methods}
		The \textit{trimmed k-means} (tk-means) algorithm  \citep{cuesta1997trimmed} robustifies k-means and relies on a ``hard-trimming'' approach to discard contaminated points -- meaning that each observation receives a binary weight to determine its inclusion into the model groups (the alternative being to estimate the probability of assigning observations to each of the groups). This is similar to the least trimmed squares (LTS) estimator for regression models \citep{Rouss84LMS}.
		Its objective function minimizes the trimmed sum, over all clusters, of the within-cluster sums of point-to-cluster-centroid distances, where an additional parameter $\alpha$ controls the trimming proportion  \citep{garcia1999robustness,garcia2003trimming}.
        More generally, \textit{TCLUST}, which is considered the state-of-the-art robust algorithm to identify heterogeneous and non-spherical clusters, is the robust counterpart of MCLUST, and it also relies on a hard-trimming procedure
       \citep{garcia2008general,garcia2010review}.
   	    Its objective function assumes a mixture of elliptical distributions:
   	    \begin{equation}
   	        Q_2 = \min_{C_j, \pi_j,  \mu_j, \Sigma_j} \prod_{j=1}^K 
   	        \prod_{\substack{i \in C_j : \\|i \in C_j| \leq n(1-\alpha) }}
   	        \pi_j \phi( \bm{x}_i ; \bm{\mu}_j, \bm{\Sigma}_j ) \nonumber
   	    \end{equation}
        where $ \phi( \cdot ; \bm{\mu}_j, \bm{\Sigma}_j ) $ are densities for multivariate normal distributions with mean vector $\bm{\mu}_j \in \mathbb{R}^{p}$ and covariance matrix $ \bm{\Sigma}_j \in \mathbb{R}^{p \times p} $,
		and it provides non-crisp assignments for the ``most typical''
		$ \lfloor  n (1- \alpha) \rfloor $ points based on mixing proportions $ \pi_j$, where $ \lfloor \cdot \rfloor $ denotes the floor function.
        Similarity between clusters' shape and variability is controlled through a restriction factor $r$ constraining the determinant or eigenvalues of $  \bm{\Sigma}_j  $'s; clearly, if they are constrained to be spherical components, one retrieves the tk-means algorithm as a special case. Note that $r$ also allows one to impose as other special cases all the well-known Gaussian Parsimonious Clustering Models by Celeux and Govaert \citep{celeux1995gaussian} (where volume, orientation and shape are controlled, leading to the classic  `VVE',`EVE',`VVV',`EVV',`VEE',`EEE',`VEV',`EEV',`VVI', `EVI',`VEI',`EEI',`VII',`EII' models).
		It is also important to remind that TCLUST is generally based on Mahalanobis distances, which provide more flexibility but at the expenses of a greater computational burden.

		Figure~\ref{fig:cont} illustrates the importance of robustness and flexibility in two rather simple contamination scenarios; our proposal (tk-merge in the figure) is introduced in Section \ref{Sec2: model}.

    \paragraph{Hierarchical Methods}
		\textit{Hierarchical clustering} produces an ordered sequence of possible groupings using divisive or agglomerative strategies \citep{johnson1967hierarchical}. 
		The latter, which are the focus of this work, typically consider each unit as a separate cluster and 
		merge them according to a pre-specified criterion, until a unique cluster containing all observations is obtained.
		The hierarchical tree may be cut at any level, based on the number of ``natural'' or plausible clusters which have to be retained. 
		Thus, the whole nested clustering structure is obtained in a single run (i.e.,~the number of clusters ranges from $n$ to 1).
		The aggregation is dictated by two aspects: (i) a pair-wise \textit{distance metric} $d(\cdot, \cdot)$ (e.g.,~Euclidean), that is used to obtain an initial $n \times n$  dissimilarity matrix $D_n$  between the $n$ points;
		(ii) a \textit{linkage function} $L(\cdot, \cdot)$ between sets of points,	which dictates their distance after the aggregation of some units (e.g.,~the single linkage considers the shortest centroids' distance).	 
		Correctly specifying these two criteria allows one to identify complex-shaped clusters. Nevertheless, hierarchical methods generally aim to cluster all points, hence they are not robust with respect to data contamination. 
		Moreover, computing and storing 
		$D_j$ dissimilarity matrices (for $j=n,n-1,\ldots,K$) may be unfeasible for large $n$'s.

    \paragraph{Hybrid Approaches}
        Probabilistic and hierarchical procedures have been recently combined to effectively identify general-shaped clusters in a computationally efficient fashion. 
        The former finds an inflated number of components,
    	and the latter merges them according to a pre-specified rule; this exploits the advantages of both methods, namely: efficiency and flexibility.
		Although in principle any partitioning algorithm can be used before the hierarchical merging phase, k-means is the preferred choice due to its computational efficiency
		\citep{peterson2018merging}.
		Similarly, the associated aggregation phase may be performed with non-hierarchical methods, but this has not been explored in the literature yet.
		In order to retain a probabilistic foundation,
		\cite{melnykov2016merging,melnykov2019clustering} proposed the use of \textit{directly estimated misclassification probabilities} (DEMP) as a distance metric.
		However, this requires computing the overlap parameter $\omega$ between mixture components, whose definition is based on misclassification probabilities \citep{maitra2010simulating},
		and this calculation can increase the computational burden \citep{riani2015simulating}.
		To date, hybrid approaches are very effective under different scenarios, especially in the presence of well-separated clusters with complex shapes.
		Nevertheless, they are not robust, since even a single atypical observation can severely affect the resulting partitions.

    \section{Methodology} \label{Sec2: model}
    
    	We focus on the identification of heterogeneous general-shaped clusters in the presence of data contamination.
      	Unlike existing robust clustering methods based on trimming, our proposal does not use restrictive parametric assumptions and is computationally more efficient, which is a key requirement in modern applications involving ever increasing sample sizes.
        
    	Our basic assumptions concern three aspects: 
    	(i) there is not an ``extremely large'' overlap between the $K$ clusters;
    	(ii) noise and/or outliers are ``distinguishable'' from uncontaminated data; (iii) the contamination percentage is between $ 0 < m/n \leq n/2 $, and when it is not known (even approximately) it is estimated from the data (however, the \textit{actual} set of outlying cases remains unknown).
	    Note that we assume no knowledge about the data generating process or clusters' shapes, but we rely on assumption (i) to control their degree of separability (as most clustering methods do).
	    Assumption (ii) is used to avoid that contaminants form strong clusters themselves or they locate too close to the true ones.
	    Assumption (iii) is also quite standard in robust statistics (see for instance \citep{garcia2008general,garcia2011exploring,torti2021semiautomatic,cappozzo2021}) and assuming that the amount of contamination is known simplifies the exposition,
	    although in practice the contamination level can be unknown and should be estimated.
	    For instance, \textit{classification trimmed likelihood curves}  can be used for this task \citep{garcia2011exploring}, and we discuss additional strategies below.
	    However, this is not the focus of the present work.

	    \begin{algorithm}[htbp] 
            \SetAlgoLined
              \Input{ dataset $\bm{X}$, number of clusters $K$, 
              inflated number of clusters $k>K$, trimming level $\alpha$, restriction factor $r$, distance  $d(\cdot,\cdot)$, linkage $L(\cdot,\cdot)$} 
              \Output{ a partition $ { \mathcal{C}_1, \ldots, \mathcal{C}_K } $ of $ \approx n ( 1 -\alpha)$ points into $K$ clusters}
              \uIf{ $r = 1$ }{
                apply tk-means with $\alpha$\% trimming to detect $C_1^*, \ldots, C_k^*$ groups\;
              } \Else{
                apply TCLUST with restriction factor $r$ and with $\alpha$\% trimming to detect $C_1^*, \ldots, C_k^*$ groups\;
              }
              initialize a $(k \times k) $ dissimilarity matrix $D_k$\;
              \uIf{$d(\cdot,\cdot)$ = $L_2$}{
                compute $D_k$ using Euclidean distance between the $k$ centroids of $C_1^*, \ldots, C_k^*$\;
              }
              \Else{
                use any other metric to compute $D_k$ (e.g.,~DEMP)\;
              }
              initialize $k^* = k$\;
              \While{$k^* \geq K$}{
              perform agglomerative hierarchical clustering using $D_K$ and a given linkage $L(\cdot, \cdot)$\;
              }
             \caption{\textbf{tk/TC-merge}$(\bm{X}, K, k, \alpha, r, d, L)$}
             \label{alg:1}
            \end{algorithm}
	    
	    Algorithm~\ref{alg:1} details our proposal.  
    	In its most general flavor,
    	it combines four ingredients:
        \begin{enumerate}
    		\item Use TCLUST (which simplifies to tk-means as a special case) to identify an inflated number of components $ k > K $ with a trimming level $\alpha \approx m/n$.

            \item Compute a $ (k \times k) $ dissimilarity matrix $D_k$ between the $k$ components identified in Step~1 based on a distance $d(\cdot, \cdot)$.				
				
            \item Perform hierarchical aggregation based on $D_k$ from Step~2 based on a linkage $L(\cdot, \cdot)$.

            \item Cut the hierarchical tree from Step~3 to identify $K$ groups.
    	\end{enumerate}
    	
 		Figure~\ref{fig:cont} shows $K=2$ elliptical and heterogeneous clusters affected by different forms of adversarial contamination, namely:
		structural (top panels) and point-mass (bottom panels).
		We consider a setting with $ n_j = 800 $ points belonging to $C_j$ (for $j=1,2)$ and $m = 100$ points are in fact contaminated.
		Here k-means provides very poor solutions as it breaks down.
		On the other hand, robust methods perform very well in detecting the true clusters, as well as contaminated points. 
		TCLUST provides optimal partitions since the underlying assumptions are perfectly met, tk-means is negatively affected by the presence of non-spherical components, and our proposal -- denoted  as \textit{tk-merge} when it relies on tk-means -- performs between these two.
		All robust methods use a trimming proportion $ \alpha \approx 6\%$.
		Despite our proposal suffers particularly in the presence of point-mass contamination, where assumption (ii) can be violated, it still performs comparably to state-of-the-art algorithms in this setting that does not comprise general-shaped clusters.

	\paragraph{Implementation Details}

        We now specialize our proposal to a few typical settings.
    	In Step~1 we tend to prefer tk-means due to its computational efficiency. Nevertheless, in some specific scenarios we rely on the greater flexibility offered by TCLUST -- which is denoted as \textit{TC-merge} -- since it can identify elliptical components of different size (see Example 3 below).
    	In the presence of general-shaped components, we empirically observed that $k=2\log(n)$ and $k=\log(n)$ are often reasonable choices for tk-means and TC-merge, respectively; the latter is reduced due to higher flexibility of TCLUST. However, specific applications might require a more careful refinement; its theoretical investigation is beyond the scope of the present paper.
    	The distance metric and linkage function for Steps 3 and 4 can be chosen based on clusters characteristics (e.g.,~shapes and overlaps).
    	We focus on Euclidean distances (between clusters' centroids) and single linkage to exploit their efficiency.
    	In our implementation (see supplementary material) we also consider DEMP to build an initial dissimilarity matrix, which provides more stable results in some settings (e.g.,~higher degrees of overlaps).
    	
    	If the trimming proportion $ \alpha $ is unknown, diagnostic methods are extremely useful for Step 1, as this choice affects the next steps.
	    Specifically, in this work we rely on a ``semi-automated'' approach which extends to TCLUST the so-called monitoring approach \cite{cerioli2018power}. The philosophy is to investigate how the results change as the trimming proportion $\alpha$ varies and was inspired by the work in \citep{torti2021semiautomatic};  see Appendix~\ref{app:application} for details.

    	Importantly, our proposal's simplicity is one of its primary advantages and it can be easily implemented using standard statistical software environments, for example~\verb+MATLAB+ and \verb+R+.
    	Our implementation, as well as the source code to replicate our simulation and application studies, builds upon the software \citet{FSDA:toolbox} and is provided in the supplementary material.

\section{Experiments}	\label{Sec4: experimentation}
    
\begin{figure}
    		\centering
    		\begin{subfigure}{.5\textwidth}
    			\centering
    			\includegraphics[width=0.8\linewidth]{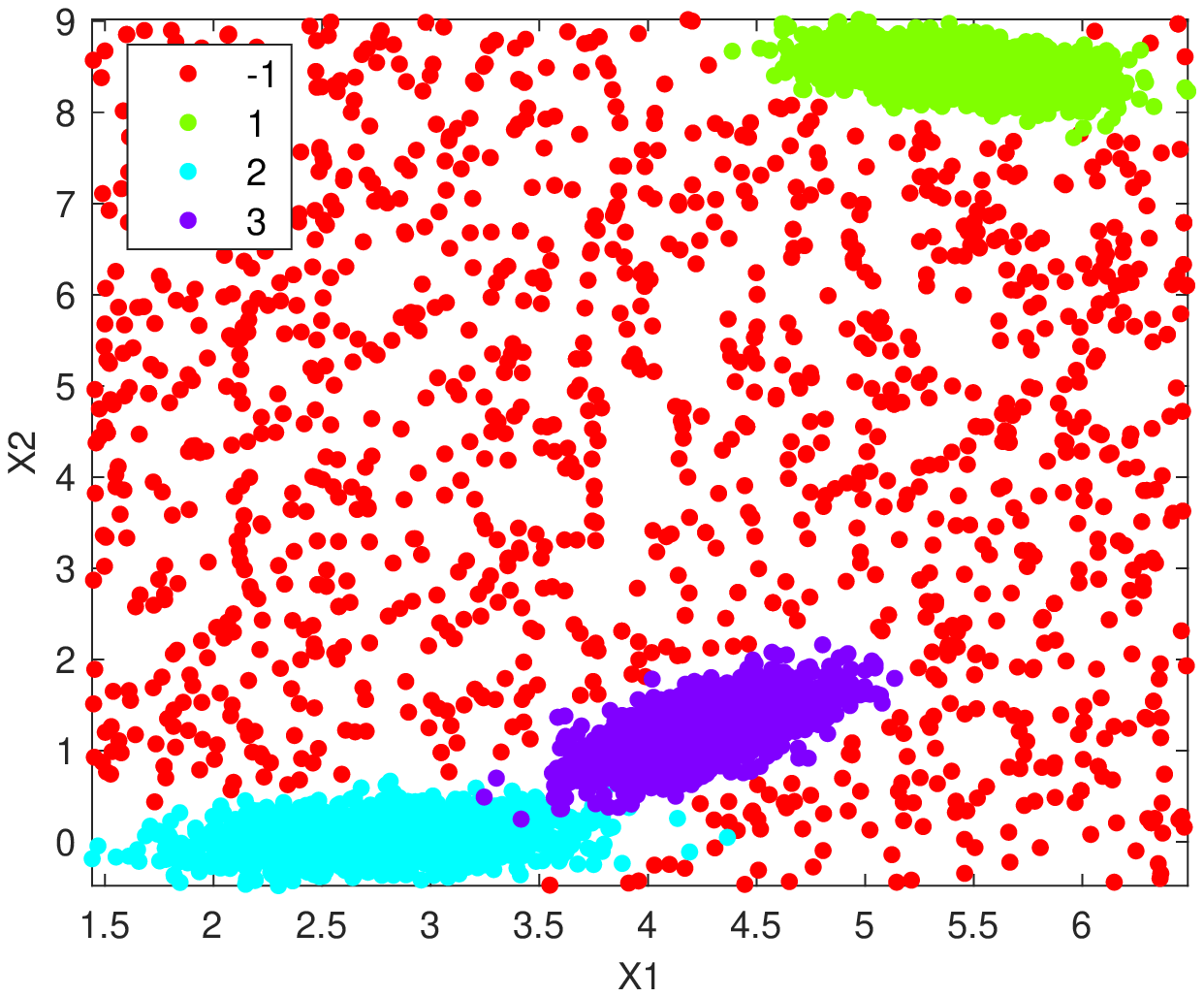}
    		\end{subfigure}%
    		\begin{subfigure}{.5\textwidth}
    			\centering
			\includegraphics[width=0.8\linewidth]{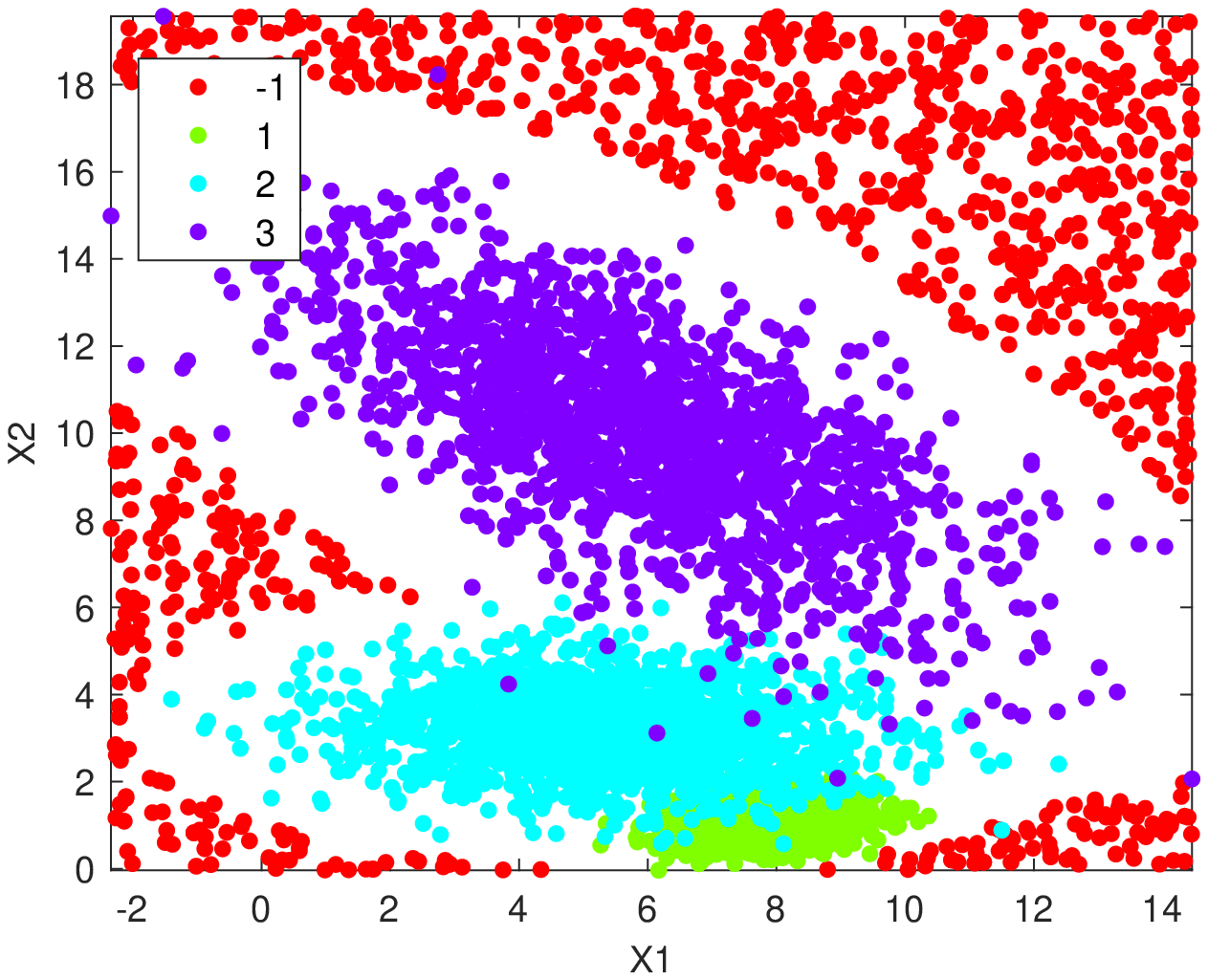}
    		\end{subfigure}
    		\caption{Left panel: example of 
    		simulation scenario 1 for $n=6000$.
    		Right panel: example of 
    		simulation scenario 2 for $\omega= 0.035$.}
    		\label{fig:ex}
    	\end{figure}

\begin{figure*}
            \centering
            \begin{subfigure}[b]{0.32\textwidth}
                \centering
                \includegraphics[width=\textwidth]{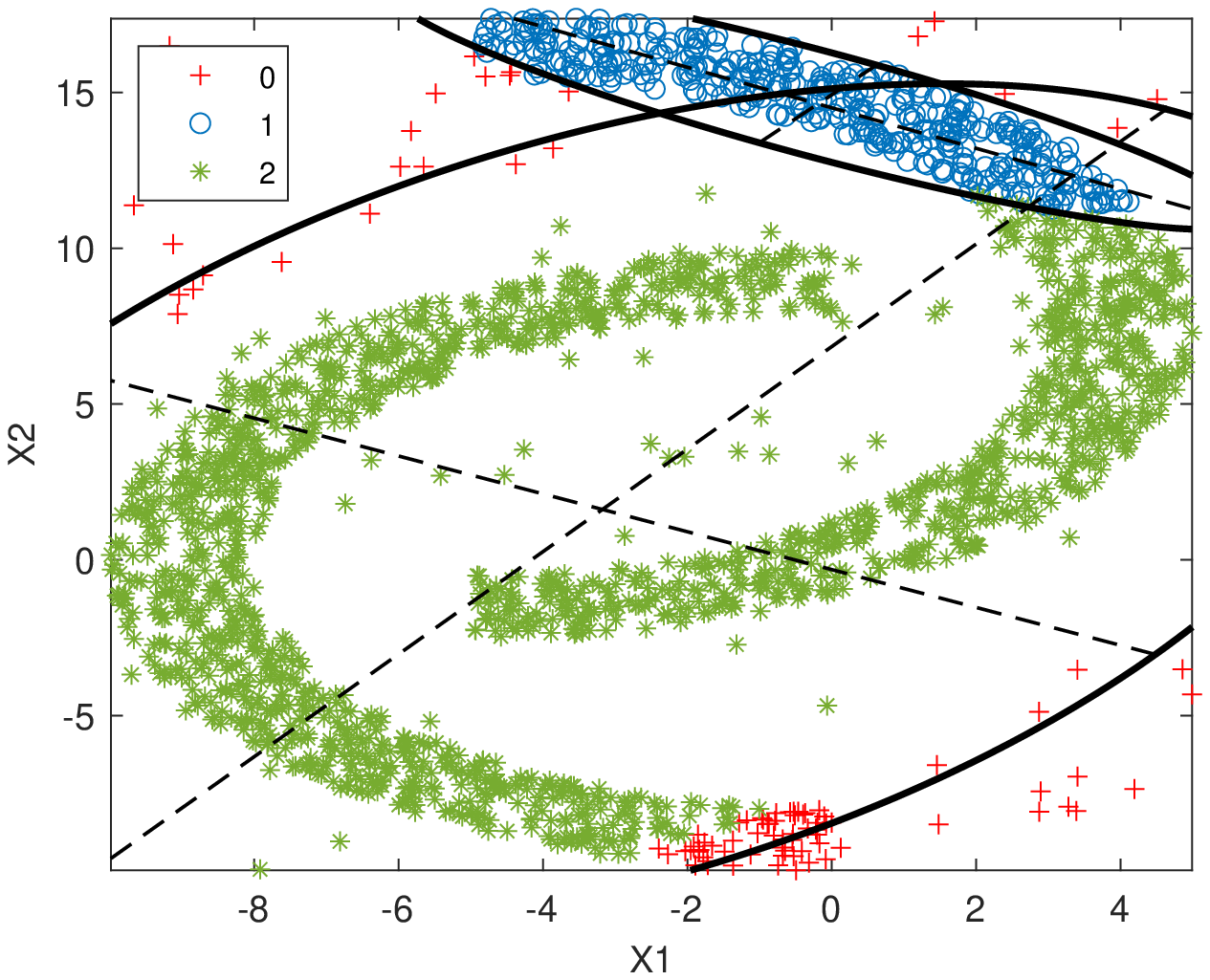}
            \end{subfigure}
            \begin{subfigure}[b]{0.32\textwidth}  
                \centering 
                \includegraphics[width=\textwidth]{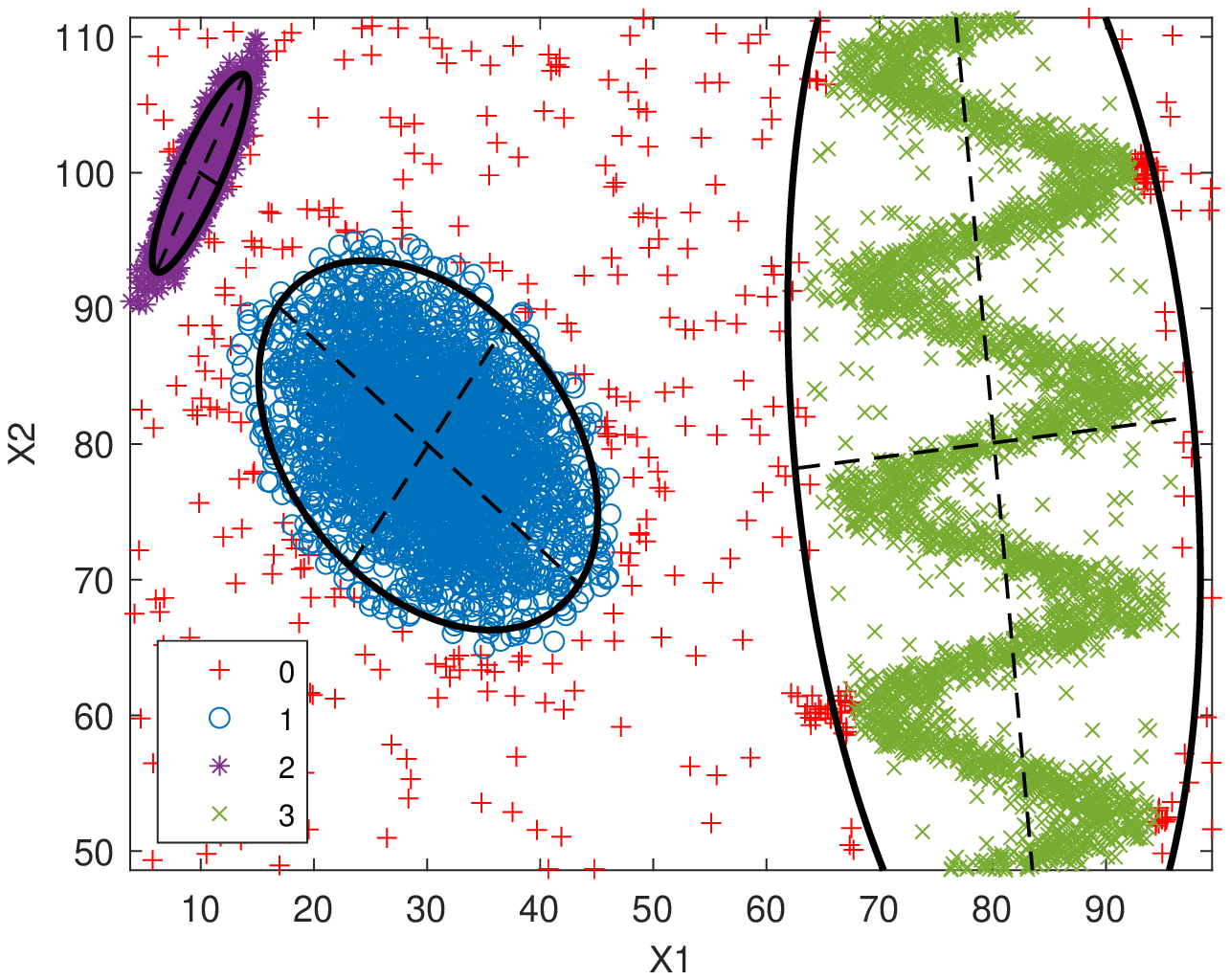}
            \end{subfigure}
            \begin{subfigure}[b]{0.32\textwidth}   
                \centering 
                \includegraphics[width=\textwidth]{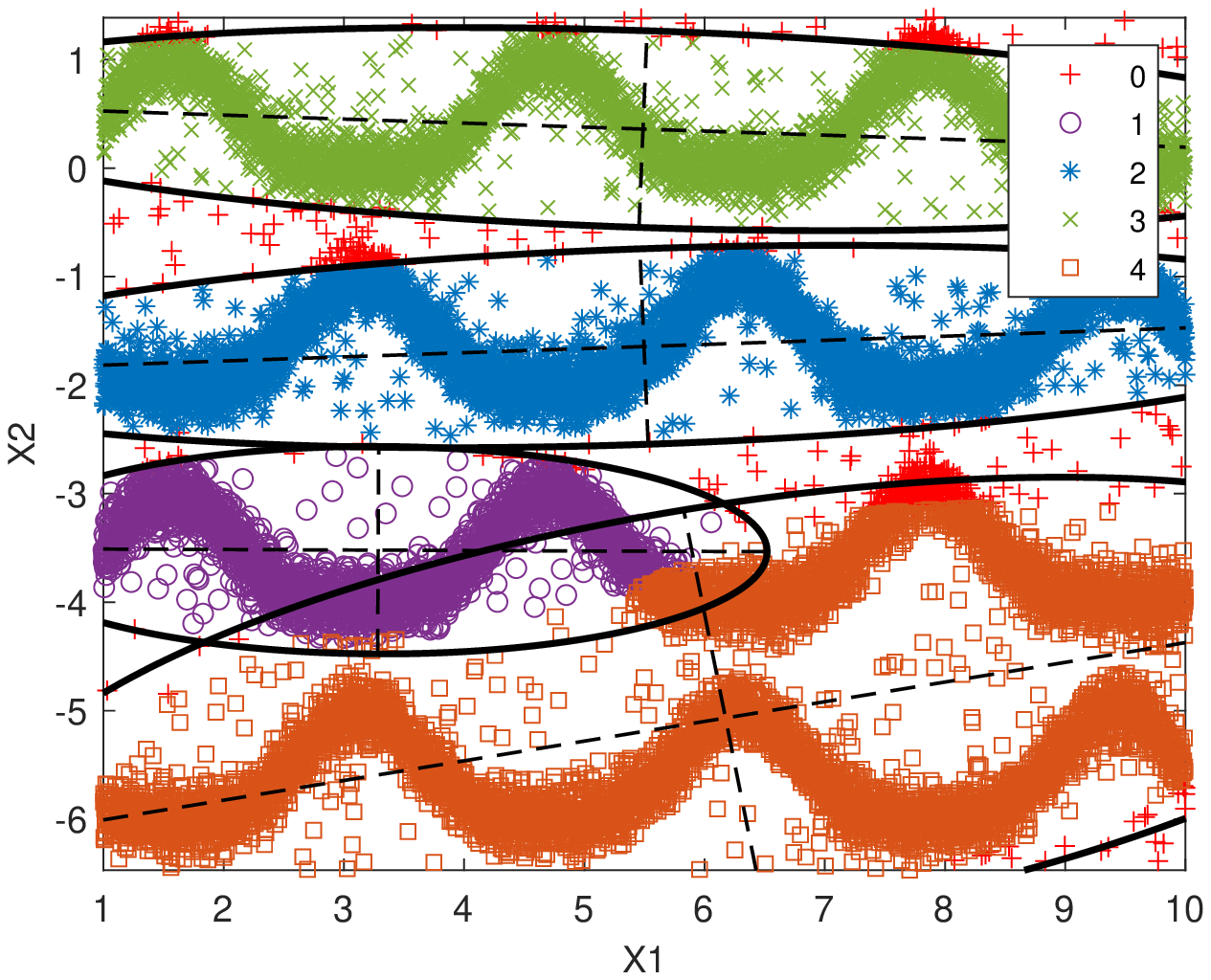}
            \end{subfigure}
            \vskip\baselineskip
            \begin{subfigure}[b]{0.32\textwidth}   
                \centering 
                \includegraphics[width=\textwidth]{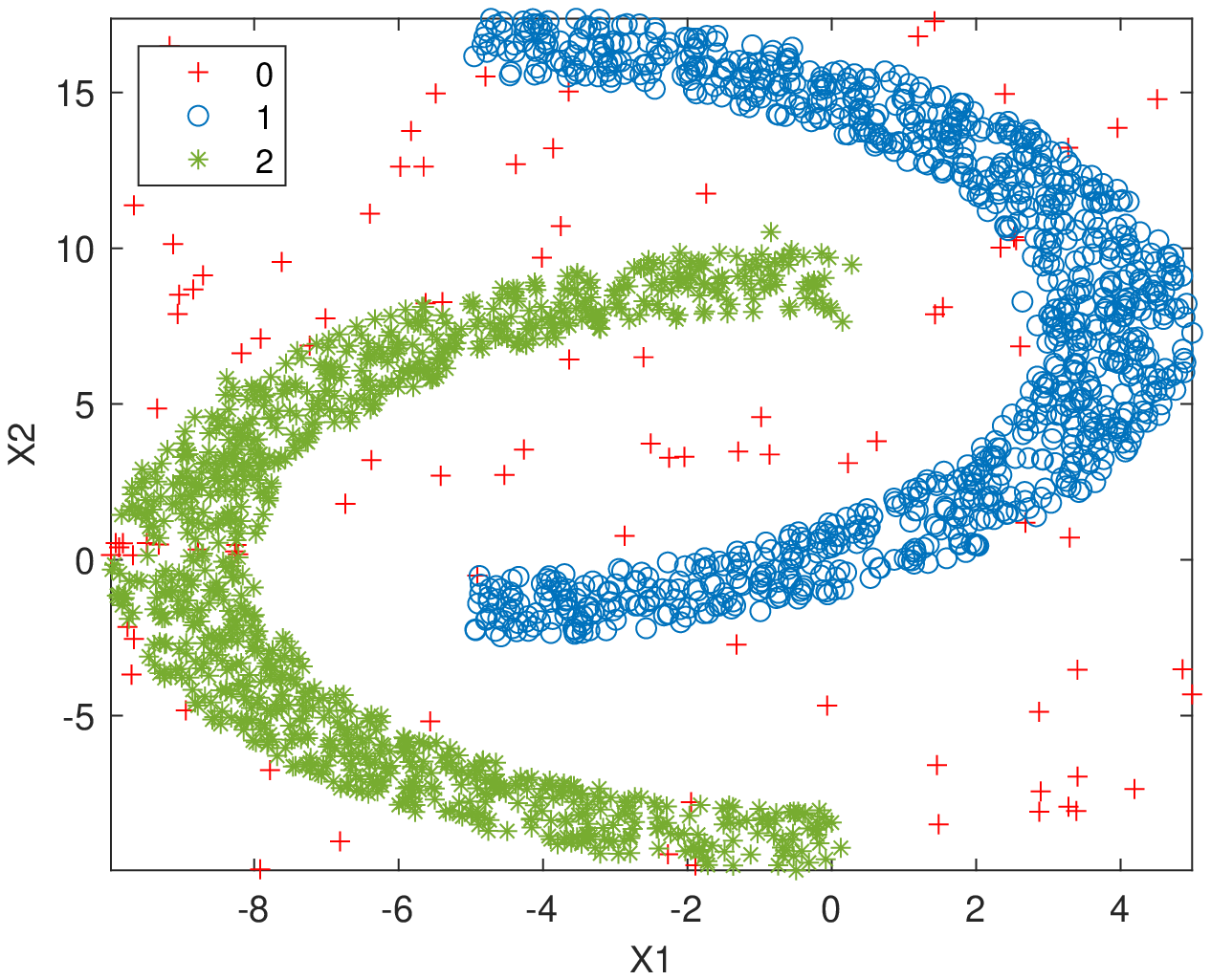}
            \end{subfigure}
            \begin{subfigure}[b]{0.32\textwidth}   
                \centering 
                \includegraphics[width=\textwidth]{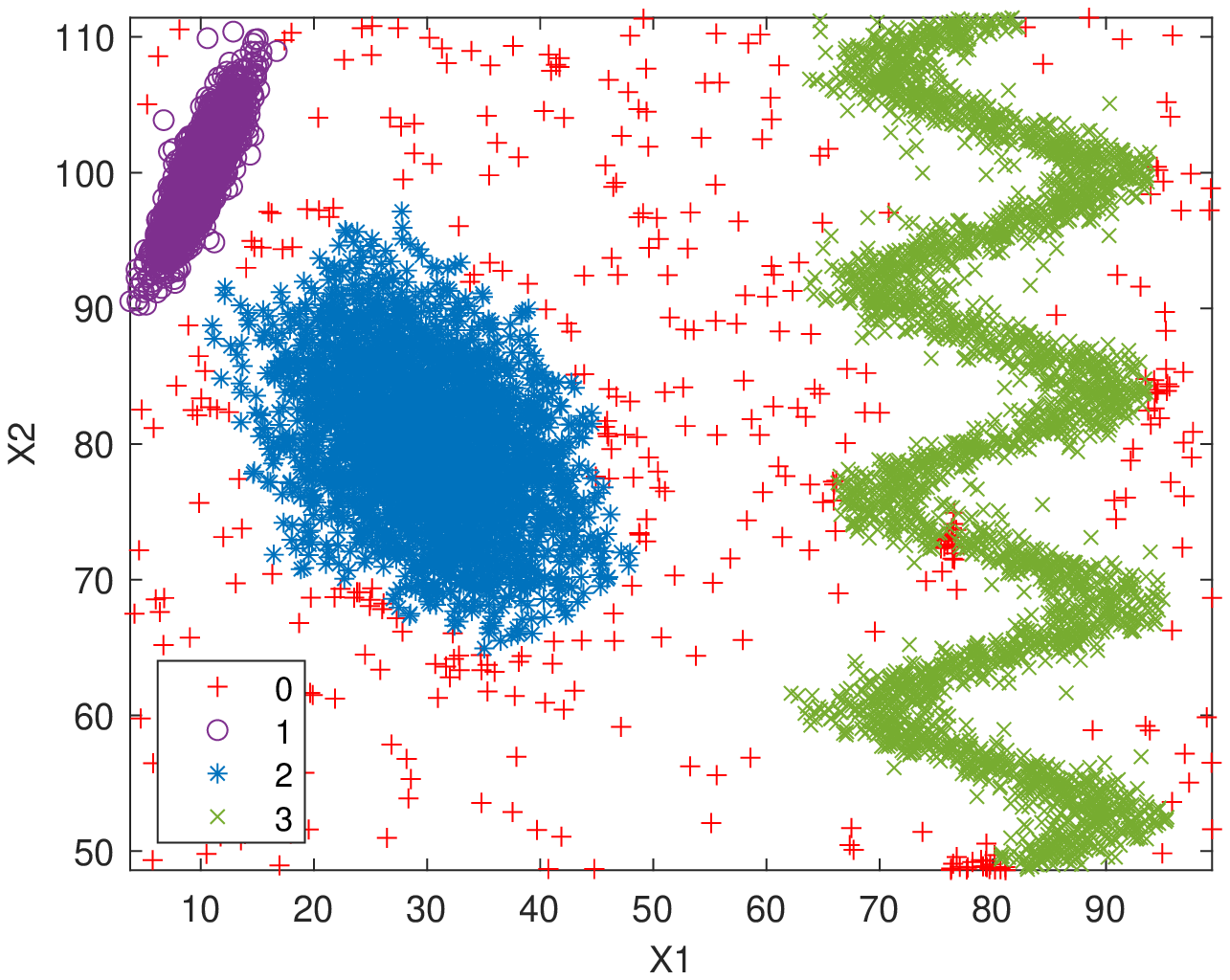}
            \end{subfigure}
            \begin{subfigure}[b]{0.32\textwidth}   
                \centering 
                \includegraphics[width=\textwidth]{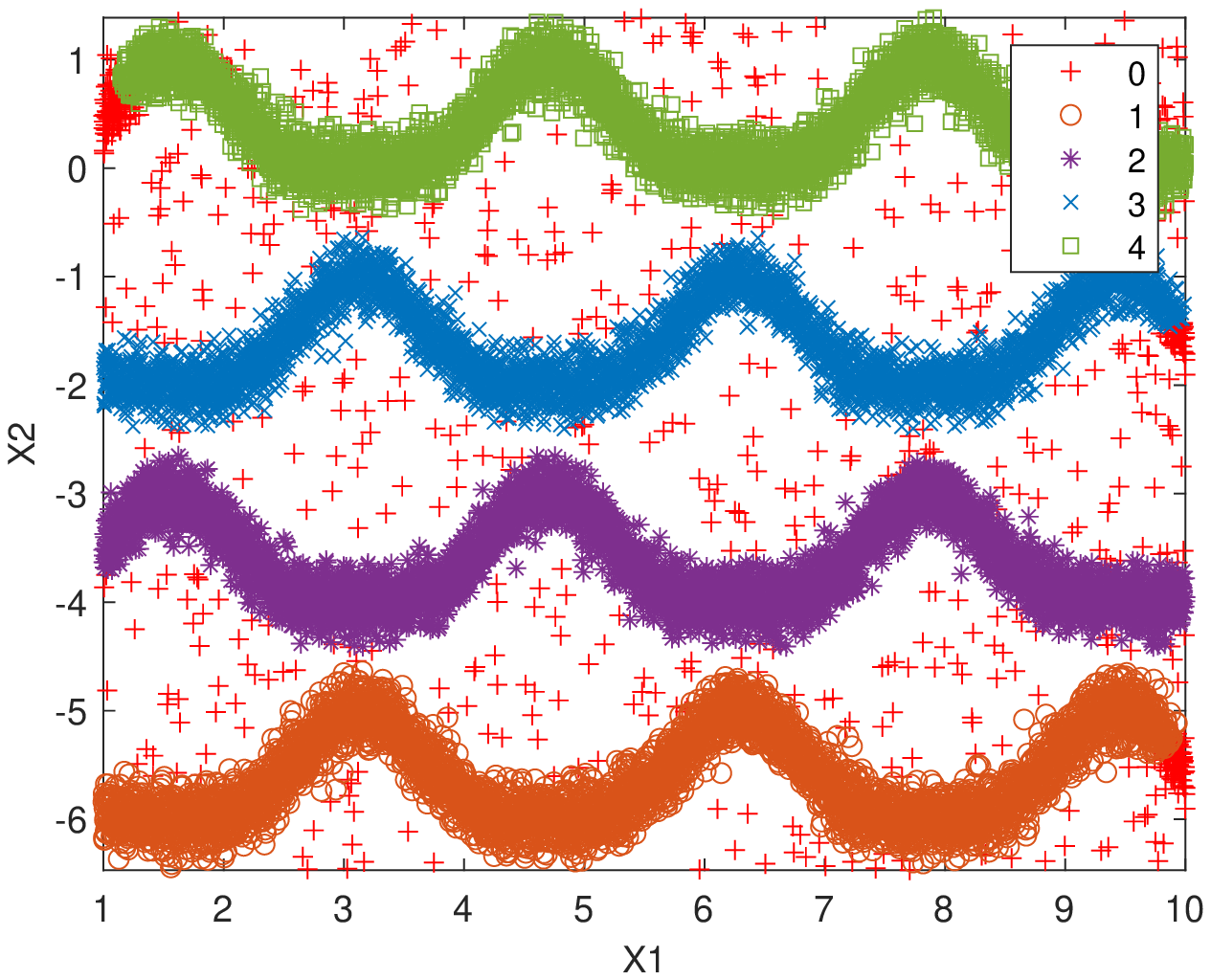}
            \end{subfigure}
            \caption{Simulation scenario  3. 
            TCLUST (top panels) and tk-merge (bottom panels) solutions
            for three simulated datasets with general-shaped clusters.} 
            \label{fig:sim3}
        \end{figure*}

    \paragraph{Experimental Setup}
    	We compare our proposal with state-of-the-art methods through Monte Carlo simulations.
        Experiments were carried out using \verb+MATLAB 2019b+ and the \verb+FSDA MATLAB Toolbox+ \citep{FSDA:toolbox}.
        The  hardware in use has an Intel Core i7-7700HQ CPU @ 2.8 GHz $\times$ 4 processors and 16 GB RAM.
    	We use the function \verb+MixSim+ to generate random data as follows.
    	We set $K=3$, $p=2$, $n_1, \ldots, n_K = n/K$, and we draw heterogeneous, non-spherical Gaussian components with mean vectors on the interval $[0, 10]$, controlling their overlap values $\omega$. Data for each scenario  are contaminated with the inclusion of $m/n \approx 0.2$ points following an uniform noise (see Figure~\ref{fig:ex}).
    	We focus on the following scenarios:
    	\begin{enumerate}
    	    \item We set $ \omega = 0.005$, and $n$ increases from 1000 to 45000 (with 10 equispaced values).
    	    \item We set $n=5000$ and $\omega$ increases from
    	    0.0005 to 0.035 (with 10 equispaced values).
    	\end{enumerate}
        Different methods are compared in terms of
    	(a) computing time in seconds, (b) clustering and outlier detection accuracy based on the \textit{adjusted Rand index} (ARI; \citep{hubert1985comparing});
    	here outliers represent an additional cluster and thus their detection affects the ARI.
    	Each simulation scenario is replicated 100 times, 
    	and results are summarized in terms of their median and $S_n$ as a robust estimator of scale
    	\cite{rousseeuw1993alternatives}.
    	We compare the following approaches:
    	(i) our tk-merge proposal, where $k = 2K  $;
    	(ii) tk-means \cite{cuesta1997trimmed};
    	(iii) TCLUST  with a liberal restriction factor on the eigenvalues equal to 1000 \citep{garcia2008general}.
    	All methods partition the data into the true number of clusters $K$ (excluding cases detected as outlying), and their trimming proportion $\alpha =m/(n+m)$ corresponds to the true amount of contamination.
    	 Moreover, 
    	 we consider additional scenarios with general-shaped clusters (see Figure~\ref{fig:sim3}) mimicking the data used for our applications in Section~\ref{Sec5: application}; namely:
    	\begin{enumerate}
	        \item[3.]
	        Each $j$-th cluster contains an equal number of points,
        	and we set: $n_j=1000$ with $K=2$ (left panels), 
        	$n_j=3000$ with $K=3$ (central panels),
        	$n_j=5000$ with $K=4$ (right panels).
        	We further include a 5\% uniform contamination $m \approx 0.05 \sum_j^K n_j$, and each method trims $\alpha \approx m /(n+m) - m / \{10 (n+m)\}  $; the last term is used since we do not rely on the function \verb+MixSim+, and thus noise overlaps with true clusters in this setting. 
        	For tk-merge we use $k \approx 2 K \log{n}$, and $r=1000$ for TCLUST.
        \end{enumerate}

    	\begin{figure}
    		\centering
    		\begin{subfigure}{.5\textwidth}
    			\centering    			\includegraphics[width=0.8\linewidth]{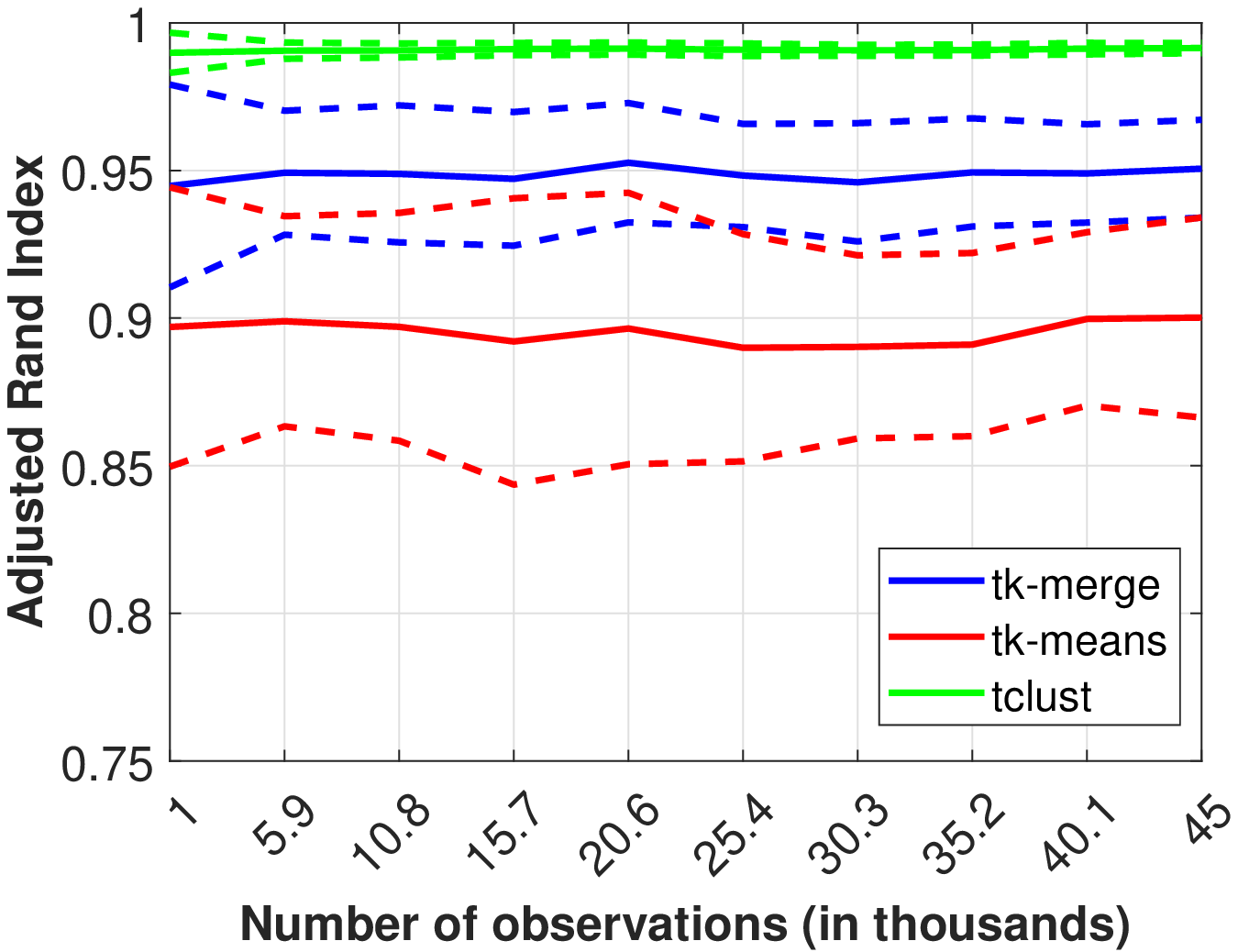}
    		\end{subfigure}%
    		\begin{subfigure}{.5\textwidth}
    			\centering
			\includegraphics[width=0.8\linewidth]{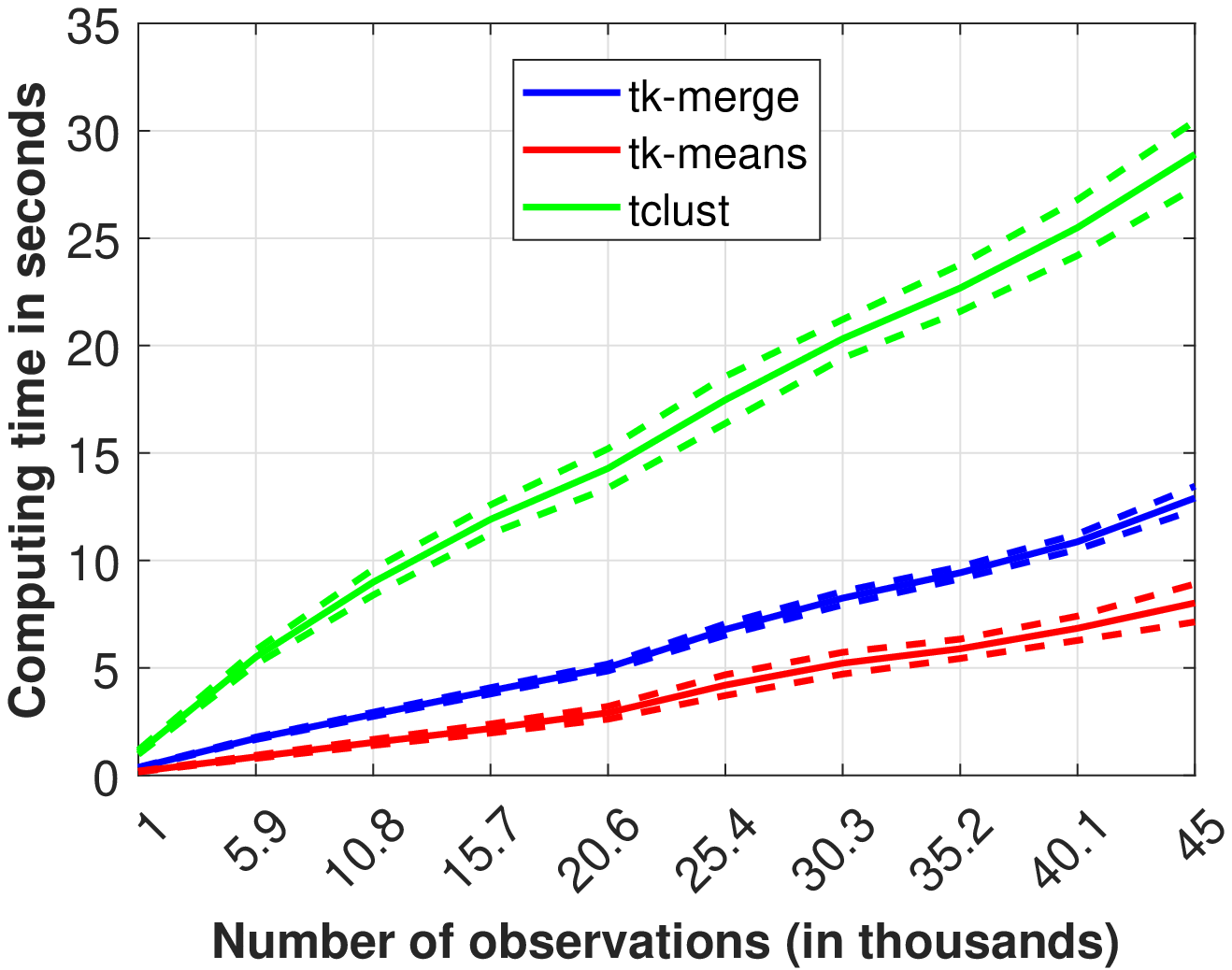}
    		\end{subfigure}
    		\caption{Simulation scenario 1. Median ARI (left) and computing time (right) for tk-merge, tk-means and TCLUST as a function of the sample size across 100 replications. 
    		Dashed lines represent $ \text{median}\pm S_n$. }
    		\label{fig:ex1}
    	\end{figure}
    	
    \paragraph{Results for Scenario 1}

    	Figure~\ref{fig:ex} (left panel) shows a typical simulation instance for $n=6000$.
    	Figure~\ref{fig:ex1} (left panel) compares ARI as a function of the sample size.
    	All methods perform well, and they are able to identify and discard noisy observations and find the real clusters.	
    	TCLUST shows nearly optimal estimates in terms of precision and stability (tight $S_n$ intervals); 
    	tk-merge improves tk-means solutions, 
    	both in terms of accuracy and stability.
    	Figure~\ref{fig:ex1} (right panel) compares computing time  as a function of the sample size.
    	As expected, the computational burden for TCLUST is significantly higher than other methods. Here tk-merge adds yet some to the computational burden of tk-means, but not a lot. 
    	Algorithmic complexity across different methods grows linearly with $n$, and  tk-merge has a lower rate then TCLUST. This holds for experiments with larger sample sizes as well; see Figure~\ref{fig:largeN} in Appendix~\ref{app:sim} for details.
    	The percentage gain in computing time for tk-merge and tk-means with respect to TCLUST varies between 50-70\% and 70-85\%, respectively (see Figure~\ref{fig:timerel} in Appendix~\ref{app:sim}).

    \begin{figure}
            \centering
            \begin{subfigure}{.5\textwidth}
                \centering
                \includegraphics[width=0.8\linewidth]{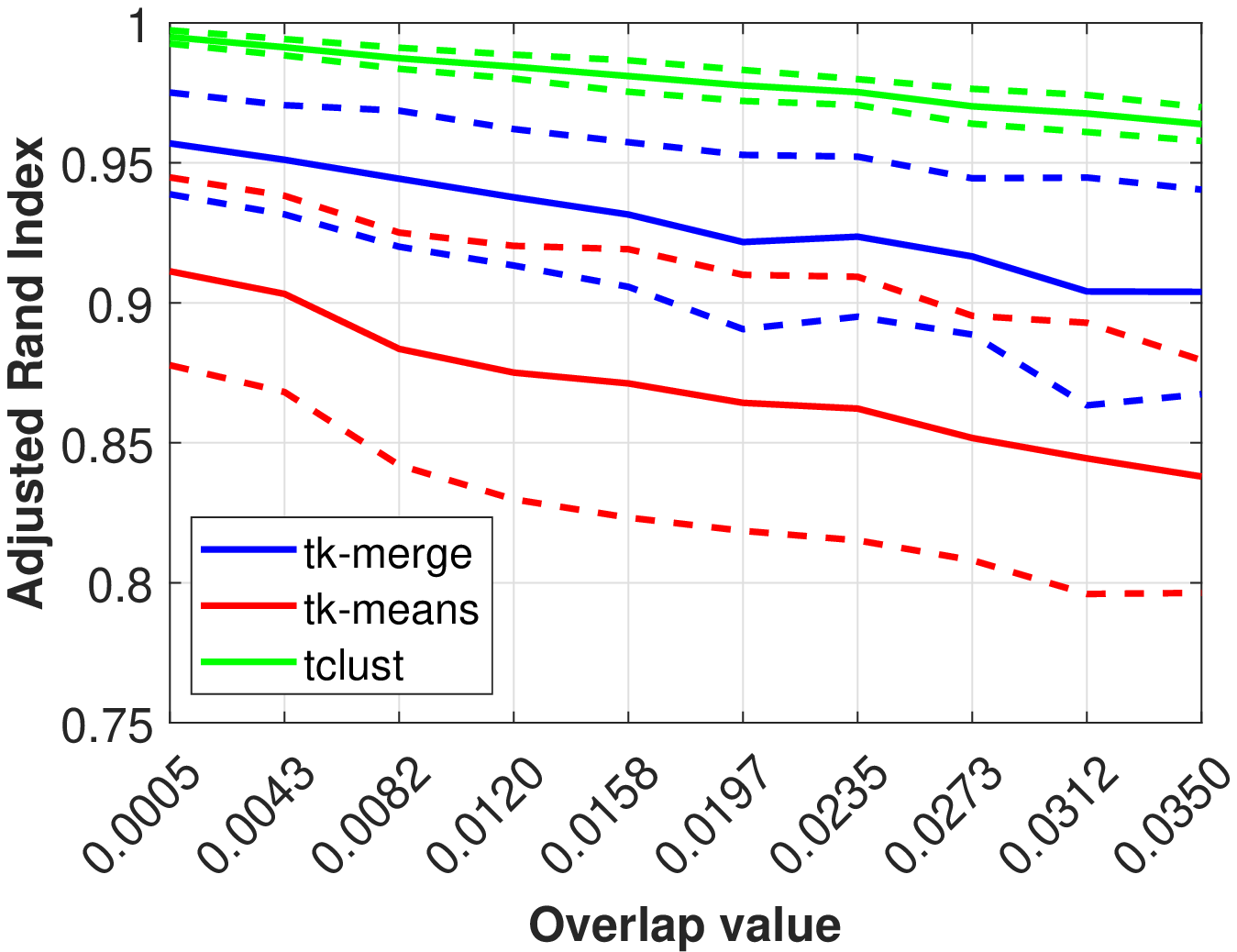}
            \end{subfigure}%
            \begin{subfigure}{.5\textwidth}
                \centering
            \includegraphics[width=0.8\linewidth]{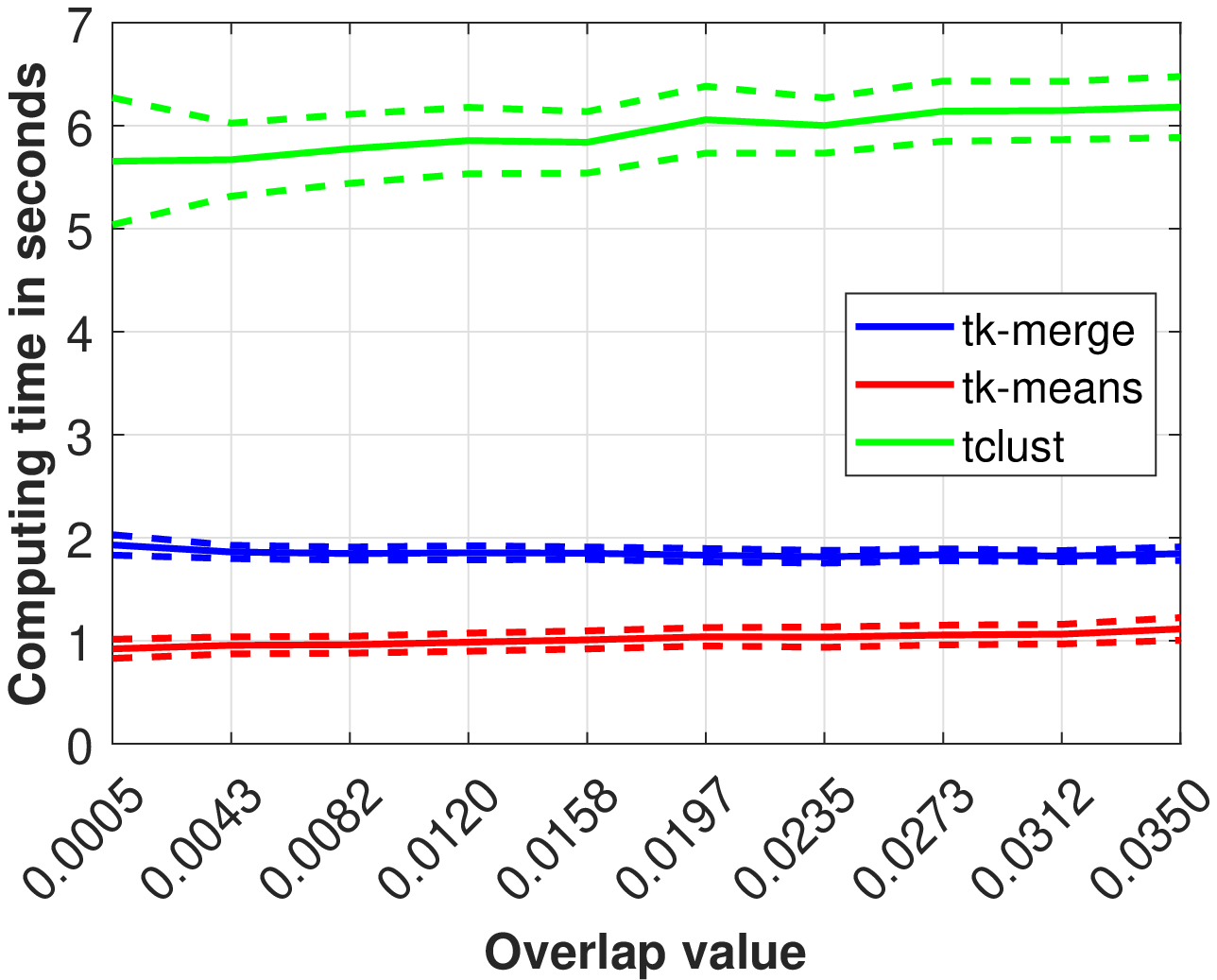}
            \end{subfigure}
            \caption{Simulation scenario 2. Median ARI (left) and computing time (right) for tk-merge, tk-means and TCLUST as a function of the sample size across 100 replications. Dashed lines represent $ \text{median}\pm S_n$.}
            \label{fig:ex2}
        \end{figure}

    \paragraph{Results for Scenario 2}
    	      Figure~\ref{fig:ex} (right panel) shows a typical simulation instance for $ \omega = 0.035$.
    	     Figure~\ref{fig:ex2} (left panel) shows that the median ARI for all methods decreases as $ \omega $ increases. 
             Here tk-merge improves again upon tk-means, but it suffers compared to TCLUST since for higher overlaps, where assumption (i) is violated, some components and/or noise are likely merged together.
             Figure~\ref{fig:ex2} (right panel) shows the computational burden across different methods,  which remains constant. Here the percentage gain for tk-merge and tk-means is approximately 70 and 85\%, respectively (see again Figure~\ref{fig:timerel} in Appendix~\ref{app:sim}).

    \paragraph{Results for Scenario 3}

    	Figure~\ref{fig:sim3} shows the clustering partitions for TCLUST (top panels) and tk-merge (bottom panels) for three synthetic datasets; 
    	results for tk-means were worse and they are not reported. 
    	In all datasets our proposal effectively detects the true clusters and discards the uniform noise therein.
    	On the other hand, TCLUST performs poorly, since its assumptions are violated (i.e.,~the presence of non-elliptical components).

    \section{Real-World Applications} \label{Sec5: application}
        
        Clustering problems are widespread across domains.
        We focus on examples characterized by general-shaped clusters, large sample sizes affected by data contamination, and where a reasonable choice for the number of existing clusters $K$ is available.
        Specifically, we analyze image, human mobility, retinal, and weather data. Note that in some cases we consider the presence of a single cluster, which might seem quite unusual. This assumption is motivated by the fact that outliers and/or noise are present in all of these applications, and they might be thought of as an additional ``cluster''.
        Importantly, a key feature of our proposal is that a single group can comprise disconnected components, and we exploit this strategy in Example~2 where a reasonable $K$ is unknown.
        Three out of the four datasets in use are openly available (see below), and no personally identifiable information or offensive content is disclosed. Source code to replicate our results can be found in the supplementary material.

        \paragraph{Example 1: Color Quantization}
            
            Crack detection is essential for structural health monitoring and inspection of buildings \cite{ozgenel2018performance,zhang2016road}.
            We analyze photographs of concrete cracks in infrastructures, and images are freely available (licensed under CC BY 4.0) at \url{https://data.mendeley.com/datasets/5y9wdsg2zt/2}.
            Our main goal is to distinguish cracks from concrete itself, as well as detecting and discarding noise and/or outliers (which arise due to different materials, weather conditions, blurring, illumination, etc.).
            Thus, pixels are clustered to reduce the original number of unique colors, and we 
            set $K=2$; ideally: black for cracks, and gray for surrounding concrete areas. 
            Each image (in \verb+jpg+ format) contains 227$\times$227 pixels with RGB channels, and the associated dataset $\bm{X}$ results in a $ 51529 \times 3 $ matrix.
            
\begin{figure}
        		\centering
        		\begin{subfigure}{.5\textwidth}
    			\centering
            	\includegraphics[width=1\linewidth]{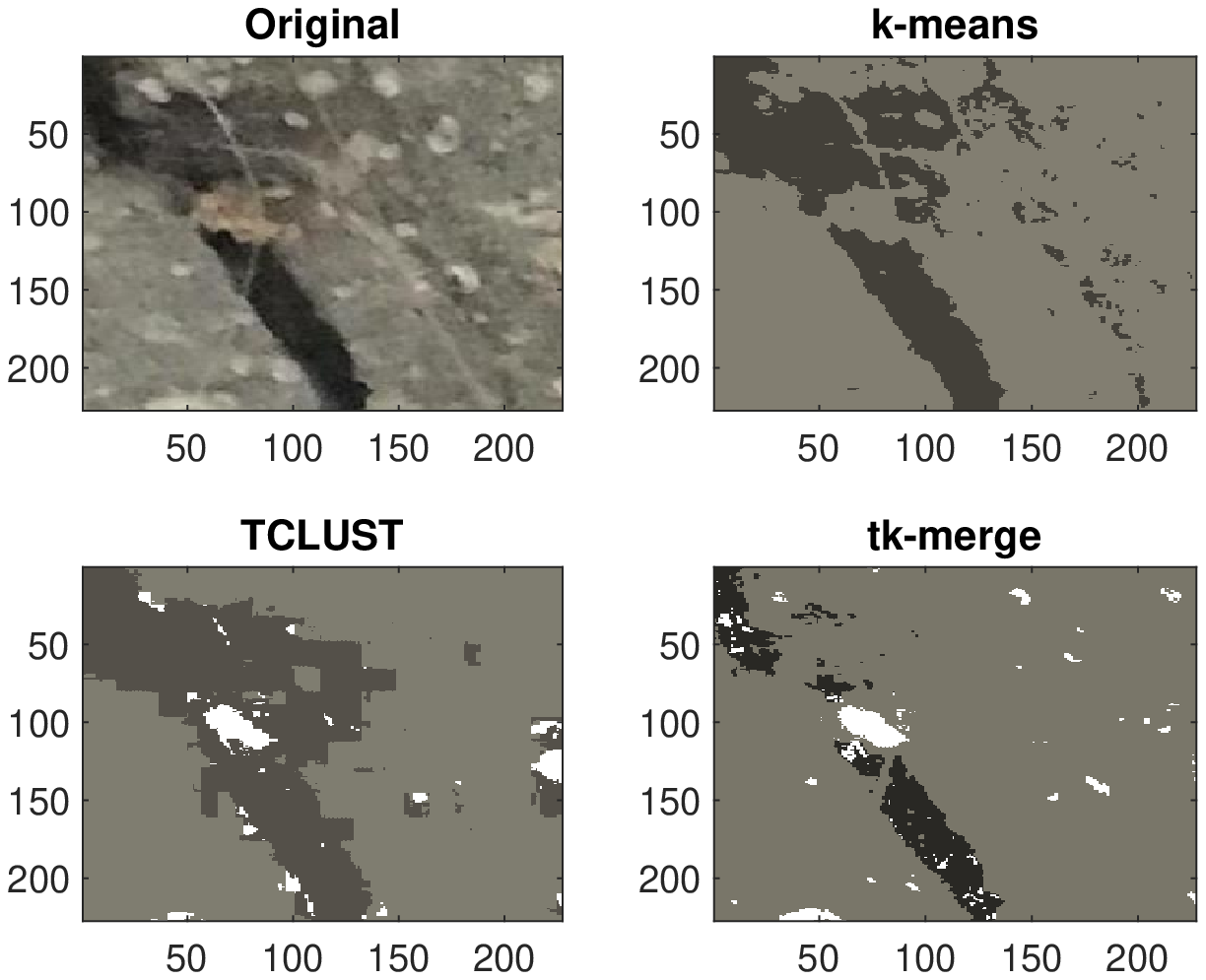}
        		\end{subfigure}%
        		\begin{subfigure}{.5\textwidth}
        			\centering
    			\includegraphics[width=1\linewidth]{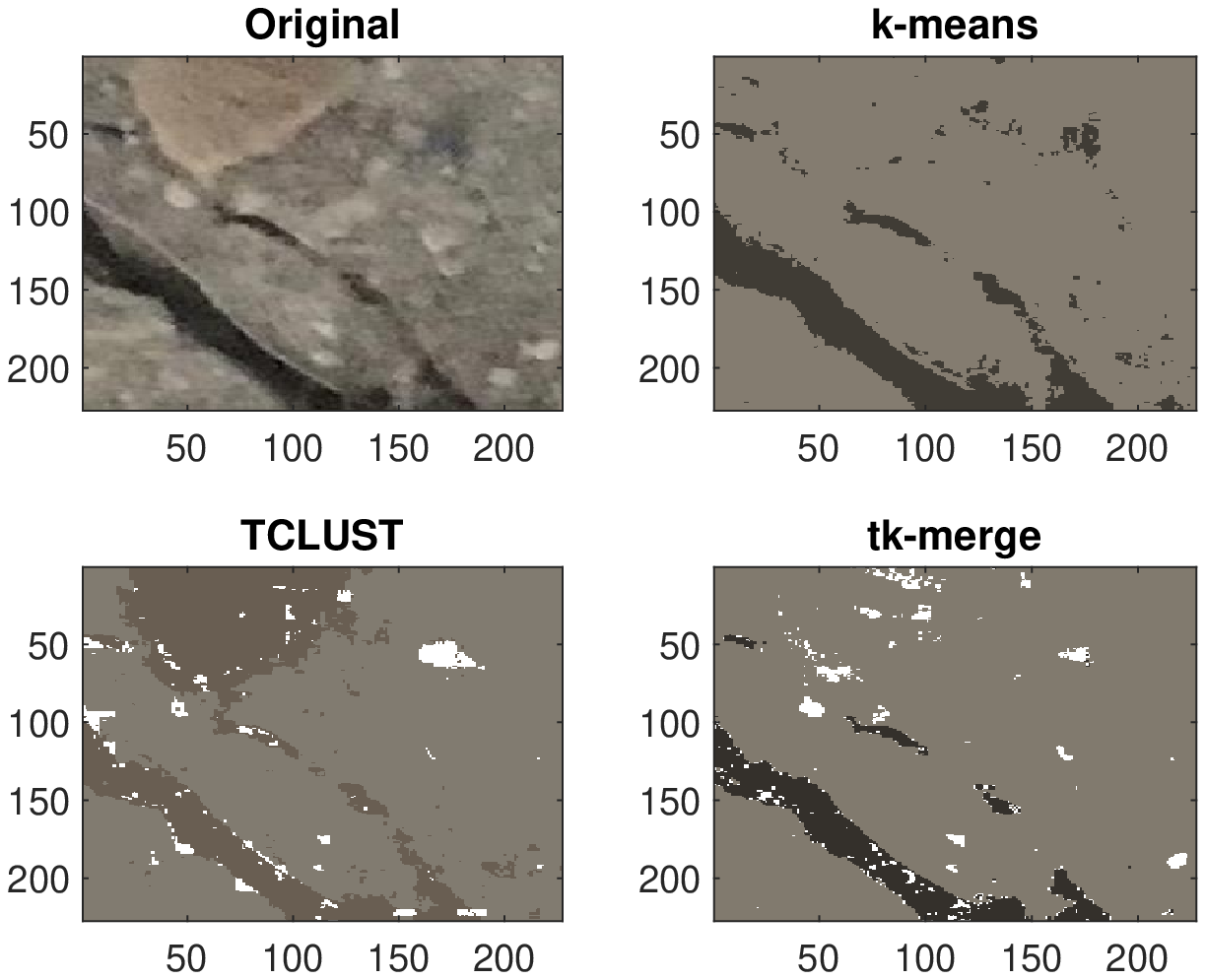} 
        		\end{subfigure}
        		\caption{Example 1. Two images of concrete cracks with clustering results for 
        		k-means, TCLUST and tk-merge,
        		using $K=2$ and $ \alpha = 2.5\%$.
        		White pixels denote the estimated noise/outlying cases.}
        		\label{fig:concrete}
        	\end{figure}

            Figure~\ref{fig:concrete} shows the partitions found by k-means, TCLUST, and tk-merge (using $k \approx 2 \log n $) for two typical images (results for tk-means were worse and not included); robust methods use $\alpha \approx 2.5\%$.
            Clustering solutions assign the value of the corresponding cluster's centroid to each pixel, where white pixels represent the points trimmed by robust methods (i.e.,~estimated as noise/outliers).
            The left panel shows that only tk-merge is able to effectively detect the black crack, where its shape and color are extremely close to the original one. On the other hand, k-means and TCLUST suffer under this setting (the shape of the crack is less defined and has a dark gray color).
            Similar conclusions hold for the results in the right panel -- where the original photo includes two cracks, and one of these is very thin and less marked.
            A similar clustering approach can be extended to several domains, such as geo-spatial images (e.g.,~detection of rivers, streets, etc.), artworks, etc.

        \paragraph{Example 2: Human Mobility}
            
            Global Navigation Satellites Systems provide extremely valuable information in several domains.
            Nowadays, positioning devices can be easily embedded in nearly any device, thus their widespread use.
            In this application we focus on human mobility data.
            Specifically, we consider taxy data for the 
            city of Shenzhen, China \citep{zhang2015urbancps}.
            GPS taxi trajectories are available (for academic research purposes only) at \url{www.cs.rutgers.edu/~dz220/data.html} as part of the \textit{Urban Data Release V2}.
            Our goal is to detect a single cluster indicating the most trafficked areas of the city. 
            The dataset in use contains trajectories of 13 unique taxicabs (which we treat as a whole), and the resulting matrix $\bm{X}$ has dimension $ 5 42129 \times 2 $ (latitude and longitude).
            
\begin{figure}
        		\centering
        		\begin{subfigure}{.33\textwidth}
        			\centering
            	\includegraphics[width=1\linewidth]{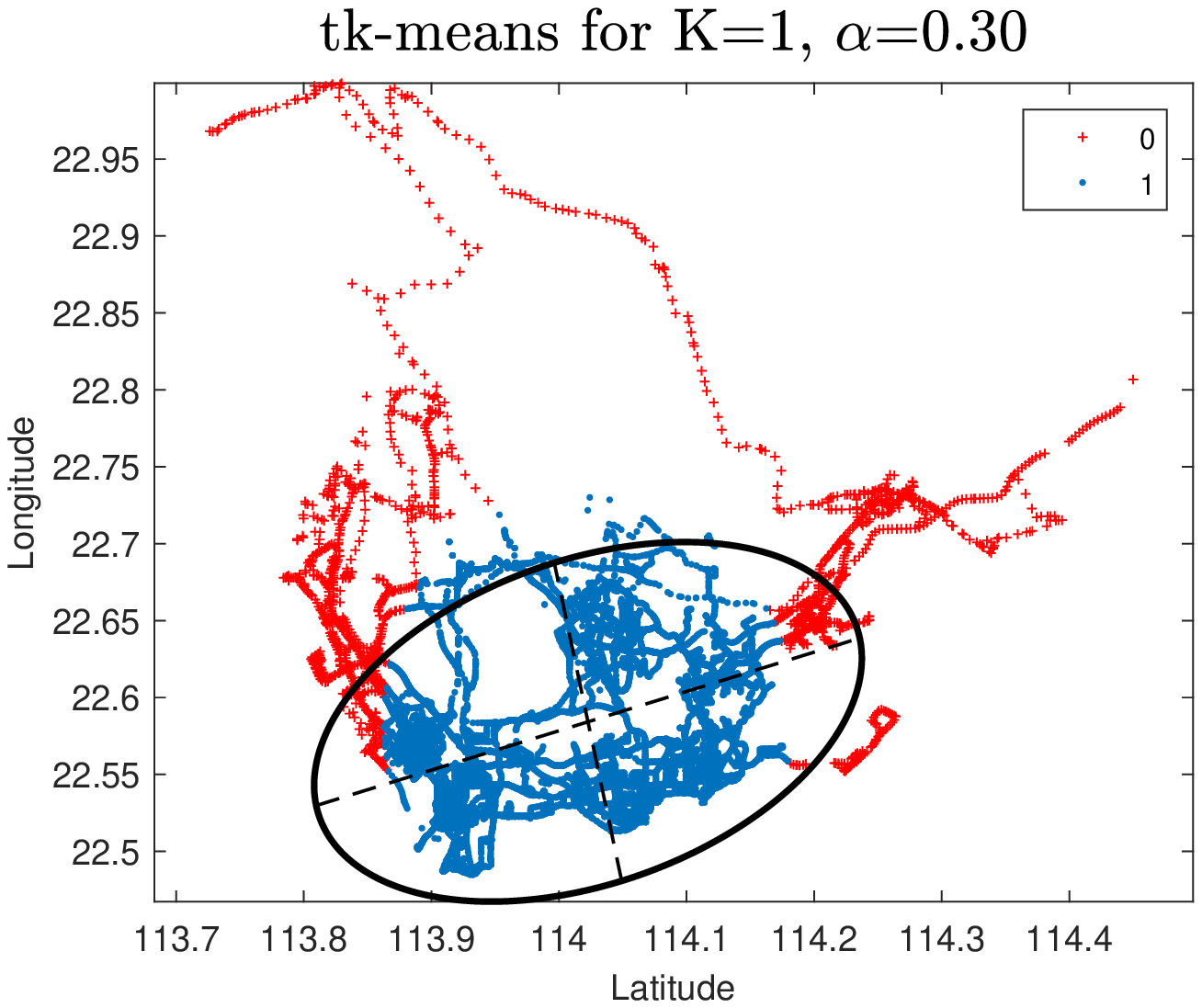}
        		\end{subfigure}%
        		\begin{subfigure}{.33\textwidth}
        			\centering
    			\includegraphics[width=1\linewidth]{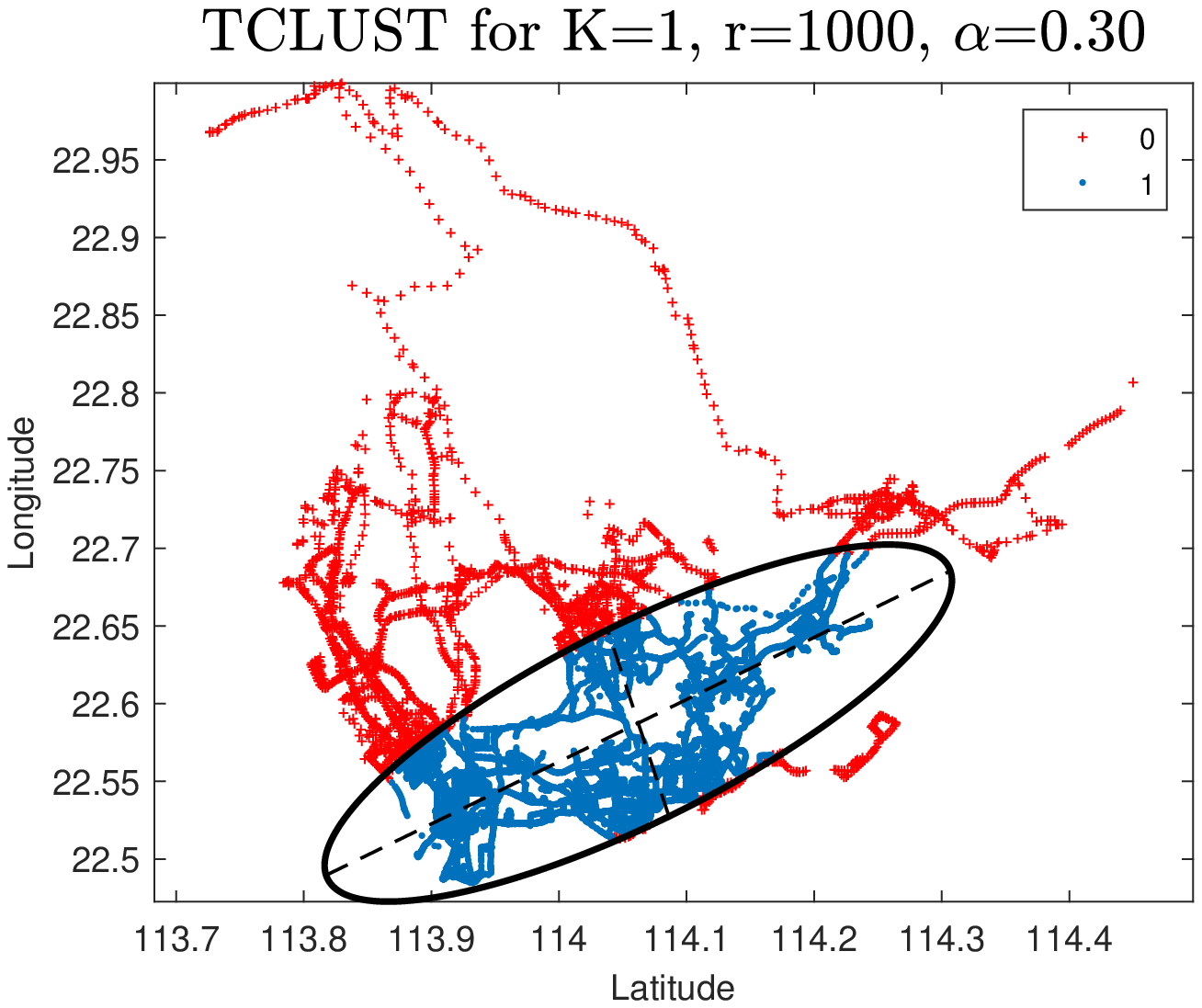}
        		\end{subfigure}
        		\begin{subfigure}{.33\textwidth}
        			\centering
    			\includegraphics[width=1\linewidth]{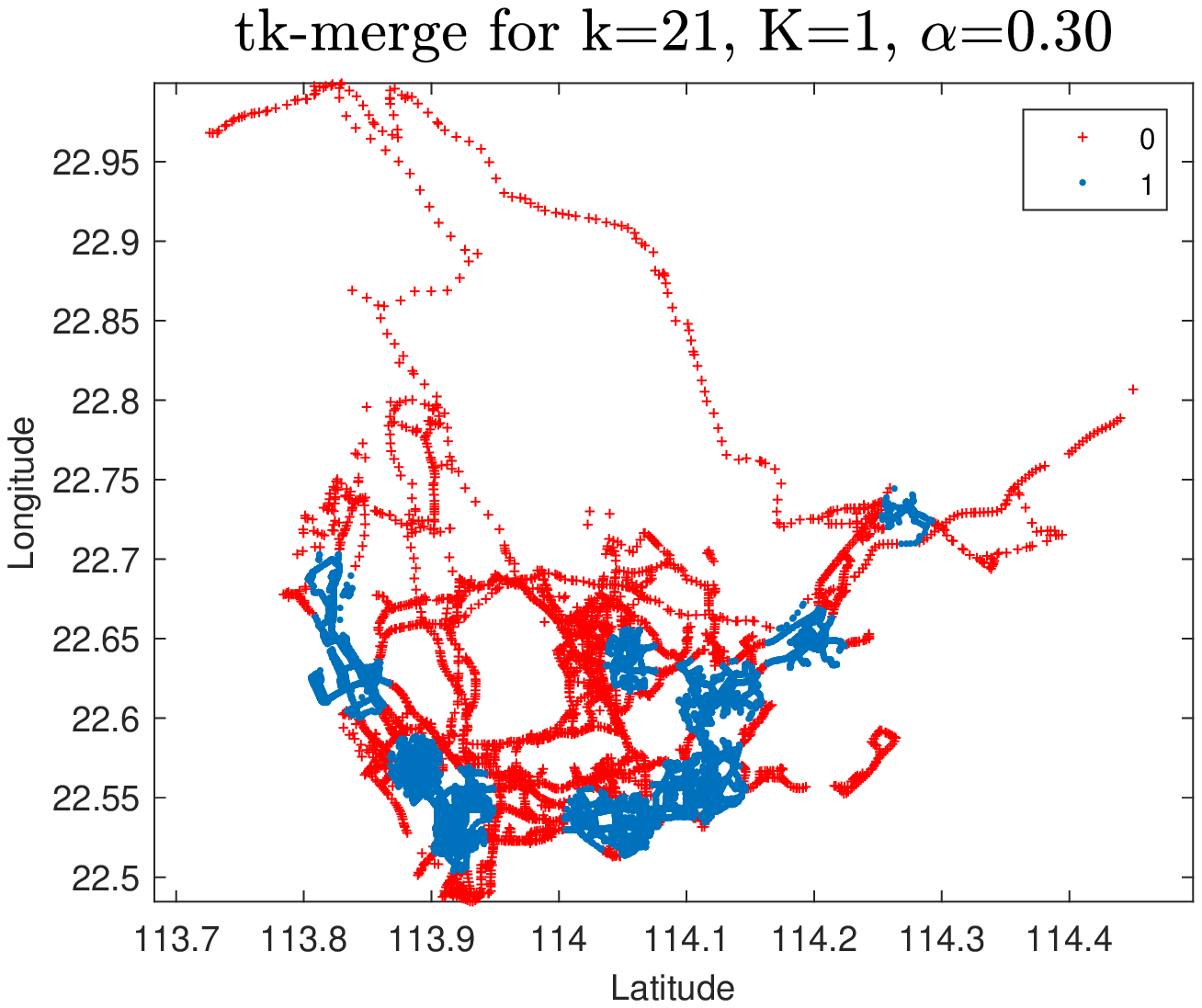}
        		\end{subfigure}
        		\caption{Example 2. Clustering partitions on taxicabs trajectories for tk-means (left), TCLUST (center) and tk-merge (right) for $K=1$ and 30\% trimming.}
        		\label{fig:mobility}
        	\end{figure}

            Figure~\ref{fig:mobility} shows the clustering partitions provided by tk-means, TCLUST and tk-merge (with $k \approx 2 \log n $), and they all use a trimming level  $ \alpha \approx 30\% $.
            This choice was dictated by the ``semi-automated'' procedure in use; see Appendix~\ref{app:application} for details.
            Despite their robustness, tk-means and TCLUST (left and central panels) suffer in this scenario, where western parts of the city are completely neglected and several less trafficked locations are clustered together. This is due to a violation of the underlying assumptions (e.g.,~the presence of non-elliptical components), as well as the use of $K=1$ since the ``true'' $K$ is unknown.
            On the other hand, tk-merge (right panel) effectively detects most hubs present in the GPS trajectories data. 
            Remarkably, unlike existing robust methods, the units assigned to the single cluster do not belong to a convex set.
            This is not unexpected due to the nature of the data, as it is more likely that the initial inflated components detected by tk-means identify some overlapping sets of outliers.
            Such a property is desirable in several applications (e.g.,~network data) where $K$ is unknown, as it allows one to identify a single cluster encompassing disconnected components.
            For instance, tk-merge solution indicates the presence of less than 10 groups, as opposed to the initial 21 clusters found by tk-means, and this solution may be refined further (e.g.,~using TCLUST with $K=10$).
            A similar clustering approach can be extended to other domains and data sources (e.g.,~social media access, car/bike sharing services, etc.), and it provides valuable insights to support strategical decisions for policymakers (e.g.,~public security), businesses (e.g.,~retailing companies), as well as households (e.g.,~housing market).
            
    \paragraph{Example 3: Retinographic Study}
        Diabetic retinopathy is one of the major  causes of blindness and vision defects in developed countries. 
        Thus, its early detection can play a significant role is mitigating these risks. 
        Several machine learning approaches have been developed to analyze 
        retinal images  and thus aid medical experts \citep{sanchez2009retinal}.
        We follow the modeling strategy described in \cite{garcia2017fitting}, which is based on  binarized retinographic images
        (we received a dataset from the authors, that is not publicly available).
        Indeed, major blood vessels arch above and below 
        the posterior pole of the retina,
        which can be approximated by parabola shapes,
        and it is important to detect lesions that appear in this area of the retina.
        While \cite{garcia2017fitting} leverage information about the data structure --  fitting parabolas which can tolerate a certain amount of data contamination based on trimming ideas -- we aim to achieve a similar goal in a model-free setting.
        This task is more difficult than the examples described above for the presence of a ``very thin'' single cluster,  which locates very closely to outlying observations and, as it is common in several biomedical studies, the sample size is relatively low. 
        The dataset $\bm{X}$ results in a matrix of dimension $ 1210 \times 2 $.

\begin{figure*}
            \centering
            \begin{subfigure}[b]{0.49\textwidth}   
                \centering 
                \includegraphics[width=0.8\textwidth]{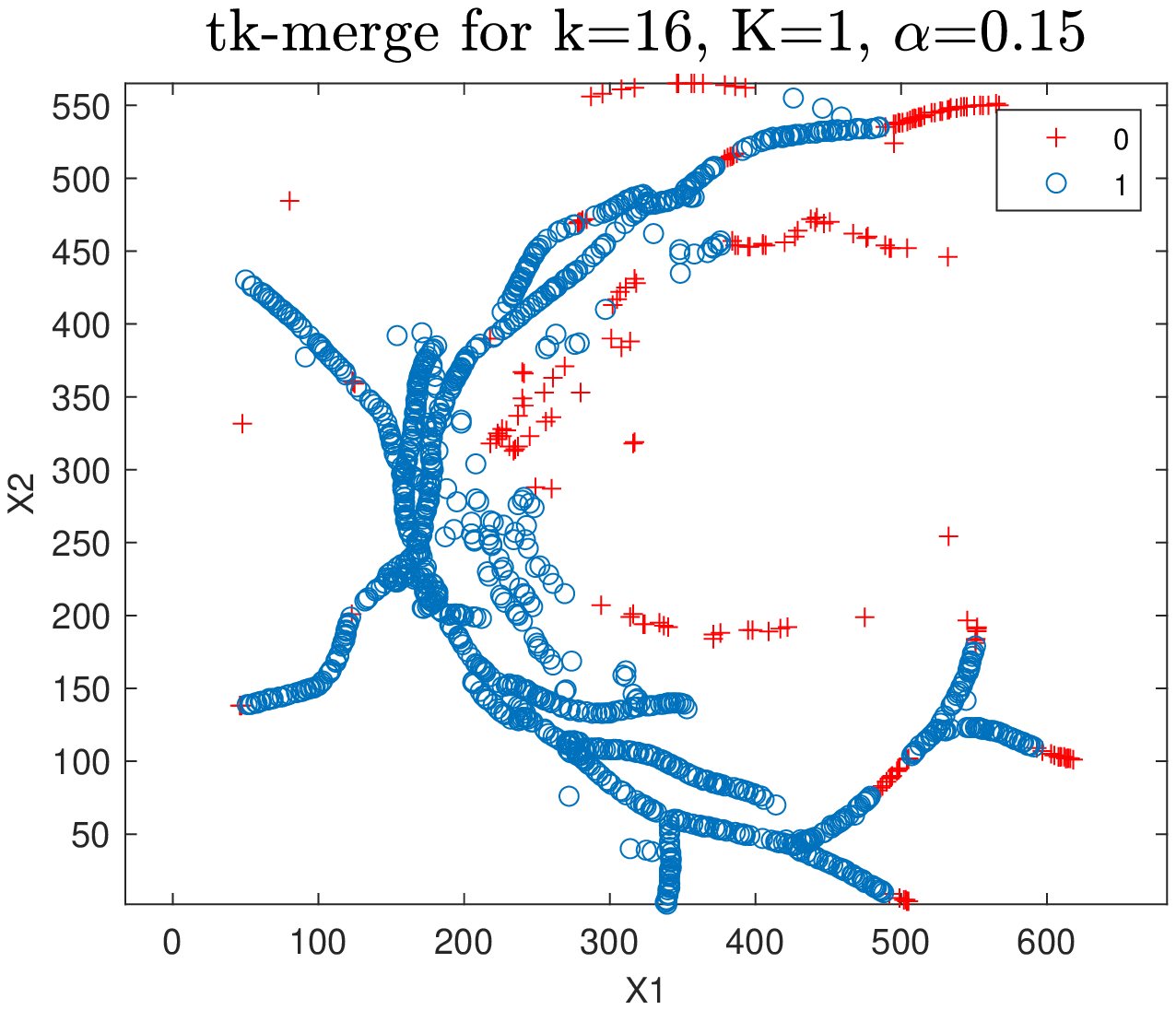}
            \end{subfigure}
            \begin{subfigure}[b]{0.49\textwidth}   
                \centering 
                \includegraphics[width=0.8\textwidth]{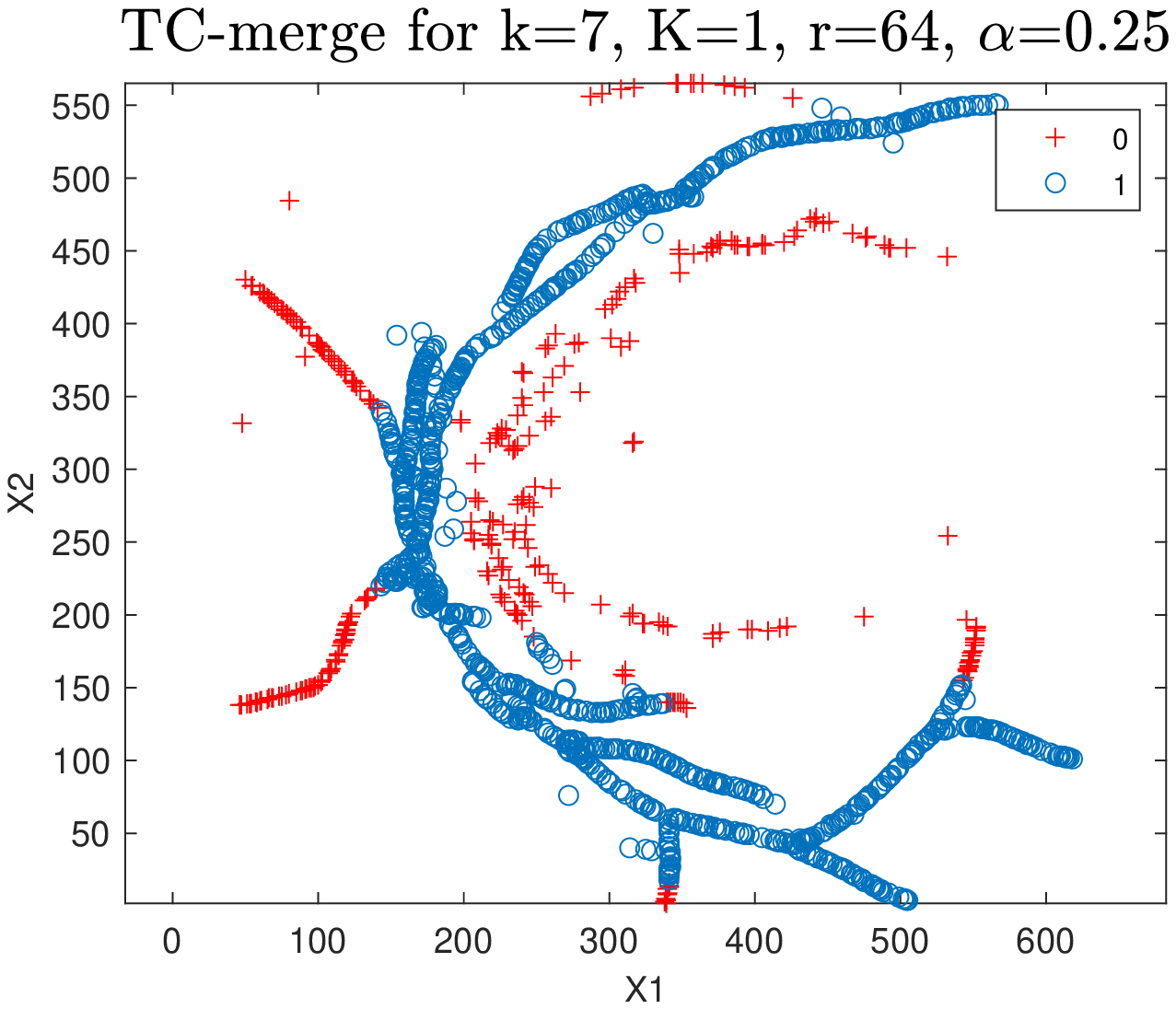}
            \end{subfigure}
            \caption{Example 3. Clustering solution of retinal data for tk-merge (left) and TC-merge (right) using 15\% and 25\% trimming, respectively, and $K=1$. Red points denote the estimated noise/outlying cases.} 
            \label{fig:retina}
        \end{figure*}

        Figure~\ref{fig:retina} compares  tk-merge (left panel) and TC-merge (right panel) solutions; results for tk-means were worse and they are not reported. 
        We set $k \approx 2 \log n $ with $ \alpha \approx 15\%  $ for the former, and $k  \approx \log n $ with $ \alpha \approx 25\% $ for the latter. The choice of these trimming levels was dictated by the ``semi-automated'' procedure in use; see Appendix~\ref{app:application} for details.
        In this challenging setting, TC-merge outperforms tk-merge due to its flexibility in accommodating a larger range of data at the expenses of a higher computing time, which is however quite negligible for the considered sample size.
        Nevertheless, tk-merge provides a fairly accurate and comparable partition.

        \paragraph{Example 4: Weather Data}

\begin{figure*}
            \centering
            \begin{subfigure}[b]{0.49\textwidth}   
                \centering 
                \includegraphics[width=0.8\textwidth]{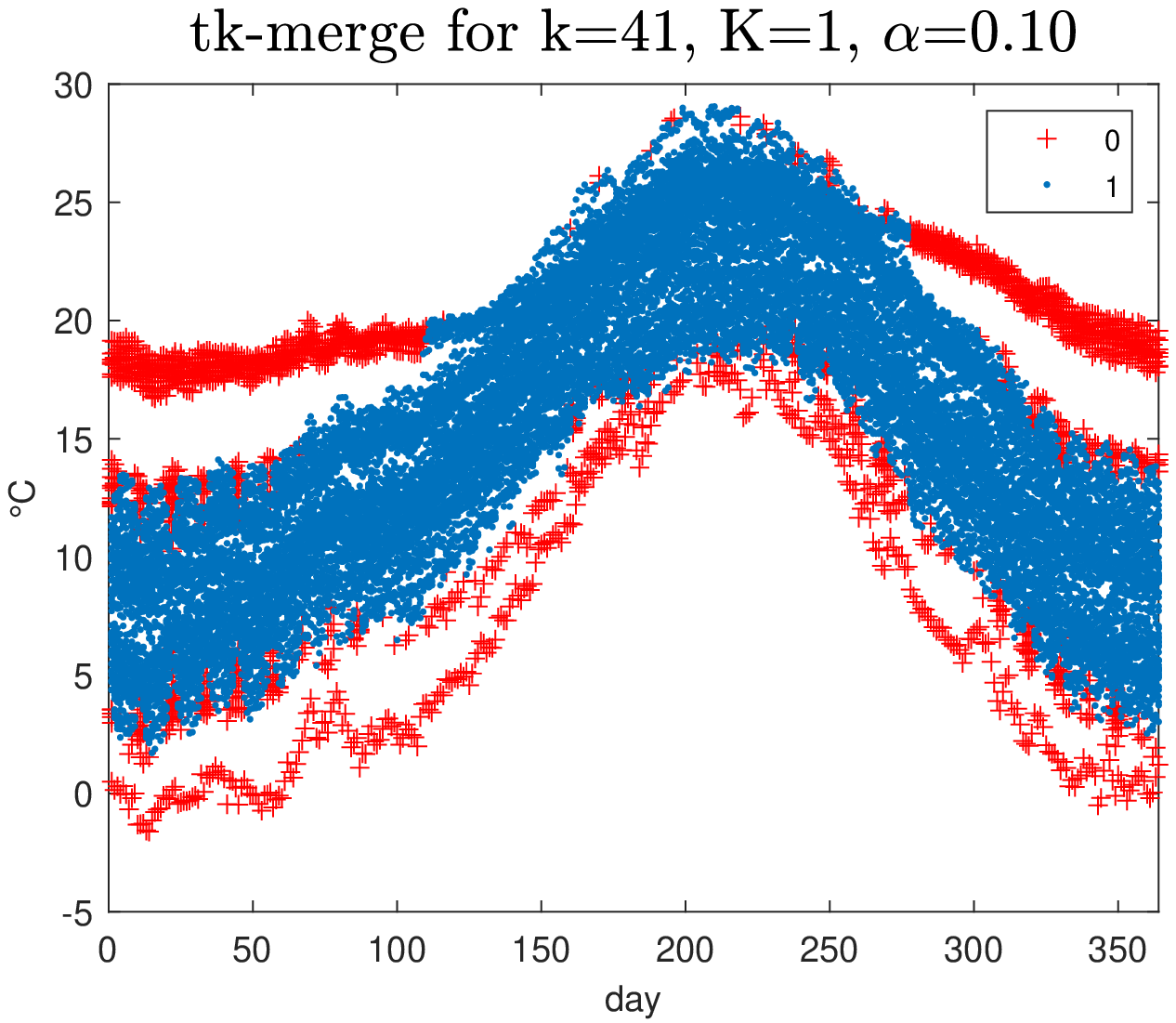}
            \end{subfigure}
            \begin{subfigure}[b]{0.49\textwidth}   
                \centering 
                \includegraphics[width=0.8\textwidth]{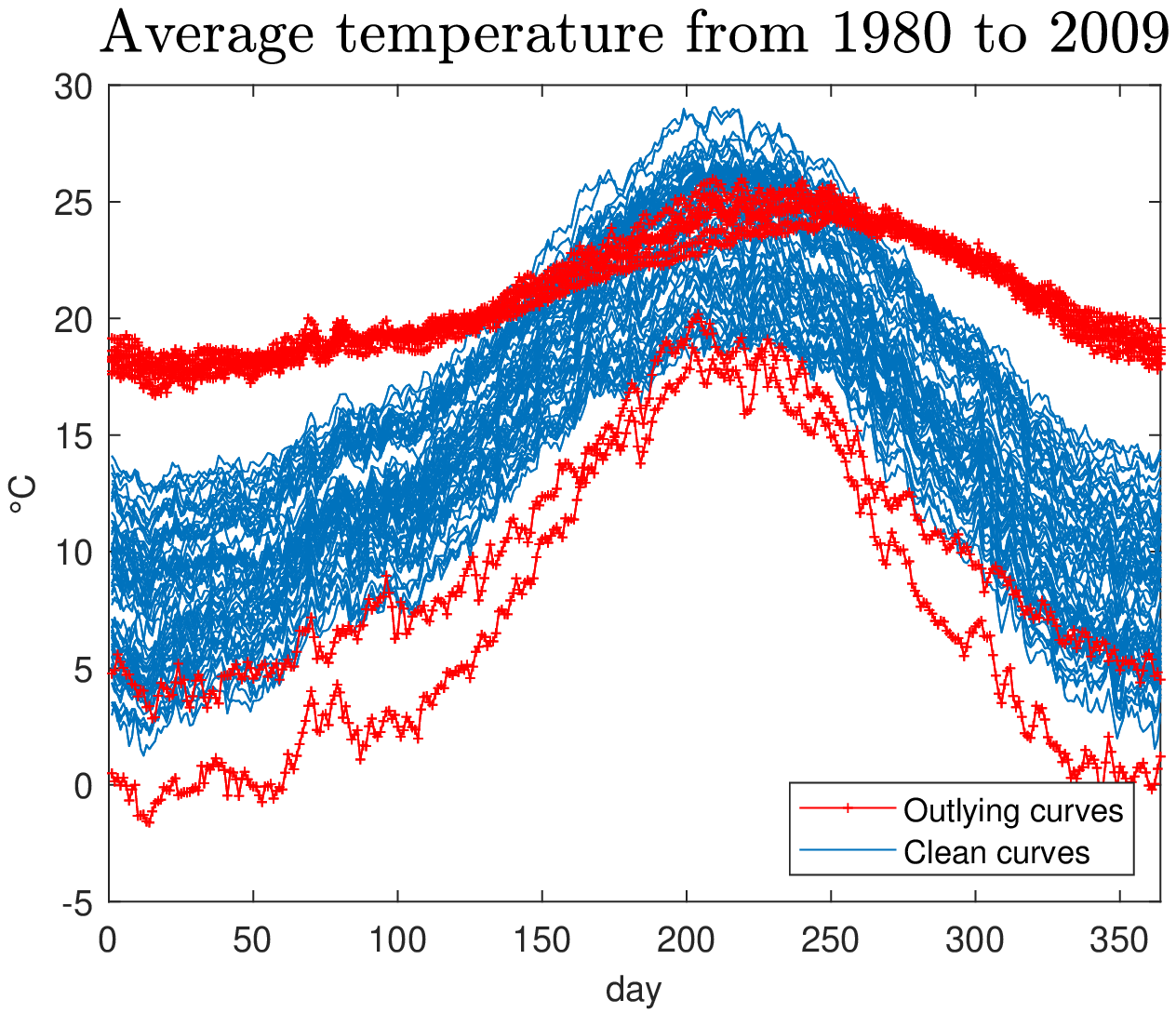}
            \end{subfigure}
            \caption{Data for the average daily temperature across 73 Spanish weather stations in the years 1980-2009. Left panel: partition obtained by tk-merge where red points denote trimmed observations. Right panel:  curves for each weather station, where red ones are detected as atypical by tk-merge.} 
            \label{fig:FDA}
        \end{figure*}

        As an example of the use of tk-merge for functional data analysis, we consider the average temperature across 73 Spanish weather stations for the period 1980-2009 \cite{febrero2012statistical}. 
        The matrix $\bm{X}$ contains $73\times365$ rows (number of stations by days) and 2 columns (temperature and day of the year).
        This dataset is freely available on \verb+CRAN+ repository as part of the \verb+fda.usc+ package in \verb+R+ (see \url{https://rdrr.io/cran/fda.usc/man/aemet.html}).
        Since the dataset contains highly dense regions, we apply tk-merge with $k \approx 4 \log n $. We set $K=1$ and $ \alpha \approx 0.1 $ in order to detect a single group of typical observations and identify aberrant points. 
        The left panel of Figure~\ref{fig:FDA} shows the partition provided by tk-merge, where red points denote trimmed observations estimated as outliers. 
        This highlights that our proposal effectively identifies the core of the data.
        The right panel of Figure~\ref{fig:FDA} represents different locations as separate curves.
        Red curves indicate the ones for which tk-merge detected at least $365/3$ days across the year as outliers (i.e.,~the red points on the left panel).
        Our results are in in line with previous studies and support the effectiveness of tk-merge to tackle some forms of functional outliers. For instance, this technique can be used as a computationally lean, pre-processing tool, since outliers might hinder the performance of non-robust methods for functional data analysis. 

    \section{Final Remarks}
    \label{Sec6: conclusions}
        
    	The proposed approach combines existing agglomerative hierarchical strategies with robust clustering algorithms based on trimming. Unlike many popular methods, our new hybrid procedure effectively detects and discards contaminated data, and identifies general-shaped clusters occurring in real applications under very general assumptions and in a computationally lean manner.
    	Also under ideal assumptions, for example in the presence of Gaussian components, its accuracy is comparable with state-of-the-art robust methods,  but it saves computing time. These properties have been demonstrated with representative case studies and simulation experiments.
    	There are theoretical and practical developments of great interest that need to be addressed to understand the main limitations induced by the procedure's assumptions (stated in Section~\ref{Sec2: model}), in view to explore more general settings. We conclude with an anticipation of our next steps.
    	
    	We are studying the effects of strong overlap between clusters. When this occurrence challenges the current assumptions, the use of DEMP in place of Euclidean distances could be effective and would provide a sound probabilistic foundation to the method similarly to \citep{melnykov2019clustering}. Also hierarchical clustering may be replaced by other techniques in this case.

    	Moreover, parameters tuning, as well as the sensitivity of our algorithm to these choices, demand further investigation. For this matter, we are currently investigating extensions of the “semi-automatic” procedure used in the present work.
    	Since our proposal is very effective in detecting  disconnected components belonging to the same cluster, we believe that it can be used to obtain an initial estimate of $K$.
    	A related but more challenging extension concerns the model and parameter adaptations needed to address contaminants from asymmetric or heavily-tailed distributions (for example \cite{PeelMcLachlan:2000} treats Student's $t$ in mixture models).

	    Our longer term challenges cover natural extensions to \textit{functional data analysis} \cite{ramsay2005functional} and \textit{clustered regression data} \cite{garcia2010robust}: in these settings the robustness and flexibility of our proposal should allow modeling (contaminated) linear and nonlinear data structures effectively and efficiently.
    	Moreover, we are interested in exploring the use in combination with bi-clustering and feature selection methods, to increase the interpretability of clustering partitions in high-dimensional problems that are ubiquitous nowadays \citep{forero2012robust}.

        In order to reach the wide user community interested in clustering challenging real data, it is necessary to conduct extensive comparisons with techniques other than model-based clustering, such as spectral clustering \citep{von2007tutorial}, DBSCAN \citep{ester1996density} and their extensions \citep{de2021birchscan,smiti2012dbscan}.

    	Finally, we remark that some ethical concerns might arise from malicious use of the proposed techniques (e.g.,~targeting minorities, mass surveillance, tracking and identifying individuals, etc.), especially if they are deployed in fully automated procedures -- which are however more difficult to develop for complex data structures. 
    	We thus believe that the role of human analysts is still a crucial element, and these risks can be mitigated by the adherence to shared ethic guidelines.

\bibliographystyle{abbrvnat}
\bibliography{biblio.bib}

\begin{thebibliography}{47}
\providecommand{\natexlab}[1]{#1}
\providecommand{\url}[1]{\texttt{#1}}
\expandafter\ifx\csname urlstyle\endcsname\relax
  \providecommand{\doi}[1]{doi: #1}\else
  \providecommand{\doi}{doi: \begingroup \urlstyle{rm}\Url}\fi

\bibitem[Banfield and Raftery(1993)]{banfield1993model}
J.~D. Banfield and A.~E. Raftery.
\newblock Model-based gaussian and non-gaussian clustering.
\newblock \emph{Biometrics}, 49\penalty0 (3):\penalty0 803--821, 1993.

\bibitem[Bock(2002)]{bock2002clustering}
H.~Bock.
\newblock Clustering methods: From classical models to new approaches.
\newblock \emph{Statistics in Transition}, 5\penalty0 (5):\penalty0 725--758,
  2002.

\bibitem[Breiman(2001)]{breiman2001statistical}
L.~Breiman.
\newblock Statistical modeling: The two cultures (with comments and a rejoinder
  by the author).
\newblock \emph{Statistical Science}, 16\penalty0 (3):\penalty0 199--231, 2001.

\bibitem[Cappozzo et~al.(2021)Cappozzo, García~Escudero, Greselin, and
  Mayo-Iscar]{cappozzo2021}
A.~Cappozzo, L.~A. García~Escudero, F.~Greselin, and A.~Mayo-Iscar.
\newblock Parameter choice, stability and validity for robust cluster weighted
  modeling.
\newblock \emph{Stats}, 4\penalty0 (3):\penalty0 602--615, 2021.

\bibitem[Celebi(2014)]{celebi2014partitional}
M.~E. Celebi.
\newblock \emph{Partitional Clustering Algorithms}.
\newblock Springer, 2014.

\bibitem[Celeux and Govaert(1995)]{celeux1995gaussian}
G.~Celeux and G.~Govaert.
\newblock Gaussian parsimonious clustering models.
\newblock \emph{Pattern Recognition}, 28\penalty0 (5):\penalty0 781--793, 1995.

\bibitem[Cerioli and Perrotta(2014)]{cerioli2014robust}
A.~Cerioli and D.~Perrotta.
\newblock Robust clustering around regression lines with high density regions.
\newblock \emph{Advances in Data Analysis and Classification}, 8\penalty0
  (1):\penalty0 5--26, 2014.

\bibitem[Cerioli et~al.(2018)Cerioli, Riani, Atkinson, and
  Corbellini]{cerioli2018power}
A.~Cerioli, M.~Riani, A.~C. Atkinson, and A.~Corbellini.
\newblock The power of monitoring: how to make the most of a contaminated
  multivariate sample.
\newblock \emph{Statistical Methods \& Applications}, 27\penalty0 (4):\penalty0
  559--587, 2018.

\bibitem[Cuesta-Albertos et~al.(1997)Cuesta-Albertos, Gordaliza, and
  Matr{\'a}n]{cuesta1997trimmed}
J.~A. Cuesta-Albertos, A.~Gordaliza, and C.~Matr{\'a}n.
\newblock Trimmed $ k $-means: An attempt to robustify quantizers.
\newblock \emph{The Annals of Statistics}, 25\penalty0 (2):\penalty0 553--576,
  1997.

\bibitem[Dempster et~al.(1977)Dempster, Laird, and Rubin]{dempster1977maximum}
A.~P. Dempster, N.~M. Laird, and D.~B. Rubin.
\newblock Maximum likelihood from incomplete data via the {EM} algorithm.
\newblock \emph{Journal of the Royal Statistical Society: Series B},
  39\penalty0 (1):\penalty0 1--38, 1977.

\bibitem[Ester et~al.(1996)Ester, Kriegel, Sander, and Xu]{ester1996density}
M.~Ester, H.-P. Kriegel, J.~Sander, and X.~Xu.
\newblock A density-based algorithm for discovering clusters in large spatial
  databases with noise.
\newblock In \emph{Proceedings of the Second International Conference on
  Knowledge Discovery and Data Mining}, KDD'96, page 226–231. AAAI Press,
  1996.

\bibitem[Febrero~Bande and Oviedo de~la Fuente(2012)]{febrero2012statistical}
M.~Febrero~Bande and M.~Oviedo de~la Fuente.
\newblock Statistical computing in functional data analysis: The {R} package
  {\verb+fda.usc+}.
\newblock \emph{Journal of Statistical Software}, 51\penalty0 (4):\penalty0
  1--28, 2012.

\bibitem[Forero et~al.(2012)Forero, Kekatos, and Giannakis]{forero2012robust}
P.~A. Forero, V.~Kekatos, and G.~B. Giannakis.
\newblock Robust clustering using outlier-sparsity regularization.
\newblock \emph{IEEE Transactions on Signal Processing}, 60\penalty0
  (8):\penalty0 4163--4177, 2012.

\bibitem[Fraley and Raftery(1998)]{fraley1998many}
C.~Fraley and A.~E. Raftery.
\newblock How many clusters? {W}hich clustering method? {A}nswers via
  model-based cluster analysis.
\newblock \emph{The Computer Journal}, 41\penalty0 (8):\penalty0 578--588,
  1998.

\bibitem[{FSDA}(2005-2021)]{FSDA:toolbox}
{FSDA}.
\newblock Flexible {S}tatistics \& {D}ata {A}nalysis toolbox for {MATLAB}, with
  extensions to {R}, {SAS} and other platforms.
\newblock Authors and developers: \url{http://rosa.unipr.it/FSDA/guide.html};
  GitHub: \url{https://github.com/UniprJRC/FSDA}; {M}atlab {C}entral {F}ile
  {E}xchange:
  \url{https://www.mathworks.com/matlabcentral/fileexchange/72999-fsda};
  On-line documentation: \url{http://rosa.unipr.it/FSDA/guide.html}, 2005-2021.

\bibitem[Garc{\'i}a-Escudero and Gordaliza(1999)]{garcia1999robustness}
L.~A. Garc{\'i}a-Escudero and A.~Gordaliza.
\newblock Robustness properties of k means and trimmed k means.
\newblock \emph{Journal of the American Statistical Association}, 94\penalty0
  (447):\penalty0 956--969, 1999.

\bibitem[Garc{\'i}a-Escudero et~al.(2003)Garc{\'i}a-Escudero, Gordaliza, and
  Matr{\'a}n]{garcia2003trimming}
L.~A. Garc{\'i}a-Escudero, A.~Gordaliza, and C.~Matr{\'a}n.
\newblock Trimming tools in exploratory data analysis.
\newblock \emph{Journal of Computational and Graphical Statistics}, 12\penalty0
  (2):\penalty0 434--449, 2003.

\bibitem[Garc{\'i}a-Escudero et~al.(2008)Garc{\'i}a-Escudero, Gordaliza,
  Matr{\'a}n, and Mayo-Iscar]{garcia2008general}
L.~A. Garc{\'i}a-Escudero, A.~Gordaliza, C.~Matr{\'a}n, and A.~Mayo-Iscar.
\newblock A general trimming approach to robust cluster analysis.
\newblock \emph{The Annals of Statistics}, 36\penalty0 (3):\penalty0
  1324--1345, 2008.

\bibitem[Garc{\'i}a-Escudero et~al.(2010)Garc{\'i}a-Escudero, Gordaliza,
  Matr{\'a}n, and Mayo-Iscar]{garcia2010review}
L.~A. Garc{\'i}a-Escudero, A.~Gordaliza, C.~Matr{\'a}n, and A.~Mayo-Iscar.
\newblock A review of robust clustering methods.
\newblock \emph{Advances in Data Analysis and Classification}, 4\penalty0
  (2-3):\penalty0 89--109, 2010.

\bibitem[Garc{\'\i}a-Escudero et~al.(2010)Garc{\'\i}a-Escudero, Gordaliza,
  Mayo-{\'I}scar, and San~Mart{\'\i}n]{garcia2010robust}
L.~A. Garc{\'\i}a-Escudero, A.~Gordaliza, A.~Mayo-{\'I}scar, and
  R.~San~Mart{\'\i}n.
\newblock Robust clusterwise linear regression through trimming.
\newblock \emph{Computational Statistics \& Data Analysis}, 54\penalty0
  (12):\penalty0 3057--3069, 2010.

\bibitem[Garc{\'i}a-Escudero et~al.(2011)Garc{\'i}a-Escudero, Gordaliza,
  Matr{\'a}n, and Mayo-Iscar]{garcia2011exploring}
L.~A. Garc{\'i}a-Escudero, A.~Gordaliza, C.~Matr{\'a}n, and A.~Mayo-Iscar.
\newblock Exploring the number of groups in robust model-based clustering.
\newblock \emph{Statistics and Computing}, 21\penalty0 (4):\penalty0 585--599,
  2011.

\bibitem[Garc{\'i}a-Escudero et~al.(2017)Garc{\'i}a-Escudero, Mayo-Iscar, and
  S{\'a}nchez-Guti{\'e}rrez]{garcia2017fitting}
L.~A. Garc{\'i}a-Escudero, A.~Mayo-Iscar, and C.~I. S{\'a}nchez-Guti{\'e}rrez.
\newblock Fitting parabolas in noisy images.
\newblock \emph{Computational Statistics \& Data Analysis}, 112:\penalty0
  80--87, 2017.

\bibitem[Hennig(2015)]{hennig2015true}
C.~Hennig.
\newblock What are the true clusters?
\newblock \emph{Pattern Recognition Letters}, 64:\penalty0 53--62, 2015.

\bibitem[Hubert and Arabie(1985)]{hubert1985comparing}
L.~Hubert and P.~Arabie.
\newblock Comparing partitions.
\newblock \emph{Journal of Classification}, 2\penalty0 (1):\penalty0 193--218,
  1985.

\bibitem[Johnson(1967)]{johnson1967hierarchical}
S.~C. Johnson.
\newblock Hierarchical clustering schemes.
\newblock \emph{Psychometrika}, 32\penalty0 (3):\penalty0 241--254, 1967.

\bibitem[Luchi et~al.(2019)Luchi, {Loureiros Rodrigues}, and {Miguel
  Varejão}]{de2021birchscan}
D.~Luchi, A.~{Loureiros Rodrigues}, and F.~{Miguel Varejão}.
\newblock Sampling approaches for applying {DBSCAN} to large datasets.
\newblock \emph{Pattern Recognition Letters}, 117:\penalty0 90--96, 2019.

\bibitem[Maitra and Melnykov(2010)]{maitra2010simulating}
R.~Maitra and V.~Melnykov.
\newblock Simulating data to study performance of finite mixture modeling and
  clustering algorithms.
\newblock \emph{Journal of Computational and Graphical Statistics}, 19\penalty0
  (2):\penalty0 354--376, 2010.

\bibitem[McLachlan and Peel(2004)]{mclachlan2004finite}
G.~McLachlan and D.~Peel.
\newblock \emph{Finite Mixture Models}.
\newblock John Wiley \& Sons, 2004.

\bibitem[Melnykov(2016)]{melnykov2016merging}
V.~Melnykov.
\newblock Merging mixture components for clustering through pairwise overlap.
\newblock \emph{Journal of Computational and Graphical Statistics}, 25\penalty0
  (1):\penalty0 66--90, 2016.

\bibitem[Melnykov and Maitra(2010)]{melnykov2010finite}
V.~Melnykov and R.~Maitra.
\newblock Finite mixture models and model-based clustering.
\newblock \emph{Statistics Surveys}, 4:\penalty0 80--116, 2010.

\bibitem[Melnykov and Michael(2019)]{melnykov2019clustering}
V.~Melnykov and S.~Michael.
\newblock Clustering large datasets by merging k-means solutions.
\newblock \emph{Journal of Classification}, 37\penalty0 (1):\penalty0 1--27,
  2019.

\bibitem[{\"O}zgenel and Sorgu{\c{c}}(2018)]{ozgenel2018performance}
{\c{C}}.~F. {\"O}zgenel and A.~G. Sorgu{\c{c}}.
\newblock Performance comparison of pretrained convolutional neural networks on
  crack detection in buildings.
\newblock In \emph{ISARC. Proceedings of the International Symposium on
  Automation and Robotics in Construction}, volume~35, pages 1--8. IAARC
  Publications, 2018.

\bibitem[Peel and McLachlan(2000)]{PeelMcLachlan:2000}
D.~Peel and G.~McLachlan.
\newblock Robust mixture modelling using the t distribution.
\newblock \emph{Statistics and Computing}, 10:\penalty0 339–348, 2000.

\bibitem[Peterson et~al.(2018)Peterson, Ghosh, and Maitra]{peterson2018merging}
A.~D. Peterson, A.~P. Ghosh, and R.~Maitra.
\newblock Merging k-means with hierarchical clustering for identifying
  general-shaped groups.
\newblock \emph{Stat}, 7\penalty0 (1):\penalty0 e172, 2018.

\bibitem[Ramsay and Silverman(2005)]{ramsay2005functional}
J.~Ramsay and B.~Silverman.
\newblock \emph{Functional Data Analysis}.
\newblock Springer, 2nd edition, 2005.

\bibitem[Riani et~al.(2015)Riani, Cerioli, Perrotta, and
  Torti]{riani2015simulating}
M.~Riani, A.~Cerioli, D.~Perrotta, and F.~Torti.
\newblock Simulating mixtures of multivariate data with fixed cluster overlap
  in fsda library.
\newblock \emph{Advances in Data Analysis and Classification}, 9\penalty0
  (4):\penalty0 461--481, 2015.

\bibitem[Rousseeuw(1984)]{Rouss84LMS}
P.~J. Rousseeuw.
\newblock Least median of squares regression.
\newblock \emph{Journal of the American Statistical Association}, 79\penalty0
  (388):\penalty0 871--880, 1984.

\bibitem[Rousseeuw and Croux(1993)]{rousseeuw1993alternatives}
P.~J. Rousseeuw and C.~Croux.
\newblock Alternatives to the median absolute deviation.
\newblock \emph{Journal of the American Statistical association}, 88\penalty0
  (424):\penalty0 1273--1283, 1993.

\bibitem[Rousseeuw and Van~Zomeren(1990)]{rousseeuw1990unmasking}
P.~J. Rousseeuw and B.~C. Van~Zomeren.
\newblock Unmasking multivariate outliers and leverage points.
\newblock \emph{Journal of the American Statistical association}, 85\penalty0
  (411):\penalty0 633--639, 1990.

\bibitem[S{\'a}nchez et~al.(2009)S{\'a}nchez, Garc{\'i}a, Mayo, L{\'o}pez, and
  Hornero]{sanchez2009retinal}
C.~I. S{\'a}nchez, M.~Garc{\'i}a, A.~Mayo, M.~I. L{\'o}pez, and R.~Hornero.
\newblock Retinal image analysis based on mixture models to detect hard
  exudates.
\newblock \emph{Medical Image Analysis}, 13\penalty0 (4):\penalty0 650--658,
  2009.

\bibitem[Smiti and Elouedi(2012)]{smiti2012dbscan}
A.~Smiti and Z.~Elouedi.
\newblock {DBSCAN-GM}: An improved clustering method based on gaussian means
  and {DBSCAN} techniques.
\newblock In \emph{2012 IEEE 16th International Conference on Intelligent
  Engineering Systems (INES)}, pages 573--578. IEEE, 2012.

\bibitem[Torti et~al.(2021)Torti, Riani, and Morelli]{torti2021semiautomatic}
F.~Torti, M.~Riani, and G.~Morelli.
\newblock Semiautomatic robust regression clustering of international trade
  data.
\newblock \emph{Statistical Methods \& Applications}, pages 1--32, 2021.

\bibitem[Vassilvitskii and Arthur(2006)]{vassilvitskii2006k}
S.~Vassilvitskii and D.~Arthur.
\newblock K-means++: The advantages of careful seeding.
\newblock In \emph{Proceedings of the 18th annual ACM-SIAM symposium on
  Discrete algorithms}, pages 1027--1035, 2006.

\bibitem[Von~Luxburg(2007)]{von2007tutorial}
U.~Von~Luxburg.
\newblock A tutorial on spectral clustering.
\newblock \emph{Statistics and Computing}, 17\penalty0 (4):\penalty0 395--416,
  2007.

\bibitem[Von~Luxburg and Ben-David(2005)]{von2005towards}
U.~Von~Luxburg and S.~Ben-David.
\newblock Towards a statistical theory of clustering.
\newblock In \emph{Pascal workshop on statistics and optimization of
  clustering}, pages 20--26. Citeseer, 2005.

\bibitem[Zhang et~al.(2015)Zhang, Zhao, Zhang, and He]{zhang2015urbancps}
D.~Zhang, J.~Zhao, F.~Zhang, and T.~He.
\newblock {UrbanCPS}: a cyber-physical system based on multi-source big
  infrastructure data for heterogeneous model integration.
\newblock In \emph{Proceedings of the ACM/IEEE Sixth International Conference
  on Cyber-Physical Systems}, pages 238--247, 2015.

\bibitem[Zhang et~al.(2016)Zhang, Yang, Zhang, and Zhu]{zhang2016road}
L.~Zhang, F.~Yang, Y.~D. Zhang, and Y.~J. Zhu.
\newblock Road crack detection using deep convolutional neural network.
\newblock In \emph{2016 IEEE international conference on image processing
  (ICIP)}, pages 3708--3712. IEEE, 2016.

\end{thebibliography}


\newpage
\appendix


\section*{Appendix}
\label{sec:app}
\renewcommand{\thefigure}{A\arabic{figure}}
\setcounter{figure}{0}
\stepcounter{section}

The supplementary material includes details and additional experiments. 
The source code to replicate our simulation and application studies is available upon request.

\subsection{Semi-Automated Monitoring Procedure}

The semi-automated monitoring procedure relies on the \verb+tclusteda+ function provided in the \verb+FSDA MATLAB Toolbox+.
Specifically, 
for tk-merge, we set the restriction factor for TCLUST equals to 1 in order to identify spherical components with approximately the same volume. For the more flexible approach TC-merge, we set the restriction factor for TCLUST equals to 64, which allows one to identify elliptical components with different orientation and volume.
In both cases, we reduce the trimming proportion $ \alpha $ from 40\% to 0\%, with a step size of 5\%. 
At each trimming level we compute the ARI, and select the partition maximizing it; see Figure~\ref{fig:mobility2} for an example.

\subsection{Further Details for the Simulation Studies}
\label{app:sim}

\begin{figure}[ht]
    		\centering
    		\begin{subfigure}{.5\textwidth}
    			\centering
    			\includegraphics[width=1\linewidth]{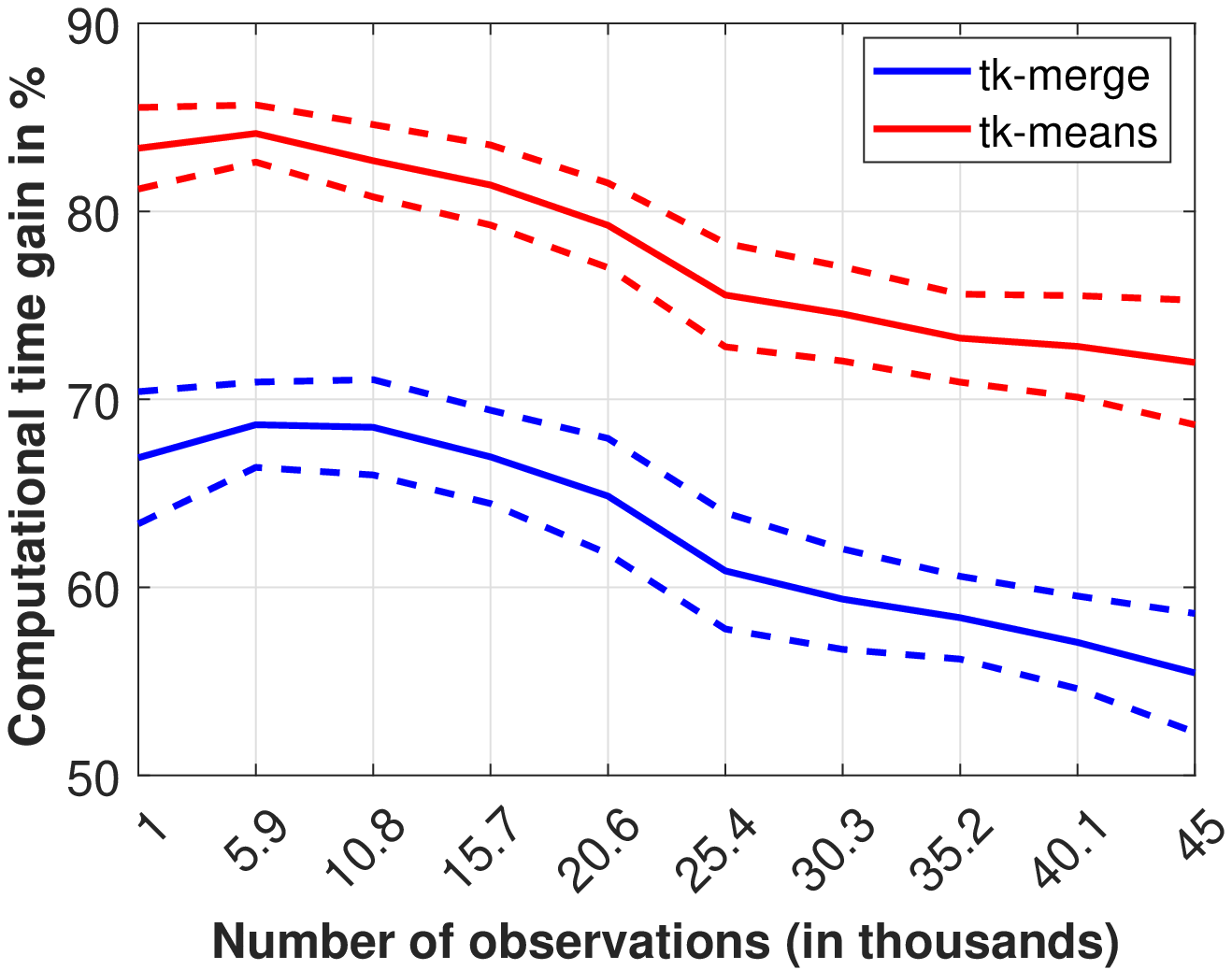}
    		\end{subfigure}%
    		\begin{subfigure}{.5\textwidth}
    			\centering
			\includegraphics[width=1\linewidth]{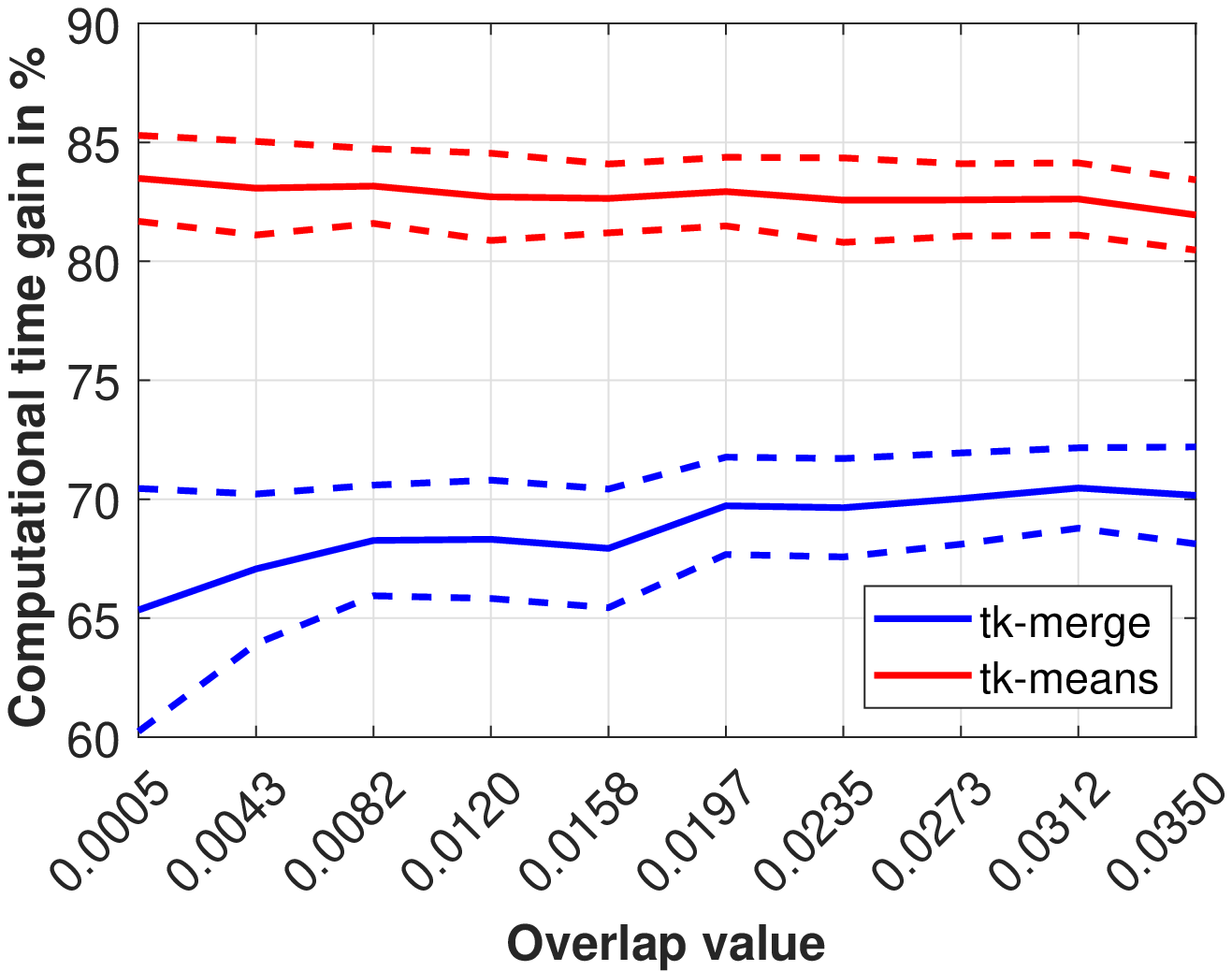}
    		\end{subfigure}
    		\caption{Median percentage computational gain for tk-merge and tk-means with respect to TCLUST, for simulation scenario 1 (left panel) and simulation scenario 2 (right panel) across 100 replications.
    		Dashed lines represent $ \text{median}\pm S_n$.
    		}
    		\label{fig:timerel}
    	\end{figure}

    Figure~\ref{fig:timerel} shows the percentage computing time gain of tk-merge and tk-means with respect to TCLUST
    for our simulation scenarios 1 (left panel) and 2 (right panel).
    Specifically, we represent medians and $S_n$'s of the percentage gain, which is computed as
    $
    | (t_i - t_T) | / t_T \times 100
    $, where $t_i$ denotes the computing time for tk-merge or tk-means, and $t_T$ is the one for TCLUST.
    	
    	      	\begin{figure}[ht]
        		\centering
        		\begin{subfigure}{.33\textwidth}
        			\centering
        			\includegraphics[width=1\linewidth]{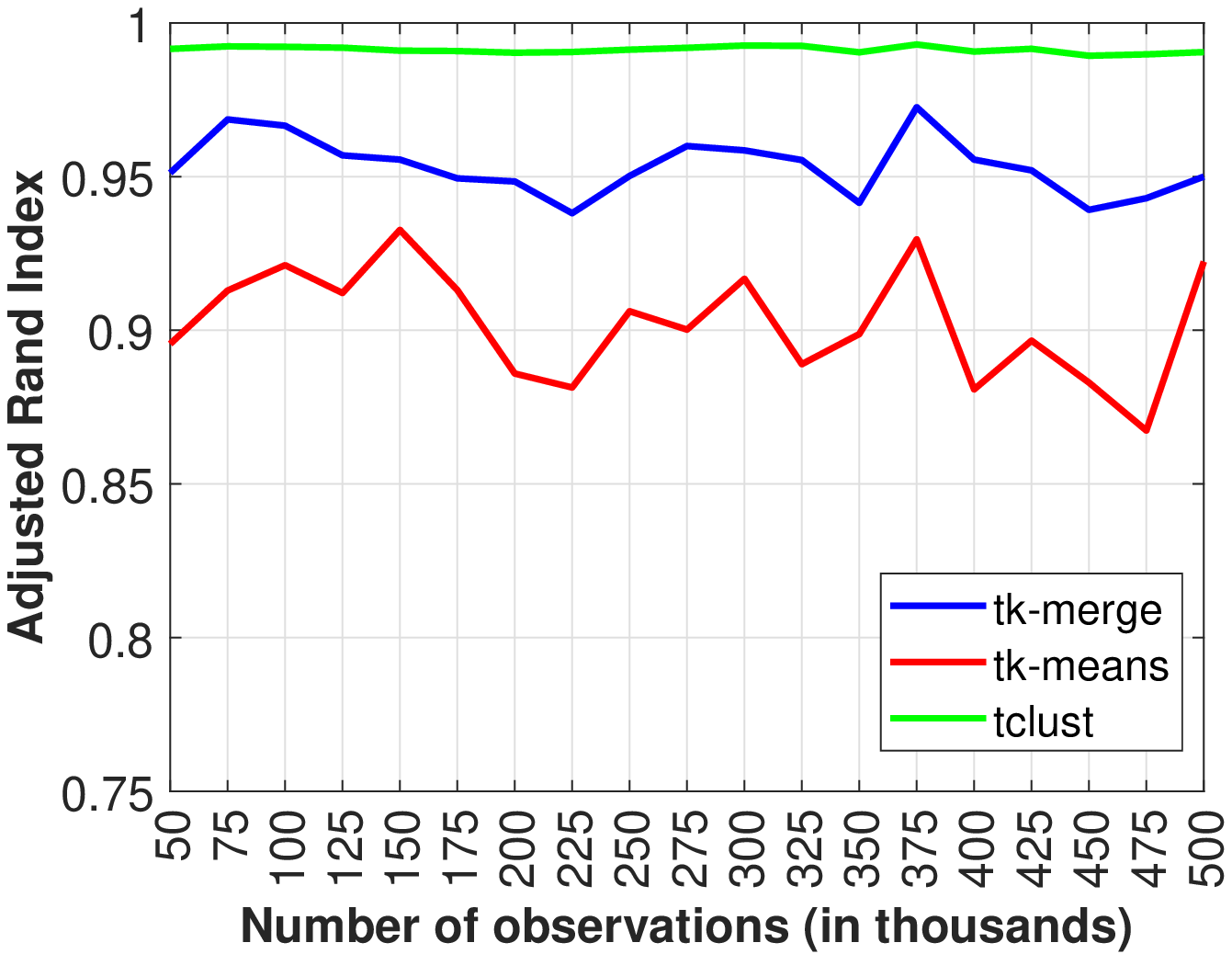}
        		\end{subfigure}%
        		\begin{subfigure}{.33\textwidth}
        			\centering
    			\includegraphics[width=1\linewidth]{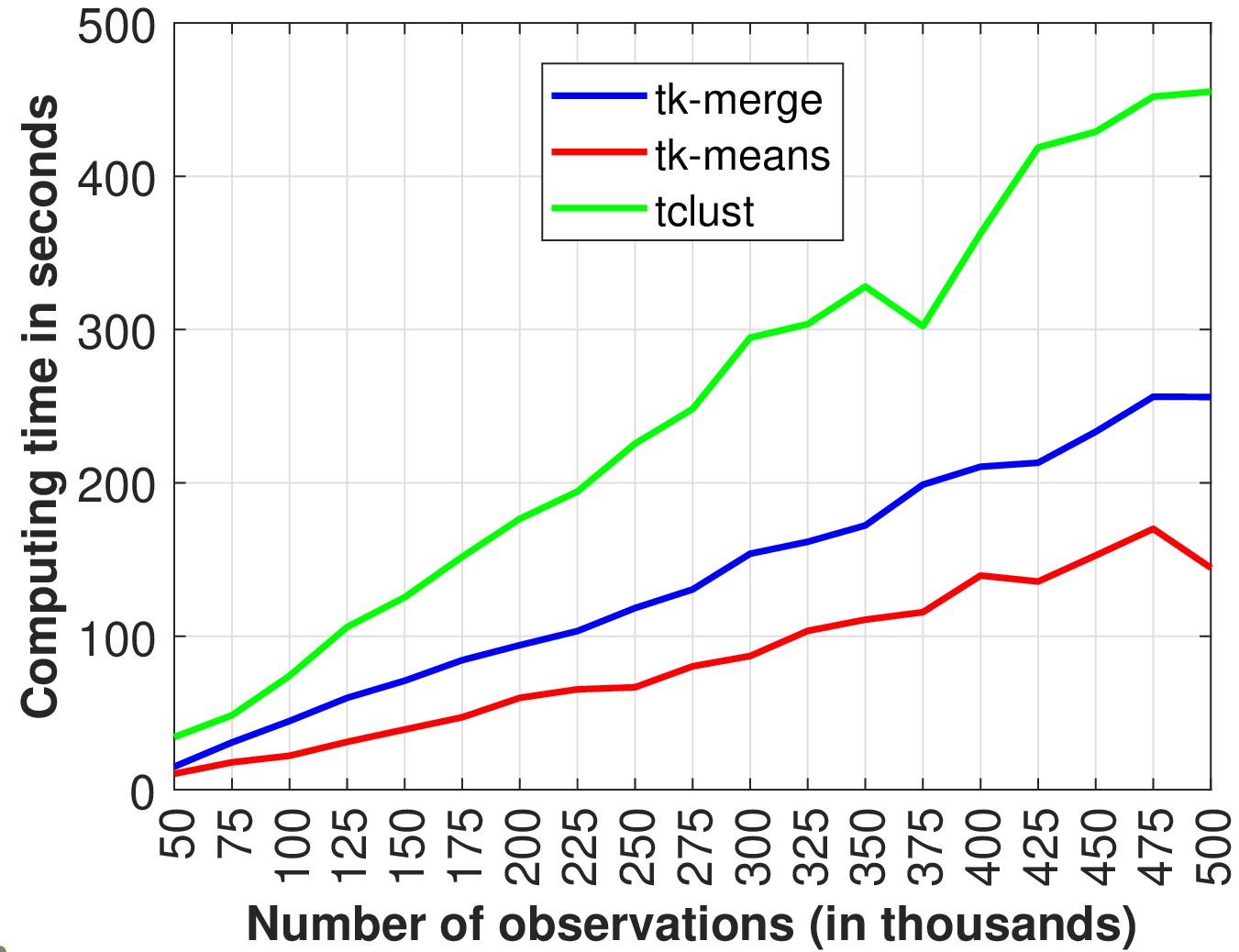}
        		\end{subfigure}
        		\begin{subfigure}{.33\textwidth}
        			\centering
    			\includegraphics[width=1\linewidth]{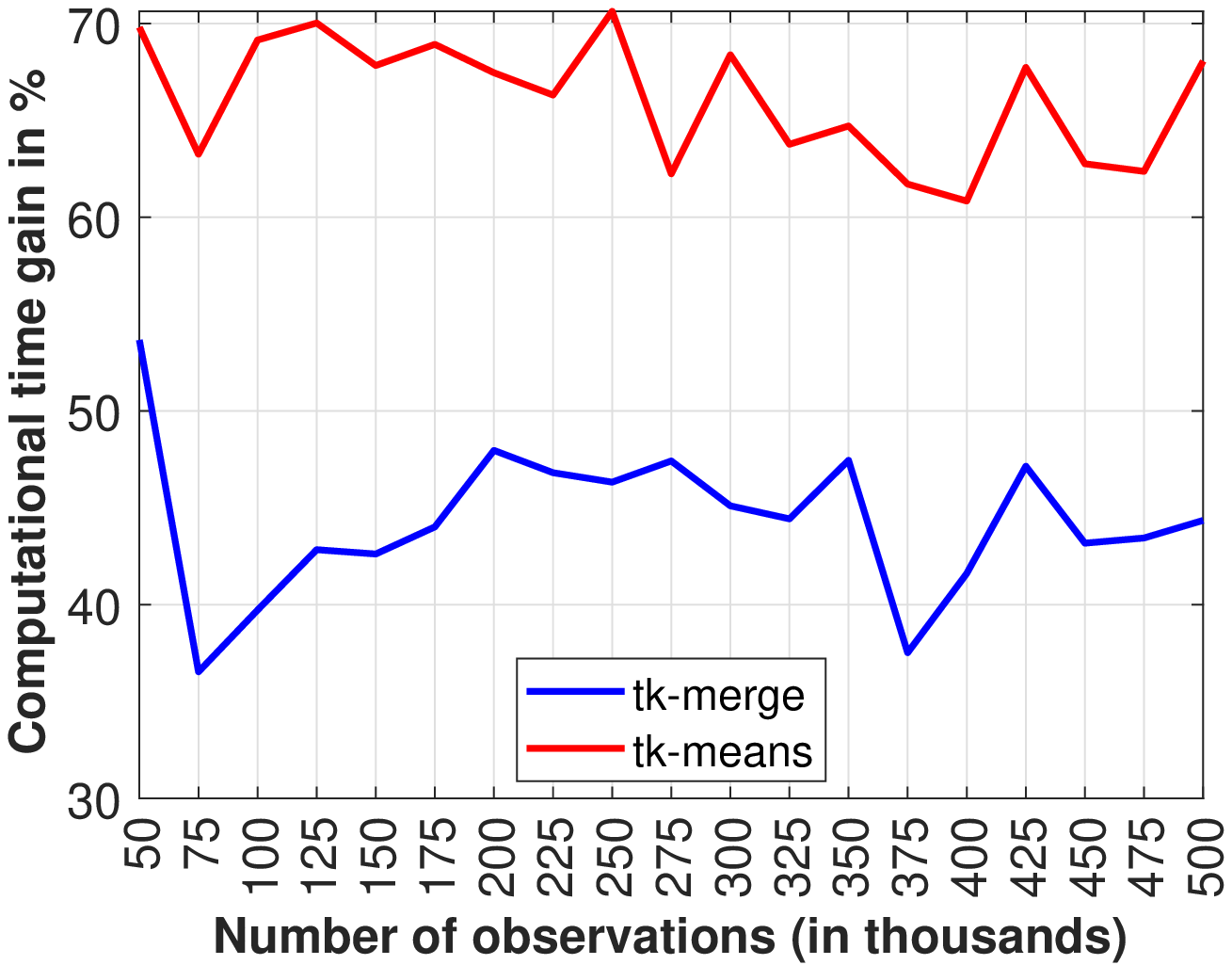}
        		\end{subfigure}
        		\caption{Median ARI (left), computing time (central), and percentage computational gain (right) across methods for simulation scenario 1 with larger sample sizes and 3 replications.}
        		\label{fig:largeN}
        	\end{figure}

\subsection{Further Details for the Application Studies}
\label{app:application}

\paragraph{Human Mobility}
    
\begin{figure}
        		\centering
        		\begin{subfigure}{.5\textwidth}
        			\centering
        			\includegraphics[width=1\linewidth]{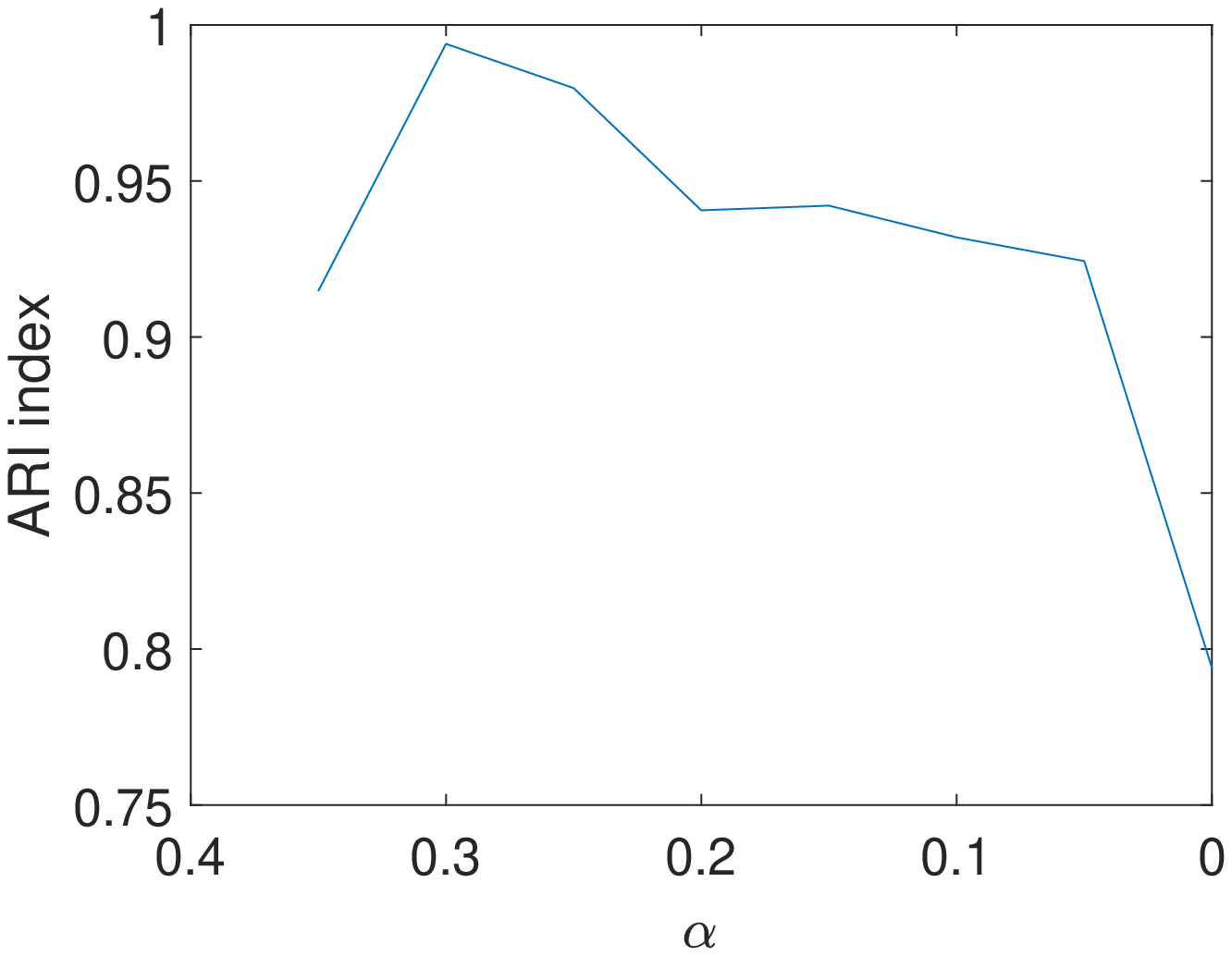}
        		\end{subfigure}%
        		\begin{subfigure}{.5\textwidth}
        			\centering
    			\includegraphics[width=1\linewidth]{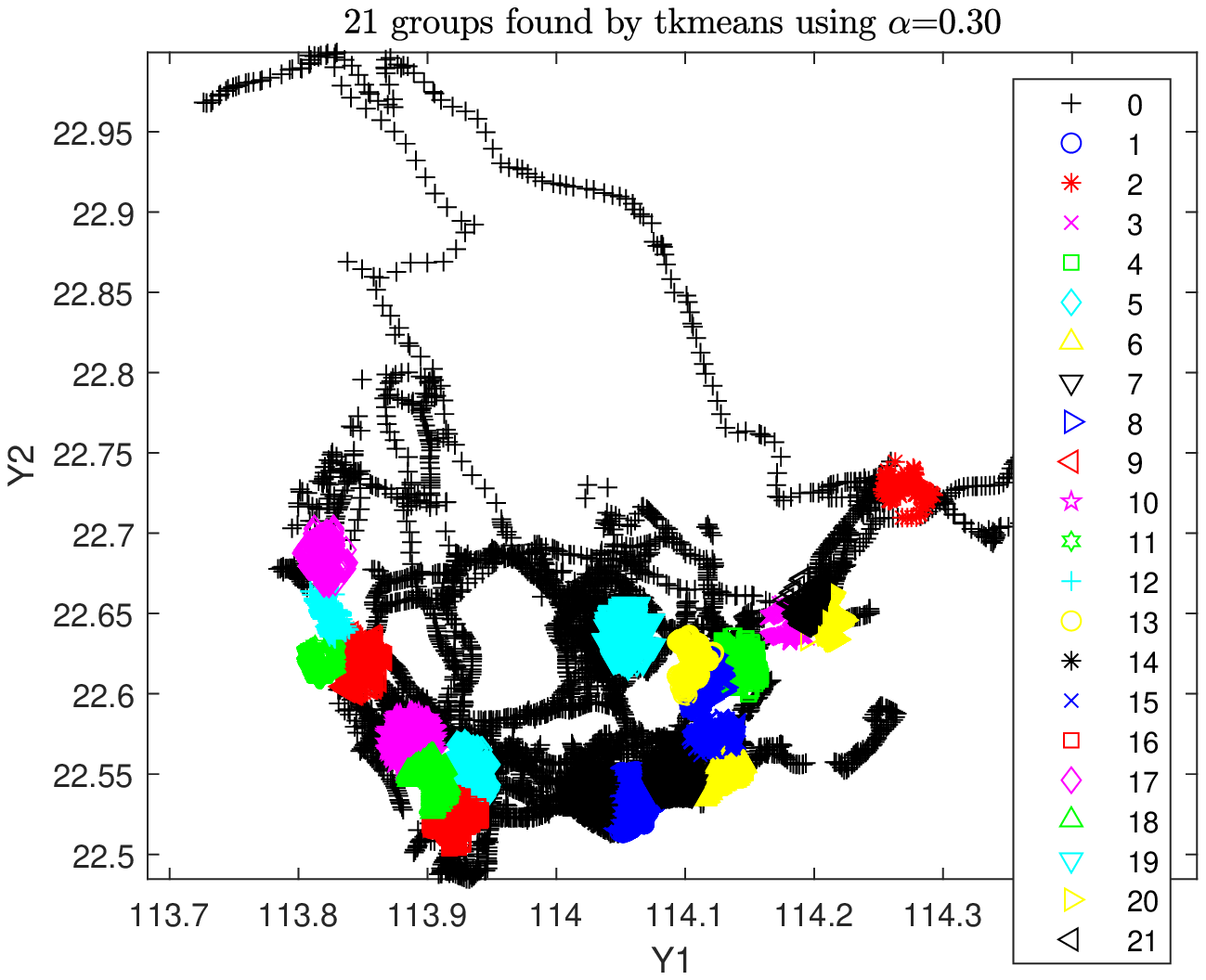} 
        		\end{subfigure}
        		\caption{Example 2. ARI (left) and tk-means components with maximum ARI (right), where outliers are labeled as 0 and marked with a plus sign.}
        		\label{fig:mobility2}
        	\end{figure}

\begin{figure}
    			\centering
    			\includegraphics[width=1\linewidth]{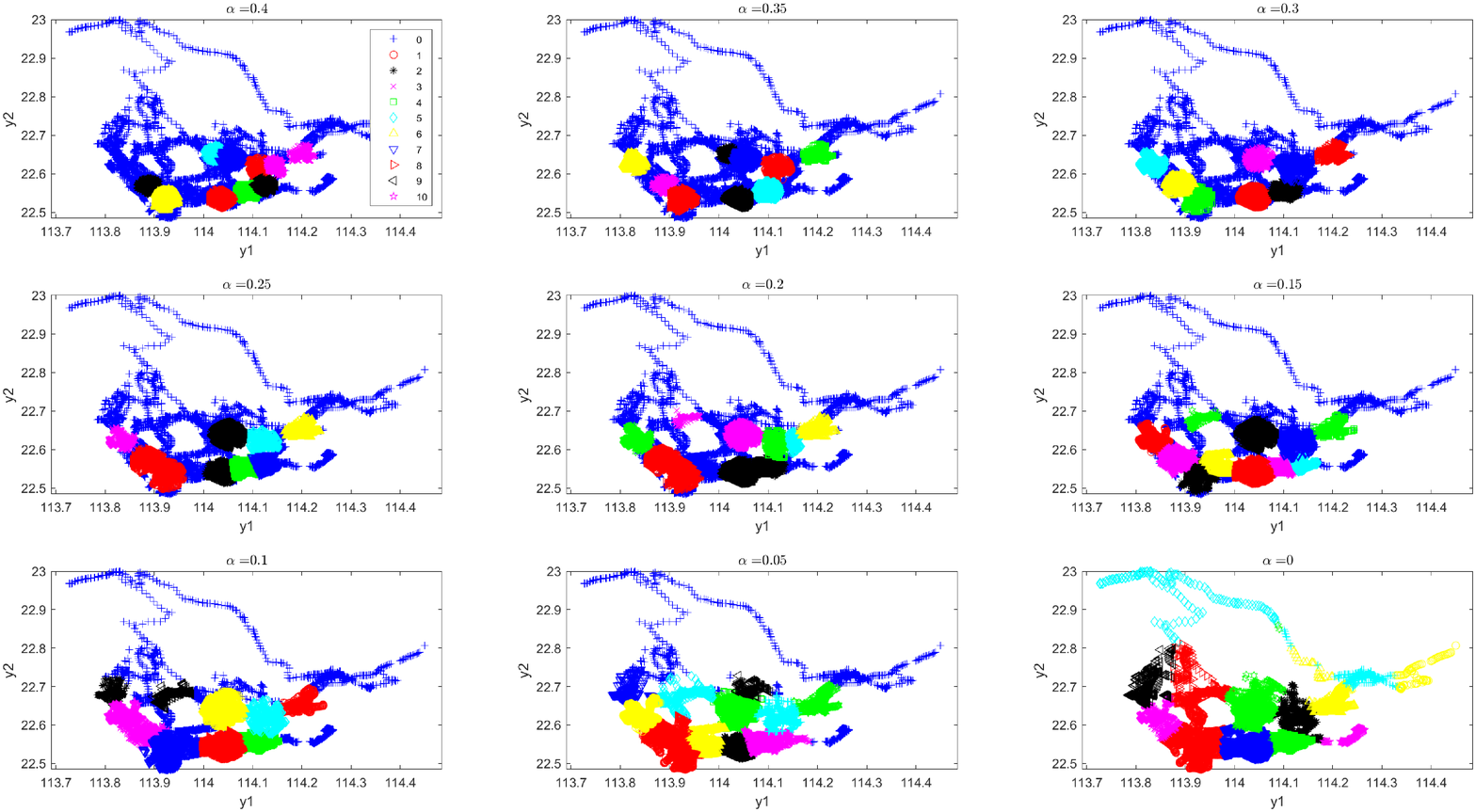}
    			\caption{Example 2. Components identified by tk-means across different trimming levels.}
    			\label{fig:mobility3}
        	\end{figure}

    The left panel of Figure~\ref{fig:mobility2} shows our monitoring procedure  across different trimming levels $\alpha$ for tk-means. 
    Here we used $k  = 21 \approx 2 \log n $ and the maximum ARI is reached for $ \alpha = 0.3$,  which is the value used in the main text, as well as to produce the result on the right panels of Figure~\ref{fig:mobility2}.
    Figure~\ref{fig:mobility3} shows the corresponding partitions obtained by tk-merge across different trimming levels.
    Its visual inspection further supports  the use of $ \alpha = 0.3$ discussed above.

\paragraph{Retinographic Study}
    
\begin{figure*}
            \centering
            \begin{subfigure}[b]{0.45\textwidth}
                \centering
                \includegraphics[width=1\textwidth]{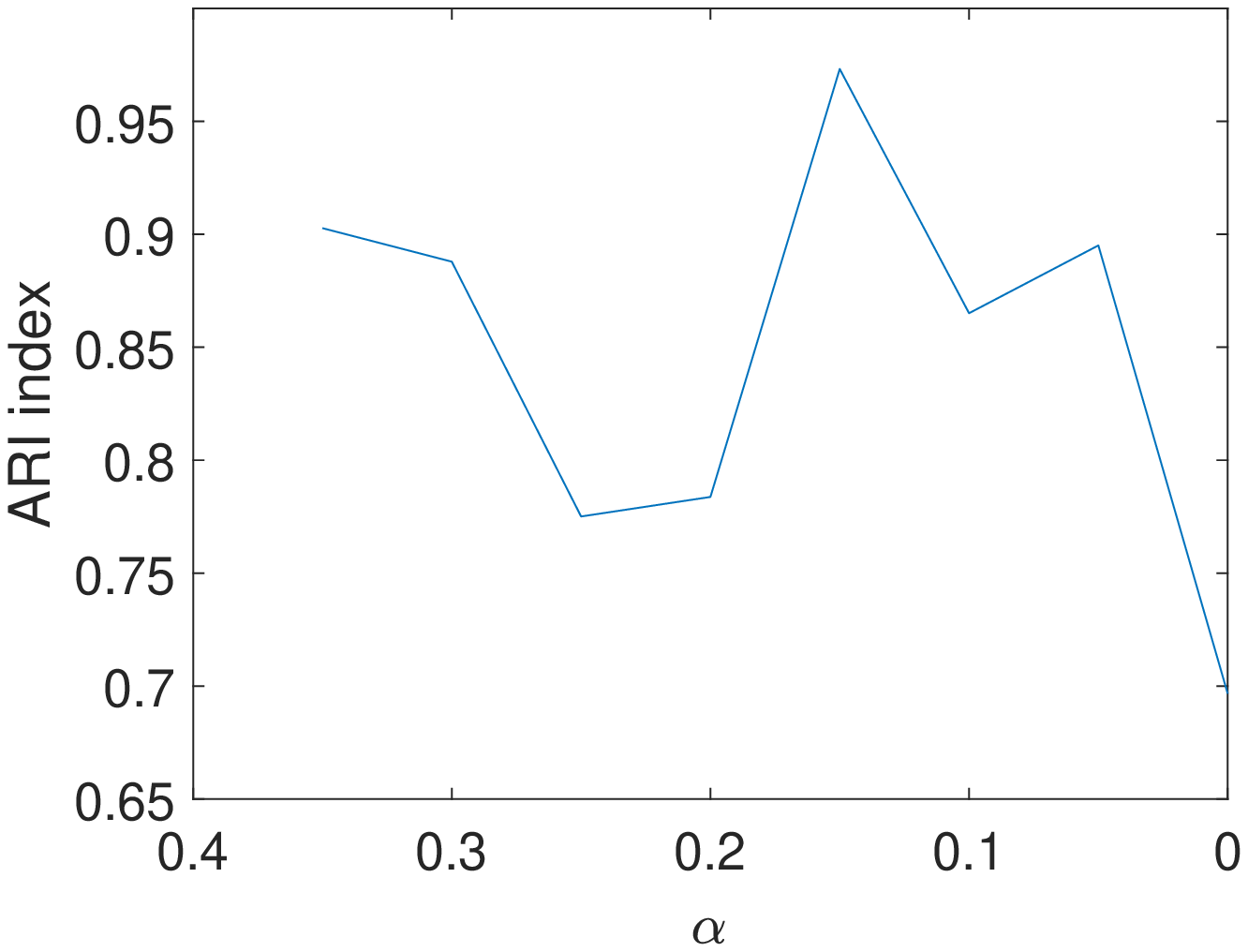}
            \end{subfigure}
            \begin{subfigure}[b]{0.45\textwidth}  
                \centering 
                \includegraphics[width=1\textwidth]{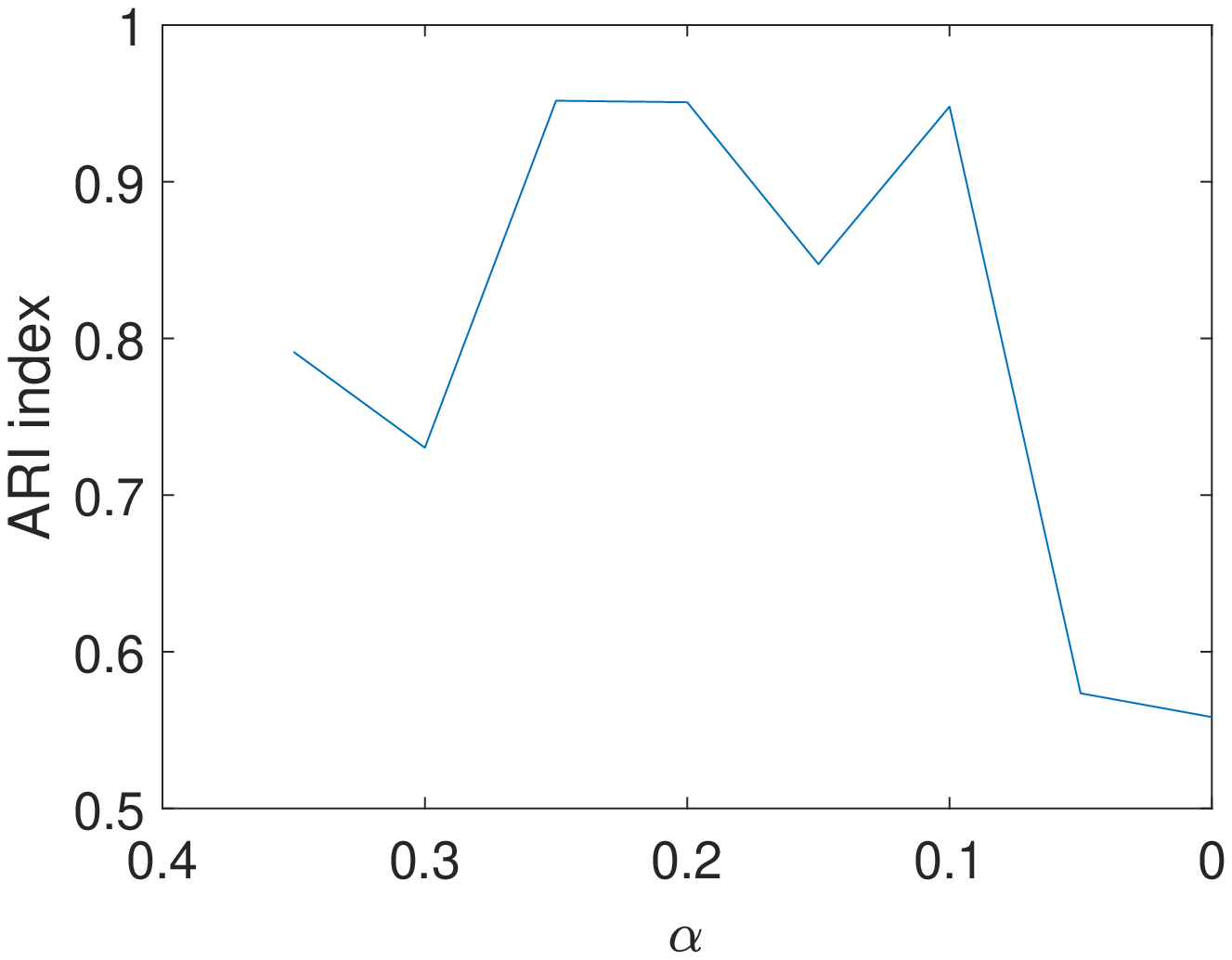}
            \end{subfigure}
            \vskip\baselineskip
            \begin{subfigure}[b]{0.45\textwidth}   
                \centering 
                \includegraphics[width=1\textwidth]{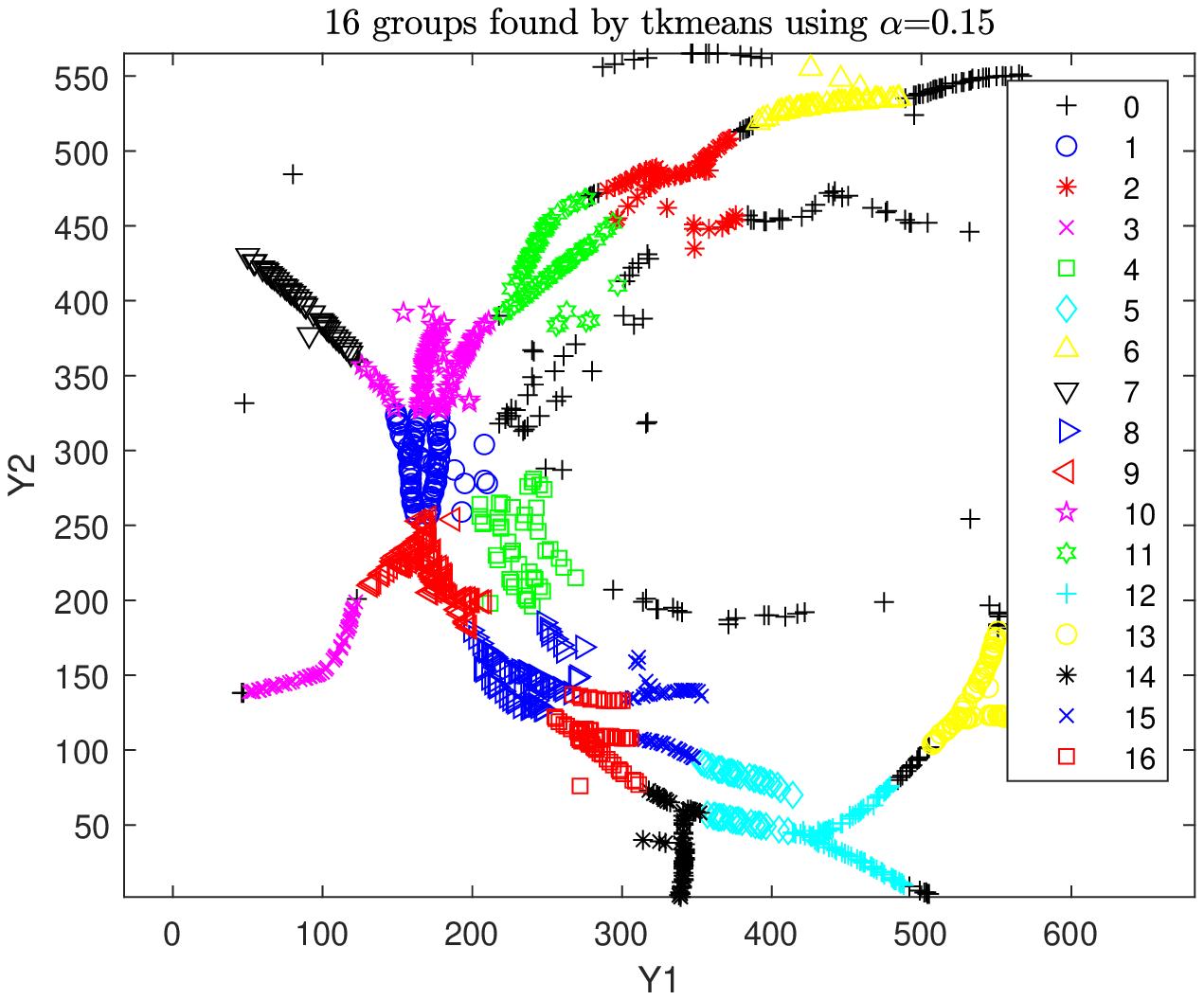}
            \end{subfigure}
            \begin{subfigure}[b]{0.45\textwidth}   
                \centering 
                \includegraphics[width=1\textwidth]{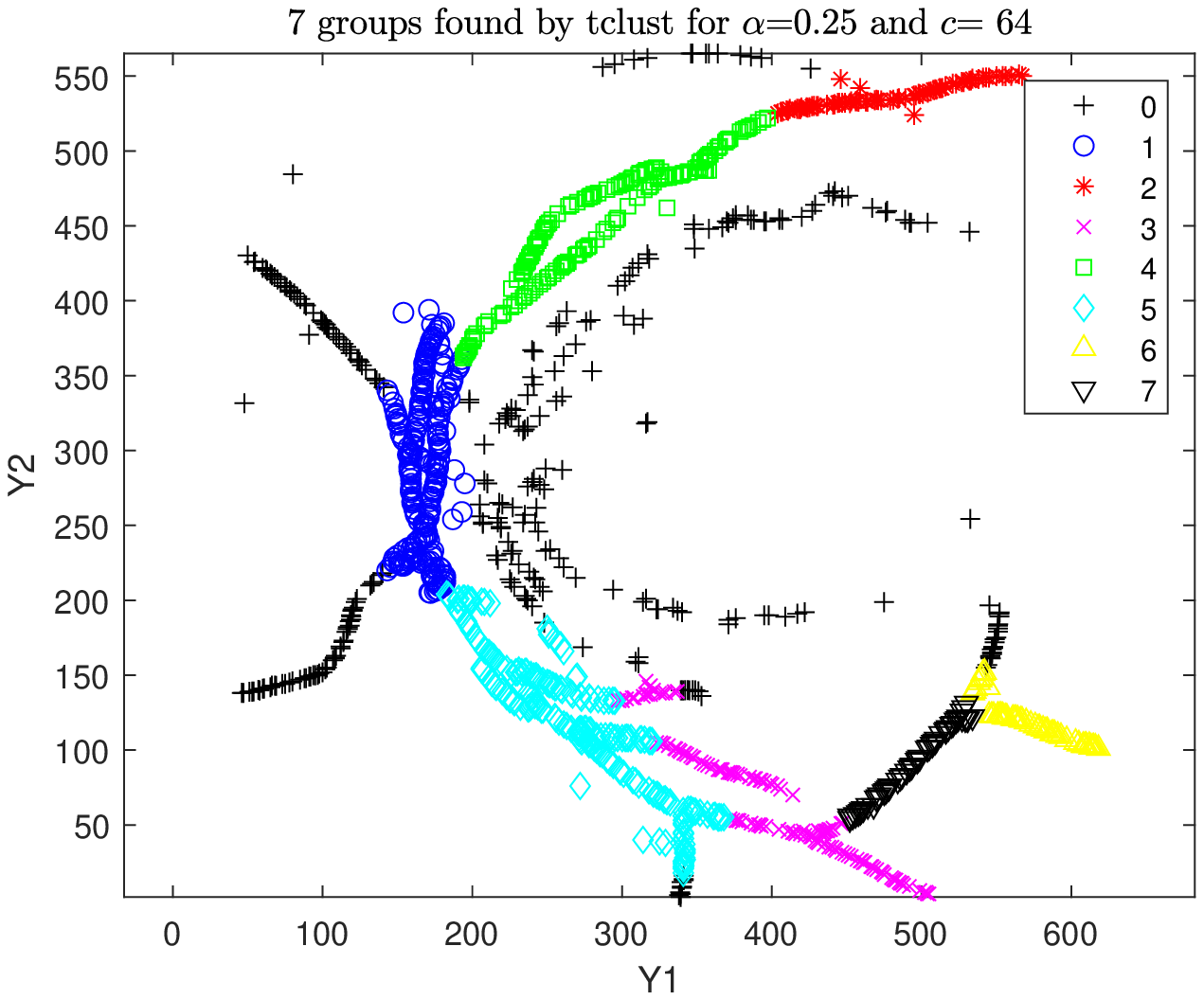}
            \end{subfigure}
            \caption{Example 3. ARI (top) and components with maximum ARI (bottom, where outliers are labeled as 0 and marked with a plus sign) for tk-means (left) and TCLUST (right).} 
            \label{fig:retina2}
        \end{figure*}

\begin{figure*}
            \centering
            \includegraphics[width=1\textwidth]{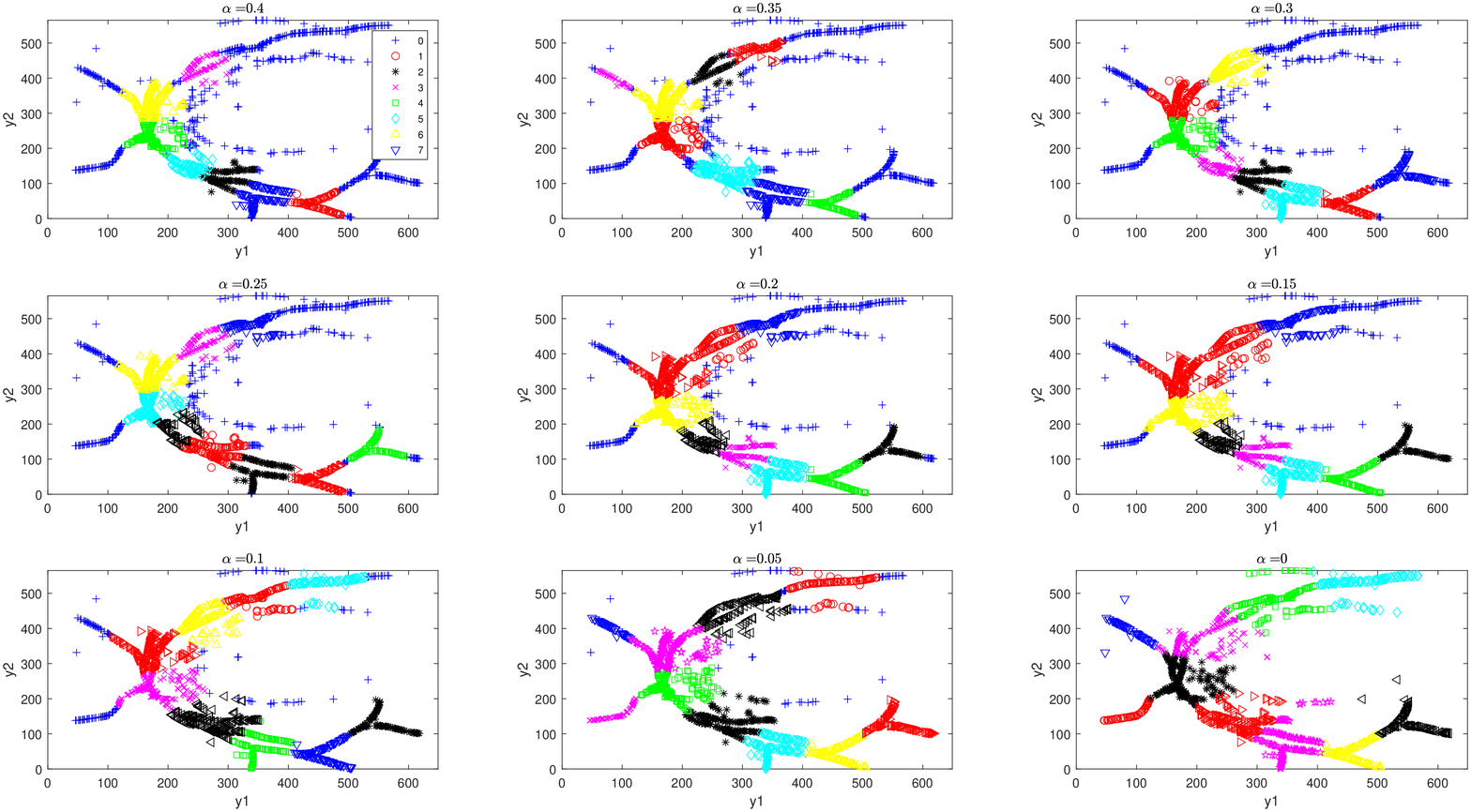}
            \caption{Example 3. Components identified by tk-means across different trimming levels.}
            \label{fig:retina3}
        \end{figure*}

\begin{figure*}
            \centering
            \includegraphics[width=1\textwidth]{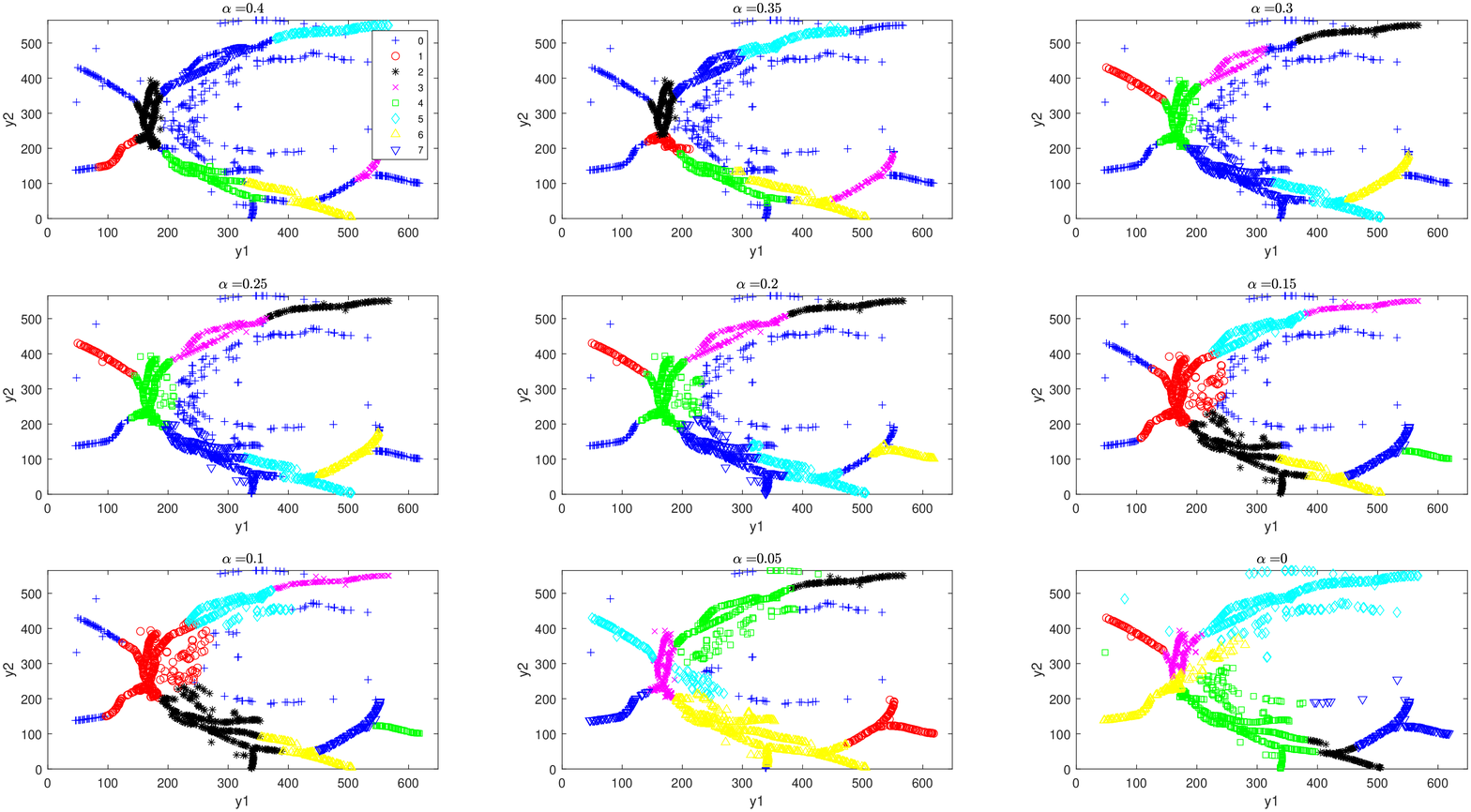}
            \caption{Example 3. Components identified by TCLUST across different trimming levels.} 
            \label{fig:retina4}
        \end{figure*}

    Figure~\ref{fig:retina2} shows ARI (top) and the partition with maximum ARI (bottom) for tk-means (left) and TCLUST (right) using $k=16 \approx 2 \log n$ with and $k=7 \approx \log n$, respectively.
    The highest ARI corresponds to $ \alpha = 15\% $ for tk-merge and 
    $ \alpha = 25\% $ for TC-merge, and thus we used these values in the main text. 
    Figure~\ref{fig:retina3} and Figure~\ref{fig:retina4} show the partitions across different trimming levels provided by tk-means and TCLUST, respectively.
    Also here a visual inspections supports the use of the solutions with maximum ARI.
        
\end{document}